%% file: thesis.tex
\documentclass[11pt,openright,twoside]{report}

\usepackage{amssymb}
\usepackage{cite}
\usepackage{epsf}
\usepackage{latexsym}
\usepackage{setspace}

\newcommand{\di}{\mbox{$i\!\!\not\!\!D$}}

\newcommand{\sib}{\mbox{$\overline{\psi}$}}
\newcommand{\ssi}{\mbox{$\langle\overline{\psi}\psi\rangle$}}

\newcommand{\ch}{$\chi$SB}

\begin{document}
\begin{titlepage}
\begin{center} 
{\Huge{\bf Topology and chiral symmetry breaking in QCD}} \vspace{12pt} \\ 
\vspace{3cm} 
{\LARGE Ujjawal Sharan} \\
{\large Keble College}\\
\vspace{2cm} 
Theoretical Physics\\
Department of Physics\\
\vspace{2cm}
\begin{center}
{\Large\font\oxcrest=oxcrest40
\oxcrest\char'01}
\end{center}
\vspace{1cm}
Thesis submitted for the degree of\\
Doctor of Philosophy\\
in the\\
University of Oxford\\
Trinity 1999\\
\end{center} 
\end{titlepage}

\thispagestyle{empty}
\cleardoublepage
\singlespacing
\include{abs}

\thispagestyle{empty}
\cleardoublepage
\pagenumbering{roman}
\vspace*{8cm}
\begin{center}
\begin{quote}
\it Live by the foma\footnote{Harmless untruths} that make you brave
and kind and healthy and happy.\\
\end{quote}
\end{center}
\hspace*{9cm}Cat's Cradle\\
\hspace*{9cm}Kurt Vonnegut\\

\cleardoublepage
\begin{center}
{\bf Acknowledgements}
\end{center}
I would like to thank my supervisor Mike Teper for his curious
inability to run out of ideas, and, his ability to hide his shock at
some of the horrors inflicted upon physics by his student. I would
also like to thank my friends at this university, especially Jose, The
Shipyard Son, Danton and The Mighty Metikas, as well as my friends
from The Other Place, especially Mole, Mao, Eggs, Jon and Frog, for
their ceaseless distraction. Last but certainly not least I wish to
thank my family, without whom, it would all seem a little pointless.
\\
{\noindent I am grateful to PPARC for their financial support (Grant
No.~96314624).}

\cleardoublepage
\doublespacing
\tableofcontents
\include{intro}
\include{model}
\include{univ}
\include{latt}
\include{unq}
\include{conc}

\appendix
\include{ovlap}
\include{jackk}

\include{biblio}
\end{document}

%% file: abs.tex
\newpage
\thispagestyle{empty}
\begin{center}
{\LARGE Topology and chiral symmetry breaking in QCD}\\
\vspace{0.5cm}
Ujjawal Sharan\\
Keble College\\
\vspace{0.25cm}
Theoretical Physics\\
University of Oxford\\
\vspace{0.5cm}
{\bf Abstract}\\
\end{center}
We study the influence of certain topological objects, known as
instantons, on the eigenvalue spectrum of the Dirac operator. We
construct a model of the vacuum based on instanton degrees of
freedom. We use this model to construct a representation of the Dirac
operator for an arbitrary configuration of instantons. The
representation is constructed for the subspace of the full Hilbert
space spanned by the zero modes of the individual objects. The model
is by necessity, approximate, though it does incorporate the important
symmetries of the underlying field theory. The model also reproduces
classical results in the appropriate limits.

We find that generic instanton ensembles lead to an accumulation of
eigenvalues around zero and hence break chiral symmetry. The
eigenvalue spectrum is divergent, however, as the eigenvalue $\lambda
\rightarrow 0$. This leads to a divergent chiral condensate in
quenched QCD, and hence, shows the theory to be pathological. In full
QCD however, we find that the parameters of the divergence are quark
mass dependent. This dependence leads to chiral symmetry breakdown
with a finite quark condensate for both $N_{f}=1$ and $N_{f}=2$. We
find the power of the divergence to be inversely related to the
density of instantons; in particular, the divergence is weak for high
density gases. Hence the importance of these results depends upon the
density of objects in the (quenched) QCD vacuum.

To investigate this, we study instanton ensembles derived by
``cooling'' lattice gauge configurations. We find chiral symmetry to
be broken as before. The spectrum, including the divergence, (and
hence, the chiral condensate) is strongly dependent upon the number of
cooling sweeps performed. Whether the problem lies with cooling or
with the identification of topological objects is yet to be resolved.
\begin{center}
Thesis submitted for the degree of Doctor of Philosophy\\
in the University of Oxford.\\
Trinity 1999
\end{center}

%% file: intro.tex
\chapter{Introduction}
\pagenumbering{arabic}

This thesis considers the effects of topological objects on the
eigenvalue spectrum of the Dirac operator, and the consequences of
this for chiral symmetry. These topological objects are known as
instantons, and, in this context, represent tunnelling between
distinct vacua of quantum chromodynamics (QCD). Tunnelling in the
quantum mechanical sense is a non-perturbative effect, it is missed
entirely by perturbation theory to all orders. The field of
non-perturbative QCD however, is renowned for the difficulty in
extracting exact analytical results. This leads practitioners to
pursue either approximations, or to perform ``brute force'' numerical
computations. We have chosen to follow both paths simultaneously.

We construct a ``simple'' model to describe instanton interactions and
their effect on the eigenvalue spectum of the Dirac operator. However,
this model is still too complicated to be tractable by analytical
means and we rely to a large extent on numerical simulations.  This
chapter comprises a brief overview of this field; we give a short
review of instanton fundamentals, how they come to influence the
spectrum of the Dirac operator, chiral symmetry (breaking) and its
implications, and how all these may be intimately
related. Chapter~\ref{ch:model} describes in detail the model we have
constructed, some of its properties, and its
limitations. Chapter~\ref{ch:univ} applies our model to generic
instanton configurations generated at random. We also carry out a
qualitative analysis of the validity (or lack thereof) of the
model. We then proceed in chapter~\ref{ch:latt} to apply our model to
``numerical snapshots'' of the quenched QCD vacuum generated by
UKQCD. These configurations exclude the effects of dynamical fermions;
there are no ``back-reactions'' from fermions on the gluonic
vaccum. In chapter~\ref{ch:unq} we incorporate the effect of fermions
within the limited scope of our model (whilst this should offer
qualitative information about the effect of light fermions upon the
spectral density, it is by no means equivalent to a full ``light
dynamical fermion QCD'' calculation). We give some conclusions on what
we have achieved, and, what has been been left undone, in
chapter~\ref{ch:conc}.

\section{Instanton fundamentals}
I am grateful, like so many other acolytes in the field of instanton
physics, for the pedagogical reviews given by Coleman~\cite{Coleman}
and Va\v{\i}nshte\v{\i}n \emph{et al.}~\cite{Vains}. I am also
indebted to the reviews on instantons in relation to chiral symmetry
breaking given by Diakanov~\cite{Diak} and Sch\"{a}fer \&
Shuryak~\cite{Shuryak-RMP}. The reader is referred to these works, and
the numerous references therein, for far greater detail than can be
accommadated in this introduction. The notation followed will be that
of Coleman.

\subsection{Instantons - A little history}

We recap a little of the folklore of instantons, before progressing to
the details. Instantons are solutions to the equations of motion of
SU(2) Yang-Mills theory in Euclidean spacetime. They were discovered
by Belavin, Polyakov, Shvarts \& Tyupkin~\cite{Belavin}, about 25
years ago. They are solutions of nontrivial topology; they have a
conserved number associated with their global, as opposed to their
local, characteristics. This number is known as the ``winding'' number
and has some important and beautiful mathematical properties.

One may question what role instantons play in nature, after all,
spacetime is Minkowskian, not Euclidean. Clarification of their
physical role was provided by, amongst others, Jackiw \&
Rebbi~\cite{Jackiw-Vac} and Callan, Dashen \& Gross~\cite{Callan}. The
physical picture is one where we have, for example, the trivial gauge
configuration ($A_{\mu}^{a} = 0$) on some spacelike hypersurface ($t =
t_{0}$). At some later time ($t = t_{1}$) we have a gauge
transformation of the initial trivial gauge configuration (so that the
stress energy tensor vanishes on the hypersurface $t = t_{1}$ as
well). This gauge transformation is a little different to that
commonly encountered. The ``small'' gauge transformations one normally
deals with, are those which may be continuously deformed to the
identity.  When we gauge fix, we partition the space of field
configurations by grouping together all configurations which are small
gauge transformations of one another. We then pick one representative
from each group. There are, however, other gauge transformations which
cannot be smoothly deformed to the identity. The field configurations
on the two hypersurfaces are related by such a ``large'' gauge
transformation. These large gauge transformations are characterised by
their winding number. So what we have is an evolution of the field
from vacuum to vacuum (by vacuum, I mean only that the field strength
vanishes on the hypersurface). The field however, cannot be vacuum
throughout the period from $t_{0}$ to $t_{1}$. This is because, for it
to remain vacuum as $t$ increases from $t_{0}$, the field
configuration needs to be a gauge transformation of the trivial
vacuum. This is, by definition, a ``small'' gauge transformation. It
cannot therefore be continuously deformed to the configuration on the
boundary at $t = t_{1}$.

So we have a non-vanishing field strength tensor in some region
bounded by the two hypersurfaces. What has in fact happened is that
the system has tunnelled between two disjoint vacua. Now tunnelling
events are associated with the theory in question being continued to
imaginary time (for instance the WKB approximation in quantum
mechanics, or even more relevantly, the double well problem using
kink--anti-kink configurations~\cite{Coleman}). The instanton is in
fact nothing other than the object in imaginary time associated with a
tunnelling event in real time.

Once it was realized that we had the possibility of tunnelling between
distinct vacua, it was apparent that the ground state of QCD was far
richer than previously imagined. By analogy with simple quantum
mechanics, the ground state became a linear combination of the
different vacua, a linear combination parameterised by a real number
$\theta$. This was the $\theta$-vacua which allowed 't~Hooft to break
the $U_{A}(1)$ axial symmetry without generating a Goldstone
boson~\cite{'tHooft}. This remains one of the great triumphs of
instanton physics. The fact that the would-be Goldstone boson, the
$\eta^{'}$, is massive, can also be related to
instantons~\cite{Witten,Veneziano}.

In time though, the initial flurry in QCD, for all things topological,
waned. The sucess of Polyakov in explaining confinement in certain
three-dimensional models using monopoles (see~\cite{Polyakov}) could
not be replicated for four-dimensional gauge theories using
instantons. Another problem lay with the fact that the instanton
weight grew with the size of the instanton (which we come to later),
and would in fact lead to a divergence when calculating the
contribution of a single instanton to the partition function. The
cutoff it would appear, is due to instanton interactions (so we do not
have objects of arbitrarily large size) but these proved difficult to
calculate.

If confinement seemed to be beyond instantons, then chiral symmetry it
seemed, was not~\cite{Caldi,Carlitz,Callan}. As we shall see, chiral
symmetry breakdown is a non-perturbative phenomenon responsible for
many of the properties of the light hadrons. In particular the reason
that the pions are light, is that they are approximate Goldstone
bosons associated with the spontaneous breakdown of an approximate
global symmetry. This breakdown is also responsible for the fact that
nearly massless quarks (we refer to the $u$ and $d$ quarks) generate a
dynamical mass perhaps two orders of magnitude greater than their
current masses. The evidence for instantons to be the mechanism for
the spontaneous breakdown of chiral symmetry is strong (though not
certain by any means). One of the aims of this thesis will be to
explore if generic models of instanton interactions break chiral
symmetry in QCD and an approximation to QCD known as quenched QCD (we
will on occasion abbreviate this to q-QCD). We explore these topics in
greater detail in the following.

\subsection{Instantons - A little mathematics}

We concentrate on instantons as objects in imaginary time. This is, as
noted previously, complementary to thinking of them as tunnelling
events in real time. The QCD action in four dimensional Euclidean
spacetime (${\mathbb R}^{4}$) for $N_{f}$ fermions is given by:

\begin{equation}
{\cal S}_{qcd}^{E} = \frac{1}{4g^{2}}\int
d^{4}x\,(F_{\mu\nu},F_{\mu\nu}) - \sum_{f=1}^{N_{f}}\int
d^{4}x\,\sib_{f}(\di - im_{f})\psi_{f}\ ,
\label{qcdact}
\end{equation}
where $F_{\mu\nu} = \partial_{\mu}A_{\nu} - \partial_{\nu}A_{\mu} +
[A_{\mu},A_{\nu}],\ A_{\mu} = gA_{\mu}^{a}T^{a}$ represents the gauge
field and $T^{a},\,a=1,\ldots,N_{c}^{2}-1$ are the generators of some
Lie group with $(T^{a}, T^{b}) = \delta^{ab}$. (In particular, $T^{a}
= -i\sigma^{a}/2$ for $SU(2)$, where $\sigma^{a}$ are the standard
Pauli matrices, and, $T^{a} = -i\lambda^{a}/2$ for $SU(3)$ where
$\lambda^{a}$ are the Gell-Mann matrices.) The Dirac operator $\di =
i\gamma_{\mu}(\partial_{\mu} + A_{\mu})$. We consider $SU(2)$
initially, and, generalize to $SU(3)$, the gauge group for QCD,
afterwards. The Dirac operator \di\ is Hermitean in this formulation,
in particular all eigenvalues are real. The partition function is
given by the functional integral over the gauge and fermion fields.

Let us consider gauge field configurations of finite action. As
pointed out by Coleman~\cite{Coleman}, we do so, not because gauge
field configurations of infinite action are unimportant, but because
we wish to do a semi-classical approximation for the partition
function. (It is clear however, that if we compute semi-classically
the effects of Gaussian perturbations around a gauge field
configuration of infinite action then the prefactor of $exp(-S/\hbar)$
for the classical configuration will trivially result in zero.) To
obtain a finite action for a given gauge field $A_{\mu}$, it must go
to zero, or a gauge transform thereof, sufficiently quickly at
infinity:

\begin{equation}
\lim_{r \rightarrow \infty} A_{\mu} = g\partial_{\mu}g^{-1} +
O(\frac{1}{r^{2}})\ ,
\end{equation}
where $g(x) \in SU(2)$. At infinity we therefore have a map from the
sphere at infinity $S^{3}$ to the gauge group $SU(2)$. We know from
homotopy theory that $\pi_{3}(SU(2)) = {\mathbb Z}$; such maps may be
labelled by an integer, and that maps associated with the same integer
may be smoothly deformed into one another, whereas no smooth
deformation takes us between maps labelled by different integers. This
integer is referred to as the winding number (or sometimes the
Pontryagin index). Na\"{\i}vely, the winding number measures the
number of times the sphere at infinity is mapped over the group
manifold. The homotopy result becomes plausible if we recall that the
group manifold of $SU(2)$ is in fact just $S^{3}$. The winding number
$Q$ for a gauge field can be computed as:

\begin{equation}
Q[A] = \frac{1}{32\pi^{2}}\int
d^{4}x\,(F_{\mu\nu},\widetilde{F}_{\mu\nu})\ ,
\label{winform}
\end{equation}
where the dual field strength tensor $\widetilde{F}_{\mu\nu} =
\frac{1}{2}\epsilon_{\mu\nu\lambda\sigma}F_{\lambda\sigma}$. The
trivial gauge field corresponds to winding number zero. Let us
consider how we may go about constructing a map of winding number
one. We wish to construct a map from Euclidean spacetime to
$SU(2)$. If we make the map independent of radial distance then we
will obtain a map from the unit sphere (or indeed the sphere at
infinity) to $SU(2)$. All we then require is that the map is a
bijection and by the na\"{\i}ve interpretation of the winding number
given above, we will have a map of winding number one. It is simple to
see that the following map admirably satisfies all our requirement.
\begin{equation}
x \longmapsto g_{1}(x) = \frac{x^{4} + i\underline{x}.\underline{\sigma}}{|x|}
\label{simpmap}
\end{equation}
It is therefore not too surprising to find that this is in fact a map
of winding number one (we can see this by noting that~\ref{winform} is
a total derivative, hence it is possible to calculate the winding
number on a sphere at infinity). A map of winding number $Q$ is given
by $(g_{1})^{Q}$, a result made plausible by spotting that this at
least satisfies the additivity in the integers of the winding number
as required. The simplest gauge field prescription satisfying our
requirements is therefore:

\begin{equation}
A_{\mu} = f(x^{2})g_{1}\partial_{\mu}g_{1}^{-1}
\end{equation}
where $g_{1}(x)$ is given by the map~\ref{simpmap}. We have a
reasonable ansatz for the form of a classical solution of winding
number one to the equations of motion; how do we solve for
$f(x^{2})$~?  One method is to simply solve the equations of motion
$D_{\mu}^{ab}F_{\mu\nu}^{b} = 0$, subtituting in our ansatz for the
gauge field. These are second order coupled partial differential
equations and hence not trivial to solve. Belavin
\emph{et al.}~\cite{Belavin} spotted that one could instead reduce the
problem to first order by using the Schwarz inequality to show that
the classical solution obeyed $F = \pm\widetilde{F}$ where the $\pm$
holds for fields with positive or negative winding number. Using this
insight they found that:

\begin{equation}
f(x^{2}) = \frac{x^{2}}{x^{2} + \rho^{2}}
\end{equation}
where $\rho$ is an arbitrary constant. A little checking confirms that
the gauge field can be written as

\begin{equation}
A_{\mu}^{a}(x) = \frac{2\eta_{a\mu\nu}x_{v}}{x^{2} + \rho^{2}}\ ,
\label{inst}
\end{equation}
where the 't Hooft symbol $\eta_{a\mu\nu}$ is given by:

\begin{equation}
\eta_{a\mu\nu} = \left\{ \begin{array}{ll}
\mbox{$\epsilon_{a\mu\nu}$} & \mbox{$\mu, \nu$} = 1,2,3\\
\mbox{$\delta_{a\mu}$} & \mbox{$\nu$} = 4\\
\mbox{$-\delta_{a\mu}$} & \mbox{$\mu$} = 4\ .
\end{array}
\right.
\end{equation}
This gauge field configuration is known as an instanton. The
corresponding gauge field configuration with winding number minus one
is called an anti-instanton and is given by the above formula but with
$\eta_{a\mu\nu}$ replaced by $\overline{\eta}_{a\mu\nu} =
\epsilon_{a\mu\nu} - \delta_{a\mu}\delta_{4\nu} +
\delta_{a\nu}\delta_{4\mu}$. We know from the homotopy result that the
winding number is additive in the integers, so for example an
approximate instanton--anti-instanton configuration may be smoothly
deformed to the trivial configuration and so on.

How do we parameterize an instanton uniquely~? A logical answer would
be to specify enough parameters to uniquely determine its gauge
field~\ref{inst}. We see immediately from~\ref{inst} that we must at
least specify a parameter $\rho$ for the object. This parameter can be
interpreted as the ``size'' of the object. Another arbitrary parameter
is the location of the centre of the object; equation~\ref{inst} is a
special case where the object is centred on the origin. An object with
centre $x_{c}$ is simply given by~\ref{inst} with $x \rightarrow
(x-x_{c})$. Equation~\ref{inst} is a special case of an underlying
principle in one further way. We see from~\ref{simpmap} that we began
with a map from a sphere at infinity to the gauge group, which for
instantons became a map between two spheres $S^{3}$. Why can we not
map from a point on one sphere to a different point on the second
i.e. a rotation~? Of course we can. (One can view it either as a
rotation of spacetime or the opposite rotation of colour space; one
can undo the effect of one, by a corresponding rotation on the other -
see~\cite{Jackiw-Con} for fascinating details.) We implement this by
the following: a point $x$ is mapped to a new $SU(2)$ group element
via $x \rightarrow g_{1}'(x) = Kg_{1}(x)$ where $K \in SU(2)$ is a
constant matrix. This leads to $A_{\mu}'(x) =
KA_{\mu}(x)K^{\dagger}$. We therefore can parameterize an instanton
with only eight real numbers, four for the location, one for the size
and three for the colour orientation.

\begin{equation}
A_{I-Classical} \equiv (x, \rho, K)
\label{coll-co}
\end{equation}
These eight numbers are referred to as the collective co-ordinates of
the instanton.

The classical instanton given by equation~\ref{inst} has action $S =
8\pi^{2}/g^{2}$. The object is therefore scale invariant (the action
is independent of the parameter $\rho$). It is also invariant under
the translations and colour rotations given above. It would be most
surprising and implausible if the action were to depend upon the
location of the single object in spacetime, or indeed, its colour
orientation. The classical action is therefore invariant under an
8-parameter family of deformations. We have to take care when we
calculate the one (anti-)instanton contribution to the partition
function, for the eigenvalues corresponding to these perturbations
must be zero.

The generalization from $SU(2)$ to $SU(N_{c})$ can be made fairly
simply. This is because of a theorem which states that, any continuous
mapping from $S^{3}$ to a general simple Lie group G can be
continuously deformed to an $SU(2)$ subgroup of G~\cite{Bott}. In
particular, there is such a thing as an $SU(3)$ instanton, and, it is
the $SU(2)$ instanton we have met earlier~! The main difference
between $SU(2)$ and $SU(3)$ concerns the number of collective
co-ordinates required to specify an instanton, and, the effect of this
on the one instanton contribution to the partition function.

The number of collective co-ordinates differs as we have more freedom
in the rotation in colour space. In particular, we require
$(N_{c}^{2}-1)$ parameters to specify the rotation matrix. However,
$(N_{c}-2)^{2}$ of those generators will not affect the ``corner''
where the $SU(2)$ instanton resides, hence we have $(4N_{c}-5)$
generators which rotate the instanton in colour space. There are,
therefore, a total of $4N_{c}$ collective co-ordinates, each of which
is associated with a zero eigenvalue when we evaluate the one
instanton contribution to the partition function.

We can evaluate the one instanton contribution to the partition
function as follows. We write the gauge field as $A_{\mu} =
A_{\mu}^{I} + a_{\mu}$, where $A_{\mu}$ is the quantum gauge field,
the classical instanton gauge field is denoted $A_{\mu}^{I}$ and
$a_{\mu}$ is a quantum fluctuation around the classical minimum. The
partition function is changed from a functional integral over all
fields $A_{\mu}$ to one over ``small'' perturbations $a_{\mu}$. By
``small'' perturbations we mean that the action is expanded to second
order only.  The first order term disappears as the instanton is the
classical minimum, leaving only a classical part
$\exp(-8\pi^{2}/g^{2})$ and the operator determinant from the second
order term. The eigenvalue spectrum (of the operator) contains
$4N_{c}$ zeroes, hence we integrate over these eigenfunction
coefficients separately (we in fact change variables from an
integration over these eigenfunction coefficients to an integration
over the collective coordinates). The net effect of all this is that
the classical formula for the weight of an instanton is modified to:

\begin{equation}
\frac{d{\cal Z}^{I}}{d^{4}x} \sim
\frac{d\rho}{\rho^{5}}(\rho\Lambda_{qcd})^{\frac{11}{3}N_{c}}
\end{equation}
The two things to note from this equation are:
\begin{itemize}
\item{The instanton weight diverges for large ${\rho}$. In practice it
is believed that instanton interactions cut off the integral.}
\item{The instanton distribution is determined accurately for small
$\rho$, in particular we note the instantons are distributed as
$\rho^{6}$ for $SU(3)$ gauge theory and $\rho^{3/2}$ for $SU(2)$ gauge
theory. This will be of concern to us when we are generating instanton
ensembles, as we wish the objects to have realistic size
distributions. We will find non-trivial effects due to the size
distribution of instantons in the vacuum.}
\end{itemize}

When we compute the functional integral in perturbation theory we are
only taking into account fluctuations around the trivial gauge field
$A_{\mu} = 0$. Instantons which represent tunnelling between distinct
vacua are missed in perturbation theory; this can be seen by the fact
that the prefactor $\exp(-8\pi^{2}/g^{2})$ which arises from the
classical instanton is zero to all finite orders of $g$.

\section{Chiral symmetry}
We turn our attention now to a seemingly unrelated topic, that of
chiral symmetry, and, why we believe it to be broken in QCD. This
symmetry is concerned with quarks in the massless limit, hence we are
mainly interested in the up and down quarks (we can extend this
symmetry to include the strange quark as well, though this is not as
good from a phenomenological point of view). We think of the $N_{f}$
flavours of light quarks as having equal mass $m_{f} = m\,\forall f$,
where we will take $m \rightarrow 0$. It is easy to see that the
action~\ref{qcdact} for massless quarks is invariant under the
following group of global transformations:

\begin{eqnarray}
G : & \psi_{f} \rightarrow & \exp(i\alpha^{a}T^{a} +
i\gamma_{5}\beta^{b}T^{b})_{fr}\psi_{r}\ ,\nonumber\\
& \sib_{f} \rightarrow & \sib_{s}\exp(-i\alpha^{a}T^{a} +
i\gamma_{5}\beta^{b}T^{b})_{sf}
\label{eq:chi_or}
\end{eqnarray}
where $T^{a},\,a = 1,\ldots,N_{f}^{2}-1$ are the generators of the Lie
group $SU(N_{f})$. Which group is this ? We note that it contains a
$SU(N_{f})$ subgroup which we denote $SU_{D}(N_{f})$ comprising
elements of the form $\exp(i\alpha^{a}T^{a})$.  The $\gamma_{5}$ part,
does not form a subgroup as it is not closed under composition. Is the
group $G$ a symmetry of QCD with massless quarks ?  We note from
equation~\ref{eq:chi_or} that the $\gamma_{5}$ part of $G$ mixes
particles with opposite parities but otherwise identical quantum
numbers (for instance we can transform the state $\sib\psi$ which
transform as ``+'' to $i\sib\gamma_{5}\psi$ which has transforms as
\mbox{``-''}). So if this symmetry holds in nature then we would
expect degeneracy of hadrons into parity doublets. This manifestly
does not occur, we find large mass splittings between particles with
opposite parities but otherwise identical quantum numbers (for
instance the splitting between the nucleon and its parity partner is
$\approx$ 600\,MeV).

We can better understand the structure of this symmetry if we
decompose the group $G$ into a direct product of groups. We rewrite
the QCD action given in~\ref{qcdact} for massless quarks in a chiral
form using the expansion $\psi = \psi_{L} + \psi_{R}$ where $2\psi_{L}
= (1 - \gamma_{5}) \psi$ and $2\psi_{R} = (1 + \gamma_{5}) \psi$:

\begin{equation}
S_{qcd}^{E} = S_{gauge}
- \sum_{f}\int d^{4}x\,\sib_{fL}\gamma_{\mu}iD_{\mu}\psi_{fL}
- \sum_{f}\int d^{4}x\,\sib_{fR}\gamma_{\mu}iD_{\mu}\psi_{fR}\ ,
\label{chgauge}
\end{equation}
where $S_{gauge}$ is the gauge part as in~\ref{qcdact}. Now this is
clearly invariant under the following independent global
transformations:

\begin{eqnarray}
SU_{L}(N_{f}) : \psi_{fL} & \rightarrow &
\exp(-i\alpha^{a}T^{a})_{fs}\psi_{sL}\nonumber\\
\sib_{fL} & \rightarrow & \sib_{tL}\exp(-i\alpha^{a}T^{a})_{tf}\nonumber\\
\nonumber\\
SU_{R}(N_{f}) : \psi_{fR} & \rightarrow &
\exp(-i\alpha^{a}T^{a})_{fs}\psi_{sR}\nonumber\\
\sib_{fR} & \rightarrow & \sib_{tR}\exp(-i\alpha^{a}T^{a})_{tf}
\end{eqnarray}
The $SU_{D}(N_{f})$ subgroup of $G$ is in fact nothing other than the
diagonal subgroup of $G$ when it is decomposed as a direct product
group. The fact that $G$ is phenomenologically not a symmetry of the
quantum theory corresponds to the breaking $G = SU_{L}(N_{f})\otimes
SU_{R}(N_{f}) \rightarrow SU_{D}(N_{f})$. We therefore expect
$N_{f}^{2} - 1$ massless bosons by Goldstone's Theorem. This is a role
played by $\pi^{\pm},\pi^{0}$ for the case of $N_{f} = 2$. These
particles are not exactly massless because the symmetry we have broken
was never an exact symmetry (the up and down quarks have a small mass
of approximately $5\,$MeV after all, which whilst being too small to
explain the splittings between parity partners, is the source for the
small mass of the pions).

The role of the order parameter for this symmetry breakdown is played
by the quark condensate:

\begin{eqnarray}
\ssi & = & 0 \quad {\rm symmetric\ phase} \nonumber\\
& \neq & 0 \quad {\rm chiral\ symmetry\ broken}\ .
\end{eqnarray}
The condensate is actually a fermion loop of a given flavour being
created and annihilated at a given point $\ssi =
\langle\sib_{f}(x)\psi_{f}(x)\rangle$. A simple calculation should
suffice to convince the reader that this condensate is zero to all
orders of perturbation theory for massless quarks (we have a closed
fermion loop with various gauge boson vertices - the crucial point is
that we always end up with an odd number of gamma matrices so that the
spinorial trace is always zero). We are forced therefore, to
non-perturbative methods if we are to understand the mechanism for
chiral symmetry breakdown (\ch). This is the first hint that
instantons may be connected to \ch, they are non-perturbative objects
after all. To further explore the links between instantons and \ch\ we
rewrite the quark condensate in terms of eigenvalues of the Dirac
operator (for non-zero quark masses and finite volume - we shall take
the appropriate limits afterwards):

\begin{eqnarray}
\ssi & = & \frac{1}{V}\langle{\rm Tr}\,\sib\psi\rangle\nonumber\\
& = & \frac{i}{VN_{f}}\frac{\partial}{\partial m} (\ln {\cal
Z}_{qcd}^{E})\nonumber\\
& = & \frac{i}{VN_{f}}\frac{\partial}{\partial m}\int {\cal
D}A\,\exp(-S_{g})\prod^{N_{f}}\det(\di[A] - im)\ ,\nonumber
\end{eqnarray}
where we have integrated out the fermion fields to generate the
determinant in the partition function ${\cal Z}_{qcd}^{E}$. To proceed
further we assume a discrete set of eigenvalues for the Dirac
operator; this holds for a finite volume (greater care should be taken
with the limits than will be done in this work: in this case however,
the end results of doing so will be the same). We recall that:

\begin{equation}
\{ \gamma_{5},\di \} = 0\ .
\label{gammasym}
\end{equation}
This has some important consequences for the the eigenvalue spectrum
of the Dirac operator.

\subsection{The spectrum is even in $\lambda$.}
All eigenvalues are real as our Dirac operator is
Hermitean. Furthermore, for any eigenfunction $\psi_{n}$ with non-zero
eigenvalue $\di\psi_{n} = \lambda_{n}\psi_{n}$, there exists a
linearly independent function $(\gamma_{5}\lambda_{n})$ such that
$\di(\gamma_{5}\psi_{n}) = -\lambda_{n}(\gamma_{5}\psi_{n})$. Hence
all non-zero eigenvalues come in pairs $\pm\lambda$ and our assertion
is proved. This will have important implications for our work, any
ostensible representation of the Dirac operator should obey this basic
requirement.

\subsection{Zero mode wavefunctions have definite chirality.}
Consider a zero mode wavefunction $\di\psi_{0} = 0$. It is easy to see
that equation~\ref{gammasym} implies that $\psi_{0}$ is also an
eigenfunction of $\gamma_{5}$, namely $\gamma_{5}\psi_{0} =
\pm\psi_{0}$. The eigenvalue is $\pm 1$ as we know that
$\gamma_{5}^{2} = 1$. We see therefore that any zero mode wavefunction
must have either positive or negative chirality (so in the continuum
the zero mode wavefunctions are Weyl spinors instead of Dirac
spinors).

Let the number of eigenfunctions with eigenvalue zero, of positive
chirality be denoted $N_{+}[A]$ and the number with negative chirality
$N_{-}[A]$ respectively. Let the total number of zero eigenvalues be
denoted $Z = N_{+} + N_{-}$. We rewrite the determinant as:

\begin{eqnarray*}
\det(\di[A] - im) = (-im)^{Z[A]}\prod_{n}(\lambda_{n}[A] -
im)\ ,
\end{eqnarray*}
where the product is taken over the non-zero eigenvalues only. A short
calculation shows that:

\begin{equation}
\det(\di[A] - im) = (-im)^{Z[A]}\exp\left(\frac{1}{2}\sum_{n}\ln(\lambda_{n}^{2}[A] +
m^{2})\right)\ .
\label{detexp}
\end{equation}
(The reason that the log term is $\log(\lambda_{n}^{2}[A] +
m^{2})$ and not $\log(-\lambda_{n}^{2}[A] - m^{2})$ is simply
because the difference between the two is a constant, which would
cancel with the same constant from the ``free'' partition function in
the denominator when we calculate any operator.) So
differentiating~\ref{detexp} we arrive at an expression for the quark
condensate:

\begin{equation}
\ssi = \frac{i}{V{\cal Z}}\int {\cal D}A\,\left(\frac{Z[A]}{m}
+ \sum_{n}\frac{m}{\lambda_{n}^{2}[A] + m^{2}}\right)
\exp(-S_{g})\prod^{N_{f}}\det(\di - im)\ .
\end{equation}
Recall that we are still working in a finite volume $V$ and at finite
non-zero quark mass $m$. The process of taking the limits in the above
expression, is a delicate one, and we will take a little more care
than elsewhere when doing so. We should first take the thermodynamic
limit ($V \rightarrow \infty$) and then the chiral limit ($m
\rightarrow 0$). As we shall see the two do not commute. When we take
the volume to infinity, the eigenvalue distribution for the Dirac
operator goes from a discrete spectrum to a continuous spectrum. We
therefore replace the sum in the integral to an integration with a
spectral density $\nu(\lambda_{0})d\lambda$, which measures the number
of eigenvalues in the interval $[\lambda_{0} - d\lambda/2,
\lambda_{0} + d\lambda/2]$. (We should of course begin with finite
intervals and then take the width of the maximum such interval to zero
in a controlled fashion, but this is assumed to be so. We also drop
the subscript as $\lambda_{0}$ is an arbitrary point.)

\begin{equation}
\ssi = \frac{i}{V}\left<\frac{Z}{m} +
\int_{0}^{\infty}d\lambda\,\frac{2m\nu(\lambda)}{\lambda^{2} +
m^{2}}\right>\ .
\label{chico}
\end{equation}
To recap our computation, we have obtained an expression for the quark
condensate in terms of the eigenvalues of the Dirac operator. The
first term arises from exact zero eigevalues, the second from non-zero
eigenvalues (whilst the integral is from zero to infinity, it is to be
understood that the exact zero eigenvalues are not included within
this integral).

\section{A Deep Link}
We have found certain non-trivial solution to the Euclidean equations
of motion, of finite action, known as instantons. We have also
discussed briefly chiral symmetry, and seen that it is broken in
nature. Furthermore, we have seen that the order parameter for \ch\
can be related to the expectation value of quantities derived from the
eigenvalue distribution of the Dirac operator. Is there some link
between instantons and the spectral density of the Dirac operator, and
hence a link between instantons and \ch\ ? Do instantons and their
interactions form a mechanism for \ch\ in QCD ?

The answer is yes. Or more accurately, maybe. The crucial relation was
found by 't Hooft in his ground breaking papers on the resolution to
the U(1) axial problem~\cite{'tHooft,Coleman}. He found that the Dirac
operator with a gauge field of a single classical instanton or
anti-instanton has an exact zero eigenvalue. It turns out that this is
an example of a more general result:

\begin{equation}
Q[A] = N_{-} - N_{+} ,
\label{atisin}
\end{equation}
where $Q[A]$ refers to the winding number of the gauge
field~(\ref{winform}) and $N_{-}/N_{+}$ are, as before, the number of
exact zero eigenvalues with negative/positive chirality
respectively. Hence an arbitrary gauge field of winding number $Q[A]$
has as least $|Q[A]|$ exact zero eigenvalues. An instanton satifies
this formula with $Q[I] = 1,\ N_{-} = 1,\ N_{+} = 0$; an
anti-instanton with $Q[\overline{I}] = -1,\,N_{-} = 0,\,N_{+} =
1$. The zero eigenfunction for an instanton is given by:

\begin{equation}
\psi_{0}(x) = \frac{\sqrt{2}}{\pi}\frac{\rho}{(\rho^{2} + x^{2})^{3/2}}u
\label{zerom}
\end{equation}
where $u$ is a constant spinor with spin and colour indices.

We know already that the spinor $u$ must have definite chirality. We
now know that the chirality must be such as to obey~\ref{atisin}. It
can be shown that for any (anti-)self-dual gauge field configuration
$(N_{-}),N_{+}$ is zero~\cite{Brown}. A simple argument we present
later extends this, and makes plausible the idea that we can always
take at least one of $N_{+}$ or $N_{-}$ to be zero for finite action
gauge fields (see~\ref{rhobar}). This implies that $|Q[A]| = |N_{-} -
N_{+}| = N_{-} + N_{+} = Z[A]$. We see therefore that the contribution
of the exact zero modes to the chiral condensate in~\ref{chico} is
$\langle |Q|/m\rangle$ and arises solely from the winding number
distribution of the gauge fields. The chiral condensate can be written
as

\begin{equation}
-i\ssi = \frac{\langle |Q|\rangle}{mV} +
\int_{0}^{\infty}d\lambda\,\frac{2m\overline{\nu}(\lambda)}{\lambda^{2} +
m^{2}}\ ,
\label{f_chico}
\end{equation}
where $\overline{\nu}(\lambda) = \lim_{V \rightarrow
\infty}\langle\nu(\lambda)\rangle/V$. 

Equation~\ref{atisin} can be derived using a variety of different
field theory methods (see~\cite{Brown}, also~\cite{Coleman,Weinberg}
and references therein) but these in turn are a special case of a yet
more general, and, much celebrated, mathematical theorem due to Atiyah
\& Singer~\cite{Atiyah}. The essential thing to note is
that~\ref{atisin} does not depend upon any property of the gauge field
apart from its winding number; in particular, it does not require the
gauge field to be a solution of the classical equations of
motion. This enables us to move beyond classical instantons to the
quantum objects that may populate the quantal vacuum of QCD.

Let us analyze~\ref{f_chico} with our new found knowledge. We use well
known arguments to try to ascertain the contribution of each of the
two terms in~\ref{f_chico}. We first concentrate on full QCD with
dynamical fermions, and then, we look at quenched QCD. It turns out
that the two cases are very different.

\section{Contribution to $\langle\sib\psi\rangle$.}
\subsection{QCD}
We require the length of the box $L = V^{\frac{1}{4}}$ to be much
greater than the Compton wavelength of the lightest particle $L \gg
m_{C}^{-1}$.

\subsubsection{$N_{f} \geq 2$}
If we have more than one flavour of fermion $N_{f} \geq 2$, then we
have chiral symmetry breakdown and almost massless Goldstone bosons:

\begin{equation}
m_{\pi}^{2} = 2m\frac{M^{2}}{f_{\pi}}\ ,
\end{equation}
where $M$ and $f_{\pi}$ are constants with dimension of mass. We
therefore have $m_{C} \propto m^{\frac{1}{2}}$, and hence, we require
$L \gg m^{-\frac{1}{2}}$ or alternatively:

\begin{equation}
mV \gg \frac{1}{m} .
\label{mV}
\end{equation}
It can be shown that if we have \ch\ then $\langle
Q^{2}\rangle^{\frac{1}{2}} \propto \sqrt{mV}$. It is intuitive that
for smooth winding number distributions we have $\langle|Q|\rangle
\sim O(\langle Q^{2}\rangle^{1/2})$. (We can rarely say much about
non-analytic quantities, and, as it is difficult to think of
distributions where these two quantities are wildly different, we will
often use this approximation.) Hence the first term of the quark
condensate is proportional to $1/\sqrt{mV} \ll m^{1/2}$ and disappears
in the chiral limit. This shows that we may legitimately ignore the
first term if we take the limits as we should for QCD with $N_{f} \geq
2$.

\subsubsection{$N_{f} = 1$}
The argument is different for $N_{f} = 1$. We have no chiral symmetry
to break (the axial $U_{A}(1)$ is anomalously broken) and hence no
Goldstone bosons. The lightest particle has a non-zero mass in the
chiral limit so we do not require our box length to diverge, only to
be larger than a fixed size (the Compton wavelength of the equivalent
of the $\eta^{'}$). In this case we will have $mV \rightarrow 0$ in
the chiral limit. As before we have $|Q| \propto \sqrt{mV}$, hence we
have a divergent contribution from the first term in the chiral limit.

We have seen that the contribution of the first term to the quark
condensate is dramatically different for different numbers of
flavours. Let us now consider the second term, which turns out to have
subtleties of its own. Writing the $\delta$-function as,

\begin{equation}
\delta(x) = \lim_{\epsilon \rightarrow 0} \frac{\epsilon}{\pi(x^{2} +
\epsilon^{2})} ,
\label{deltafn}
\end{equation}
allows us to conclude that

\begin{equation}
\lim_{m \rightarrow
0}i\int_{0}^{\infty}d\lambda\,\frac{2m\overline{\nu}(\lambda)}{\lambda^{2}
+ m^{2}} = i\pi\overline{\nu}(0) .
\label{bc}
\end{equation}
This is the Banks-Casher relation~\cite{Banks} which shows that \ch\
in the physical world (where $N_{f} = 2$ is a good approximation) is
directly related to the accumulation of eigenvalues at $\lambda =
0$. One should apply this formula with care however; for instance, it
is entirely possible that for any given finite quark mass, we have a
divergence in the spectral density as $\lambda \rightarrow 0$, yet we
still obtain a finite quark condensate (a simple example of such a
spectral density is $\overline{\nu}(\lambda) = (m/\lambda)^{d}\ \ d
\in (0,1)$). The subtlety arises as the expectation of the spectral
density is itself dependent upon the quark mass through the fermion
determinant(s), and hence, it may not be legitimate to substitute a
delta function into the left hand side of equation~\ref{bc} as we have
done above. We therefore choose to use the integral form
of~\ref{f_chico} rather than the Banks-Casher relation~\ref{bc} when
calculating the quark condensate in the case of dynamical fermions (we
do computations at different quark masses and extrapolate
appropriately). We also exclude the effects of the first term in
calculating the chiral condensate. As we have seen, these will either
be pathological or irrelevant.

To illustrate the lack of commutativity in the order of the limits, we
now consider what happens if we take the quark mass to zero in a
finite volume $m \rightarrow 0, V$ fixed. In this case we expect no
\ch\ and $\langle Q^{2}\rangle \propto m^{N_{f}}V$. The first term
therefore contributes nothing for $N_{f} > 2$; we get a finite
contribution for the case of $N_{f} = 2$ (which we can reduce by
increasing the volume of the system), and, a divergent contribution
for $N_{f} = 1$. In this case, we obtain a discrete spectrum of
eigenvalues; in particular, we have a gap in the eigenvalue spectrum
at $\lambda = 0$ of $O(V^{-1})$. Consequently we find that the
expectation of the spectral density also has a gap for small
eigenvalues. The contribution of the second term therefore is always
zero.

\subsection{Quenched QCD}
If QCD is an accurate model for the strong interactions, then ideally
one should be able to derive hadron properties (such as masses, cross
sections etc.) from first principles using the theory. Doing so in
practice has been very difficult. One of the most successful methods
for obtaining knowledge of hadron masses, verification of confinement,
high temperature effects etc., from first principles, has been to
formulate QCD in terms of degrees of freedom which can be simulated on
a computer. In practice, this involves discretizing spacetime and
formulating the theory in terms of fermions which live on the
``sites'' of the lattice and gauge fields which live on the links (as
we would expect from thinking of gauge fields as ``connections''). As
we now have finite degrees of freedom, one can try to evaluate the
partition function numerically using importance sampling. This is
normally implemented via a Monte Carlo routine. This theory is known
as Lattice QCD. One can obtain continuum QCD by taking the lattice
spacing to zero whilst holding the total volume sufficiently large to
keep finite size effects small. (In momentum variable terms, we have
introduced an ultraviolet cutoff for momentum and a discrete set of
possible momenta. When we take the lattice spacing to zero we are
removing the ultraviolet cutoff whilst making the spacing between
possible momenta vanish.)

Suffice it to say that the field of lattice calculation is mature
enough to merit its own Los Alamos archive~! We will not need to know
the details of this field, only some of the problems faced by its
practitioners. The difficulties of generating lattice configurations
which incorporate the fermion determinant in the weighting are well
known (see for instance~\cite{Sharpe-Prog} for a review of lattice
simulations and the difficulties associated with them,
and~\cite{Gupta} for a recent introduction to the field). It is only
relatively recently that computer power has advanced to the stage
where such calculations are feasible, and even now, the volumes are
relatively small and the lattice spacings, relatively large. It has
been estimated that computer power will need to increase by a factor
of 30-40 before full QCD calculations with all errors under control
are possible~\cite{Sharpe-Prog}. It is primarily the difficulties of
full QCD simulation which have led to the widespread use of the
``quenched'' approximation to full QCD. This is simply the gauge
theory, where there are no fermion interactions at all during the
generation of the configurations. The weighting is simply the gauge
weighting. The fermions are ``test fermions'', they are propagated
through these background gauge fields.

The results obtained from quenched QCD are remarkably good however,
hadron masses are correct at around the 10\%-20\%
level~\cite{Sharpe-Prog,Aoki}. Can we obtain quenched QCD as some
limit of full QCD ? One way to think of this is to consider full QCD
but to give the test fermions a finite mass ($m_{t}$) and the virtual
fermions a far larger mass ($m_{d}$). QCD is the theory with $m_{t} =
m_{d}$. The theory with $m_{d} \gg m_{t}$ is known as partially
quenched QCD. In partially quenched QCD we think of the test fermions
being propagated through a vacuum composed of gauge bosons and heavy
virtual fermions. In the limit of the virtual fermions having infinite
mass, we would have decoupling of the virtual fermions and the result
would be quenched QCD ($m_{d} \rightarrow \infty$). (The decoupling is
nothing exotic, it is only saying that a large mass in the fermion
determinant would make the eigenvalues of the Dirac operator
irrelevant and so the determinant would be simply an infinite constant
in the partition function.) It is important to note that (partially)
quenched QCD is not actually a physical theory; the Hamiltonian for
(partially) quenched QCD is not Hermitean~\cite{Gupta}. It is entirely
possible that whilst we get fairly good results for some quantities we
may get much worse results for others.

It is of interest to consider whether chiral symmetry is broken in the
quenched theory, and whether instantons are the mechanism for this
breaking. This is not only because we hope similar mechanisms may
apply to full QCD but because we can view quenched QCD as a theory in
its own right. Crucially, is the chiral condensate a quantity which
behaves well in quenched QCD~?

The chiral condensate is an especially interesting quantity to study
in quenched QCD because of the influence of topology on the fermion
determinant. We know that in full QCD, configurations with non-trivial
topology are suppressed by light quarks (for massless quarks we have
total suppression of configurations with $Q[A] \neq 0$ as the fermion
determinant is zero). This suppression is lost in quenched QCD as we
have no fermion determinant. It is therefore plausible that the
answers for the two theories may be radically different from one
another.

We can visualize the vacuum for quenched QCD to be composed of
instantons and anti-instantons placed at random throughout our box (we
have only gauge degrees of freedom so we think of these as being
composed of instantons - more on this later). If the box has volume
$V$ then we expect $|Q| \propto \sqrt{V}$ (this is nothing but the
standard deviation of $V$ Bernoulli trials where we may pick $\pm1$
charges with equal probability). If we choose the volume to obey
equation~\ref{mV} when we take the chiral limit for our external
quark, then the first term contributes $1/m\sqrt{V} \ll O(1)$. The
second term is again given by the Banks-Casher formula~\ref{bc}, where
we no longer have to worry about the mass dependence of the spectral
density.

If however we choose to take the chiral limit in a fixed volume, the
first term gives a divergent contribution, the second is zero as in
the case for full QCD. In our analysis we will always take the limits
as in the physically applicable case and concentrate on the
contribution to the quark condensate from the second term. This is
what will be of primary significance in the physical world.

\section{Instanton mixing and $\overline{\nu}(0)$}
\label{rhobar}
It seems we have a handle on the contribution to the quark condensate
coming from the winding number distribution term. What about the
contribution from the non-zero eigenvalues ? We know from the above
that it is the low lying eigenvalue spectrum which is of primary
importance to the quark condensate. If ensembles of instantons are
thought to be the mechanism to break chiral symmetry in nature, then
they must somehow contribute to $\overline{\nu}(\lambda)$ for small
$\lambda$.

We can in fact see that they contribute, via a process of
``mixing''. Let us consider the case of a gauge configuration given by
one instanton and one anti-instanton. Whilst an
instanton--anti-instanton pair is not a solution to the equations of
motion for Yang-Mills theory, we can consider a gauge potential which
is that of an instanton in one region and an anti-instanton in another
region. The simplest way of doing so would be to express the instanton
and anti-instanton in ``singular gauge'' and then just linearly add
their gauge potentials. Singular gauge refers to a discrete
transformation of the conformal group, namely co-ordinate inversion
$x^{\mu} \rightarrow (1/x)^{\mu} = x^{\mu}/x^{2}$.  This has the
effect of shifting the topological charge from infinity to the origin
(see~\cite{Jackiw-Con} for details on the conformal properties of
instantons).

What can we say about their respective zero modes ?  We know through
the additivity of winding numbers that the total winding number for
this configuration is $Q = 0$. Atiyah-Singer does not preclude the
possibility that $N_{-} = N_{+} = 1$ i.e. each of the two objects was
associated with a zero eigenvalue and nothing changes when we put the
objects together. However, this is not what occurs in most cases. What
happens is that the two would-be zero modes split symmetrically about
zero, by an amount determined by the overlap of their would-be zero
mode wavefunctions:

\begin{eqnarray}
\lambda_{s} & = & \langle\psi_{A}|\di|\psi_{I}\rangle\nonumber\\
& = & \int d^{4}x\,\sib_{A}(x)\di\psi_{I}(x)\ ,
\label{split}
\end{eqnarray}
so we get no exact zero eigenvalues $N_{-} = N_{+} = 0$, but
eigenvalues $\pm\lambda_{s}$ (the eigenvalues come in pairs due to the
$\gamma_{5}$ symmetry as they must). (Note the presence of the Dirac
operator, otherwise chirality would force the matrix element to zero.)
If we recall the form of the zero mode wavefunction, then it is
evident that for large separations $R$ (beween the center of the
instanton and anti-instanton, in comparison to their sizes),
$\lambda_s \sim 1/R^{3}$. We therefore recover the two exact zero
eigenvalues as the objects become infinitely separated. However, for
finite separation, the interaction between the objects precludes any
exact zero eigenvalues, and we get a splitting from zero due to the
mixing of wavefunctions. Is it possible to have two exact zero
eigenvalues with the objects at finite separation~? The answer to this
is, unfortunately, yes. It is possible to orient the two objects (in
colour space) in such a manner that the overlap integral~\ref{split}
is zero. This is equivalent to picking a single direction with
precision on a sphere. Such configurations are therefore a set of
measure zero in instanton configuration space. However we should not
exclude them for this reason (otherwise, why are we working with
finite action fields in the first place~?). We exclude these
configurations because their quark mass suppression is enhanced;
ultimately, we are interested in the chiral limit after all.

We shall go on in the next chapter to generalize this procedure for
arbitrary numbers of instantons and anti-instantons interacting with
one another, suffice it to say that for a collection of $N_{I}$
instantons and $N_{A}$ anti-instantons with (w.l.o.g.)  $N_{I} >
N_{A}$ we obtain the following spectrum from the $(N_{I} + N_{A})$
would-be zero modes:

\begin{eqnarray}
\lambda_{0}^{1},\ldots,\lambda_{0}^{N_{I}-N_{A}}\nonumber\\
\pm\lambda_{1},\ldots,\pm\lambda_{N_{A}}
\end{eqnarray}
where $\lambda_{0}$ remain exact zero eigenvalues (sufficient in
number to obey the Atiyah-Singer index theorem and all of negative
chirality $N_{-} = Q,\ N_{+} = 0$), and the remaining non-zero
eigenvalues come in pairs and have split from zero. We see therefore
that instantons might in principle generate a spectrum of eigenvalues
near zero i.e. they can produce a non-zero $\overline{\nu}(\lambda)$
for small $\lambda$ which is what we require to break chiral symmetry.
Any model based on instantons should generate such a spectrum from the
mixing of would-be zero modes. There will be other eigenvalues but we
postulate that these (arising from mixing with the non-zero eigenmodes
associated with each object) will be larger, and therefore less
interesting for chiral symmetry. We can contrast the effect of
instantons on the spectrum of the Dirac operator with the perturbative
spectrum. The free Dirac operator has a spectrum which grows as
$\lambda^{3}$ i.e. the contribution for small $\lambda$ is zero. If
instead we look at an ensemble of instantons and anti-instantons then
we obtain a non-zero spectrum near zero, hence it is more hopeful to
consider instanton configurations than perturbative configurations.

We have obtained some understanding of the contribution of the exact
zero modes to the quark condensate. Our understanding of the
contribution of the $\overline{\nu}(\lambda)$ term is more limited
however. The aim of this work is to try to predict the qualitative
form of $\overline{\nu}(\lambda)$ for small $\lambda$, due to
instanton interactions for quenched QCD and full QCD, using ideas such
as those given above. In order to do this we will construct a
simplified model of the vacuum which will be based solely on instanton
degrees of freedom - the only parameters which we will use will be a
list of positions and sizes of the (topological) charges populating
the vacuum (we will even ignore their colour orientations for the sake
of simplicity).

\subsection{Why instantons, why a model ?}

The first question that arises is, ``Does this make sense at
all~?''. If we consider the path integral for QCD, is is possible to
decompose arbitrary finite action gauge field configurations into
ensembles of finite numbers of instantons and anti-instantons~? Will
there be an instanton configuration which is ``optimal''~? Thankfully,
we shall not have to answer these difficult questions, we shall look
at the problem from a simpler angle altogether. We begin with the
question, ``If I choose to view the vacuum as being composed of
instanton degrees of freedom, then, what if anything can I say about
QCD or quenched QCD~?'' As we shall see, even simple models can give
rise to unexpected richness and structure.

The first reason for looking at instantons as the relevant degrees of
freedom for chiral symmetry breaking is that, in a finite volume, we
expect a finite number of such objects. This makes the calculation
tractable; it is certainly far simpler than an analytical calculation
of the path integral with an (uncountably) infinite number of degrees
of freedom. A corresponding lattice Dirac operator on a $16^{4}$
lattice in $SU(3)$ gauge theory is a 786432 dimensional matrix
($16^{4}$ sites, each site with a fermion with $4$ spins and 3 colour
degrees of freedom). In contrast, if the volume represented is about
about $10fm^{4}$ (which is common) then we would expect $O(100)$
instantons i.e. if we look at only would-be zero modes then our model
will be an $O(100)$ dimensional matrix. We are obtaining greater
simplicity by discarding underlying degrees of freedom which we
believe to be superfluous for the questions we will be
asking. Nevertheless, lattice calculations are well within the realms
of computing power (we will be using the results of $SU(3)$
calculations on a $32^{3}64$ lattice in chapter~\ref{ch:latt}), so why
bother with a model if all we are gaining is a little time~?

The fundamental reason is that lattice calculations have a few
drawbacks:
\begin{itemize}
\item{Lattice artefacts.}

Lattice calculations have errors associated with discretizing
spacetime. We should recover continuum physics when we shrink the
lattice spacing to zero. However, in practice, how ``close'' you are
to the continuum limit depends upon the problem you are
studying. Consider the following example. An instanton of size $\rho
\gg a$ (where $a$ is the lattice spacing) is discretized and placed on
a lattice. Now as the object is large and smooth in comparison to the
lattice spacing, we would expect something close to a zero eigenvalue
in the corresponding spectrum of the Dirac operator. We now shrink the
instanton smoothly (in the continuum) so it becomes of size $\rho \ll
a$, and, we place it at the centre of a lattice hypercube so it is far
from any of the gauge links. On the lattice it now resembles a pure
gauge object and we expect no zero eigenvalue. The problem stems from
the fact that we are on a lattice and topological laws no longer
apply. In practice however, this problem may not be of significance,
after all, small objects are suppressed as $\rho^{6}$ in $SU(3)$
theory. However, current lattice calculations have lattice spacings of
$a \approx 0.1fm$ (see chapter~\ref{ch:latt}) and so there may be
objects of only a couple of lattice spacings. We would not expect to
get a zero eigenvalue for these objects. Furthermore, mixing of such
objects would not yield the correct spectrum either. In practice we
are concerned with the region of small eigenvalues which we believe to
be dominated by mixing of would be zero modes. This is problematic for
the lattice if the would be zero modes are not would be zero modes at
all.
\item{Dynamical fermions}

We have already mentioned the difficulties faced with simulations
involving fermions. The problem is particularly acute for light
fermions which is the limit we are aiming for.
\item{Chiral symmetry and the Nielsen \& Ninomiya~\cite{Nielsen} theorem.}

There is also a famous problem associated with chiral symmetry on a
lattice. We expect the na\"{\i}ve lattice discretization of the Dirac
action to be chirally symmetric for massless quarks. This is indeed
what one finds, however, one also find extra species of fermions (the
lattice fermion propagator for massless quarks has 16 poles instead of
1). A common way of removing these extra fermions is to give them a
mass which is inversely proportional to the lattice spacing. This
forces them to decouple in the continuum limit. The addition of a mass
term however, explicitly breaks chiral symmetry at any non-zero
lattice spacing. The Nielsen \& Ninomiya theorem proves that (under
general conditions), lattice actions which possess chiral symmetry
must also be afflicted by fermion doublers.
\end{itemize}
Recent work using novel lattice fermion formulations such as
domain-wall fermions~\cite{Fleming,Mawhinney,Chen} and related lattice
fermions~\cite{Edwards,Neuberger,Niedermayer} shows promise that the
difficulties with obtaining exact chiral symmetry on a lattice may be
overcome. (This relies on obtaining something which is not quite
chiral symmetry - hence evading Nielsen \& Ninomiya - but close enough
for many purposes.)

Our model is in some sense a generic instanton model, it has
information about nothing but instantons. If we find that we cannot
break chiral symmetry within our model, then it difficult to see how
instantons could hope to do so in reality. This is in contrast to
lattice calculations, where the results are clouded by various
problems such as those listed above.

%% file: model.tex
\chapter{A toy model of the vacuum}
\label{ch:model}

We wish to calculate the low lying eigenvalues of the Dirac operator
with a gauge field composed of instanton and anti-instanton degrees of
freedom. We will refer to objects generically as ``instantons'' when
there is no fear of confusion. We have already made an approximation
which can be quantified as:

\begin{equation}
A \approx \sum_{i=1}^{N_{A}}A_{i}^{+}(x_{i}^{+},\rho_{i}^{+},K_{i}^{+}) +
\sum_{j=1}^{N_{I}}A_{j}^{-}(x_{j}^{-},\rho_{j}^{-},K_{j}^{-}) \,
\label{decomp}
\end{equation}
where $N_{I}/N_{A}$ are the number of instantons and anti-instantons
in the configuration respectively, and $A_{i}(x,\rho,K)$ represents an
(anti-)instanton with centre $x$, size $\rho$ and colour orientation
$K$. We have a zero mode associated with each object:

\begin{equation}
\di[A_{i}^{\pm}]|\psi_{0i}^{\pm}\rangle = 0 .
\label{eq_motion}
\end{equation}
(The $\pm$ superscript refers to the chirality of the object, ``+''
for an anti-instanton, ``-'' for an instanton.) The idea is simple; we
wish to construct a matrix representation of the Dirac operator using
these would-be zero modes as a basis. The eigenvalues for this matrix
which would be the contribution to the spectral density from this
configuration of objects. We will therefore get the following
$(N_{A}+N_{I}) \times (N_{A}+N_{I})$ matrix representation:

\begin{equation}
\begin{array}{ccc}
\hspace{1.8cm}\overbrace{\hspace{2.7cm}}^{N_{A}} & \hspace{0.0cm}\overbrace{\hspace{2.7cm}}^{N_{I}} & \\
\di \doteq D = 
\left(
\begin{array}{l}
\langle\psi_{i}^{+}|\di|\psi_{j}^{+}\rangle = 0 \\
\langle\psi_{i}^{-}|\di|\psi_{j}^{+}\rangle = M_{ij}^{\dagger}
\end{array}
\right. &
\left.
\begin{array}{l}
\langle\psi_{i}^{+}|\di|\psi_{j}^{-}\rangle = M_{ij}\\
\langle\psi_{i}^{-}|\di|\psi_{j}^{-}\rangle = 0
\end{array}
\right) &
\begin{array}{l}
\Big\}\ N_{A}\\ \Big\}\ N_{I}
\end{array}
\end{array}
\label{dirm}
\end{equation}
where the matrix elements $M_{ij}$ are given by some suitable function
involving the collective co-ordinates of the objects in the
ensemble. (This is in fact where equation~\ref{split} comes from; we
have constructed a matrix representation using the single zero mode
from each of the objects to form a $2 \times 2$ matrix.) It is
important to note that in all this work, we ignore the detailed
spinorial structure of the zero mode wavefunction (the constant spinor
$u$ is equation~\ref{zerom}). This implies that we lose the relative
colour orientation of the objects (so one consequence is that we
cannot have two exact zero eigenvalues for an
instanton--anti-instanton pair at finite separation through colour
orientation). This should not affect any results as long as instantons
are oriented at random in the vacuum; if however, there are dynamical
effects which for instance increase or decrease overlaps between
objects then our answers will be slightly incorrect. (A similar effect
will be seen in chapter~\ref{ch:latt} where we find evidence that
instanton positions are not random in the vacuum but occur so as to
increase overlaps between objects of opposite chirality.) As our study
is exploratory, and as little is known of instanton orientation in the
vacuum, we ignore this slight concern. It should also be noted (as we
shall see) that should such information become known, then the
modifications required to incorporate colour information are fairly
trivial. This simplification allows us to keep our representation
real, the matrix $D$ is symmetric as opposed to Hermitean. The
formul\ae\ given in this chapter will be for the more general Hermitean
case (so should the need to go to a full complex representation arise,
then (hopefully~!) no modifications need be made). So to summarise, in
our work, we have no information $K^{\pm}$ in the decomposition given
by equation~\ref{decomp}.

There are however, a number of more serious concerns associated with
viewing $D$ as a matrix ``representation'' of the Dirac operator; we
will come to these later. First we consider the properties of such a
matrix representation which make us hopeful that our model may be
related to reality in some way.

\section{A few simple results}
A few points about notation. The matrix representation of the Dirac
operator is a map $D : {\mathbb R}^{N_{A}+N_{I}} \rightarrow {\mathbb
R}^{N_{A}+N_{I}}$ ,

\[
\left(
\begin{array}{cc}
0 & M\\
M^{\dagger} & 0
\end{array}
\right)
\left(
\begin{array}{r}
\underline{e}\\
\underline{f}
\end{array}
\right) =
\left(
\begin{array}{r}
M\underline{f}\\
M^{\dagger}{\underline{e}}
\end{array}
\right) .
\]
The chiral structure of the Dirac operator allows us to work with two
smaller maps, namely $M : {\mathbb R}^{N_{I}} \rightarrow {\mathbb
R}^{N_{A}}$ and the transposed map $M^{\dagger} : {\mathbb R}^{N_{A}}
\rightarrow {\mathbb R}^{N_{I}}$. We will find this very convenient in
all that we do.

\subsection{Our representation obeys the Atiyah-Singer theorem.}
It is simple to prove that for a general matrix with the above
structure $Q = N_{-}-N_{+}$, where $Q$ is the winding number of the
gauge field ($Q = N_{I} - N_{A}$) and $N_{\pm}$ are the number of zero
eigenvalues with positive/negative chirality as always.

\subsubsection{Proof}
W.l.o.g. assume that $N_{A} > N_{I}$. Let the kernel of the map
$M^{\dagger}$ be denoted $K$. By the standard Rank-Nullity theorem of
linear algebra we have ${\rm dim}(K) \geq (N_{A}-N_{I})$. Therefore
there exist linearly independent vectors
$\underline{e}_{1},\ldots,\underline{e}_{N_{A}-N_{I}}$ such that
$M^{\dagger}\underline{e}_{i} = 0$. Hence:
\[
\left(
\begin{array}{cc}
0 & M\\
M^{\dagger} & 0
\end{array}
\right)
\left(
\begin{array}{r}
\underline{e}_{i}\\
\underline{0}
\end{array}
\right) =
\left(
\begin{array}{r}
0\\
0
\end{array}
\right)\quad i = 1,\ldots,N_{A}-N_{I} .
\]
So a configuration with winding number $Q$ has at least $|Q|$ exact
zero eigenvalues as required. Furthermore, all these eigenvectors have
the correct chirality i.e. in the above, all the eigenvectors have
positive chirality, hence $N_{+} = |Q|,\ N_{-} = 0$ as required.

Of course we cannot say that there are not further eigenvectors with
zero eigenvalues, all we are sure of is that there are at least the
required number. Any further ``accidental'' zero eigenvalues are
dependent upon the choice of wavefunction we use in constructing
$M$. Accidental zeroes are associated with isolated objects, there
should be no isolated objects in a finite volume unless we have chosen
an artificial wavefunction. (We will in fact choose one such
wavefunction, namely a hard sphere. We normally work with a dense
enough gas such that the number of these accidental zeroes is a small
fraction of the total number of eigenvalues.)

\subsection{Our representation obeys the $\gamma^{5}$ symmetry}
We can also show that for such a matrix representation, the
$\gamma^{5}$ symmetry is obeyed, so that all non-zero eigenvalues
occur in pairs $\pm\lambda$.

\subsubsection{Proof}
The proof consists of constructing an independent eigenvector with
eigenvalue $-\lambda$. This of course amounts to nothing more than
applying the $\gamma_{5}$ symmetry explicitly:

\[
\left(
\begin{array}{cc}
0 & M\\
M^{\dagger} & 0
\end{array}
\right)
\left(
\begin{array}{r}
\underline{e}\\
\underline{f}
\end{array}
\right) =
\left(
\begin{array}{r}
M\underline{f}\\
M^{\dagger}{\underline{e}}
\end{array}
\right) =
\lambda\left(
\begin{array}{r}
\underline{e}\\
\underline{f}
\end{array}
\right) .
\]
Therefore for $\lambda \neq 0$ ($\Rightarrow \underline{e}$ and
$\underline{f} \neq 0$):
\[
\left(
\begin{array}{cc}
0 & M\\
M^{\dagger} & 0
\end{array}
\right)
\left(
\begin{array}{r}
\underline{e}\\
-\underline{f}
\end{array}
\right) =
\left(
\begin{array}{r}
-M\underline{f}\\
M^{\dagger}{\underline{e}}
\end{array}
\right) =
-\lambda\left(
\begin{array}{r}
\underline{e}\\
-\underline{f}
\end{array}
\right) .
\]
so that we have independent eigenvectors with the required
eigenvalues.

\subsection{A stitch in time ...}
We turn now to ideas which allow us to greatly increase the efficiency
of our numerical code. Ideally, we do not wish to deal with the
representation of \di\ given by~\ref{dirm} but with the matrix
squared:
\begin{equation}
(\di)^{2} \doteq D^{2} = 
\left(
\begin{array}{cc}
MM^{\dagger} & 0\\
0 & M^{\dagger}M
\end{array}
\right)
\label{dirmsq}
\end{equation}
As we shall see, $MM^{\dagger}$ and $M^{\dagger}M$ both contain all
the information contained within $D$. Furthermore, as the
dimensionality of these matrices is $N_{A} \times N_{A}$ and $N_{I}
\times N_{I}$ respectively, we will be able to work with a matrix with
at most $1/4$ of the entries of $D$. It will become apparent shortly
that we have to do far more than just calculate eigenvalues, so this
saving will make the difference between days and weeks on a computer~!
In order to use these smaller matrices, we need to relate their
eigenvalues to the eigenvalues of the original. It is of course
trivial to say that the eigenvalues of $D^{2}$ are the squares of the
eigenvalues of $D$, what we wish to say goes a little further:

\subsection{Eigenvalues of $MM^{\dagger}$ and $M^{\dagger}M$.}
Our assertion is that, if $\lambda \neq 0$ is an eigenvalue of $D$,
then $\lambda^{2}$ is an eigenvalue of $MM^{\dagger}$ {\em and of}
$M^{\dagger}M$.

\subsubsection{Proof}
If $\lambda$ is an eigenvalue of $D$ then:

\begin{eqnarray*}
M\underline{f} = \lambda\underline{e} \\
M^{\dagger}\underline{e} = \lambda\underline{f}
\end{eqnarray*}
hence for $\lambda \neq 0$ we have:
\begin{eqnarray}
MM^{\dagger}\underline{e} = \lambda^{2}\underline{e}\nonumber\\
M^{\dagger}M\underline{f} = \lambda^{2}\underline{f} .
\label{mm-evec}
\end{eqnarray}
So we have a very simple spectrum for such symmetric matrices $D$. The
squares of all non-zero eigenvalues will occur as eigenvalues of
$MM^{\dagger}$ and $M^{\dagger}M$ so we can work with whichever of the
two is smaller. The one which is larger has the same non-zero
eigenvalues as the smaller and also $|Q|$ exact zero eigenvalues. We
shall see later on that we can also reconstruct the eigenvectors of
the original matrix from the eigenvectors of the smaller matrices.

Our ostensible representation of the Dirac operator therefore shares
some of the symmetries of the true Dirac operator, namely the chiral
structure and the index theorem structure. It is possible to further
endow our representation with many of the features from the underlying
theory; for instance, in chapter~\ref{ch:latt} we actually use data
from lattice calculations for the positions and sizes of the
instantons in the configurations (so that the
decomposition~\ref{decomp} has been made of actual lattice gauge
configurations). We can further add to these qualities; one obvious
freedom we have is in the choice of the would-be zero mode
wavefunction we use in calculating matrix elements. As we shall see
later, this freedom is a two-edged sword; on one hand it allows us to
replicate some classical results, on the other, we do not wish all our
results to be wavefunction dependent, for what is the correct
wavefunction for the quantal vacuum ?

Before we come to these topics, we should address a few of the
reservations to our claims of representing the Dirac operator. Two
main reservations spring to mind.

\section{Good, but\ldots}

\subsection{The zero mode wavefunctions do not span}
The situation we have is the following. We have a linear operator
acting on a vector space defined by its actions on some ``basis'' of
vectors:
\begin{equation}
\di|e_{j}\rangle = D_{ij}|e_{i}\rangle .
\label{lin_map}
\end{equation}
However the ``basis'' of vectors
$\left\{|\psi_{1}^{+}\rangle,\ldots,|\psi_{N_{A}}^{+}\rangle,|\psi_{1}^{-}\rangle,\ldots,|\psi_{N_{I}}^{-}\rangle\right\}$
we have chosen, does not span the vector space in question. (If for
instance we take our vector space to be the space of square integrable
wavefunctions $L^{2}({\mathbb R}^{4})$, then this is an infinite
dimensional Hilbert space - it most certainly cannot be spanned by any
finite set of wavefunctions~!) Is this a problem~?

It clearly is a problem if we wish to calculate all the eigenvalues of
this linear operator (the Dirac operator). It is not a problem if we
wish to do what we have stated, calculate the low lying eigenvalues
coming from would-be zero modes. This problem is directly analogous to
the symmetrical double well potential problem (mentioned previously)
in 1-d quantum mechanics. If we wish to calculate all the eigenvalues
of the Hamiltonian for the double well then we would need a complete
set of wavefunctions. An obvious choice would be some subset of
wavefunctions calculated from each of the wells separately. If however
we only wish to calculate the splitting of the ground state then we
need only the ground states of the two wells taken separately - the
two lowest lying states for the double well are given by the sum and
the difference of the ground states for the single wells
respectively. Note: In this example the low lying (split) states are
given by linear combinations of the would-be lowest level states. This
is precisely what we are doing in our calculation.

\subsection{$D$ is not a representation of \di}
There is a more serious problem however. It is trivial to see from
equation~\ref{lin_map} that the matrix $D = \langle
e_{i}|\di|e_{j}\rangle$ iff the basis is orthonormal:

\begin{equation}
D_{ij} = \langle e_{i}|\di|e_{j}\rangle \Longleftrightarrow \langle
e_{k}|e_{l}\rangle = \delta_{kl}
\end{equation}
In our case though, we have:
\begin{eqnarray}
\langle\psi_{i}^{+}|\psi_{j}^{+}\rangle & = & \sigma_{ij}(s_{ij}^{+},
\rho_{i}^{+}, \rho_{j}^{+}, K_{ij}^{+})\qquad 1 \leq i, j \leq
N_{A}\nonumber\\
\langle\psi_{i}^{-}|\psi_{j}^{-}\rangle & = & \omega_{ij}(s_{ij}^{-},
\rho_{i}^{-}, \rho_{j}^{-}, K_{ij}^{-})\qquad 1 \leq i, j \leq
N_{A}\nonumber\\
\langle\psi_{i}^{\pm}|\psi_{j}^{\mp}\rangle & = & 0\ ,
\label{non-orth}
\end{eqnarray}
where the overlap matrix elements are dependent upon the separation
$s$, the sizes $\rho_{1,2}$ and the relative colour orientation $K$ of
the objects. (The actual matrices are denoted $\sigma$ and $\omega$
for later use, do not confuse these with the Pauli matrices or other
such objects, their matrix elements are given by the overlaps of
wavefunctions which we are free to choose.) The cross overlaps are
zero due to the chiral structure of the eigenvectors. Hence we see
that $D = \langle e_{i}|\di|e_{j}\rangle$ is not a representation of
\di\ as defined by~\ref{lin_map}. Can we salvage the situation~? We
can, though it will involve a little more work.

\section{Orthonormalization}
We shall implement the well known Gram-Schmidt orthonormalization
procedure to create a new set of basis vectors
$\left\{|\widetilde{\psi}_{1}^{+}\rangle,\ldots,|\widetilde{\psi}_{N_{A}}^{+}\rangle,|\widetilde{\psi}_{1}^{-}\rangle,\ldots,|\widetilde{\psi}_{N_{I}}^{-}\rangle\right\}$,
which are orthonormal:

\begin{eqnarray}
\langle\widetilde{\psi}_{i}^{\pm}|\widetilde{\psi}_{j}^{\pm}\rangle &
= & \delta_{ij}\nonumber\\
\langle\widetilde{\psi}_{i}^{\pm}|\widetilde{\psi}_{j}^{\mp}\rangle &
= & 0 ,
\label{ortho}
\end{eqnarray}
We define the change of basis matrices by:
\begin{eqnarray}
|\psi_{i}^{+}\rangle = \Sigma_{ij}|\widetilde{\psi}_{j}^{+}\rangle \quad 1
\leq j \leq i \leq N_{A}\nonumber\\
|\psi_{i}^{-}\rangle = \Omega_{ij}|\widetilde{\psi}_{j}^{-}\rangle \quad 1
\leq j \leq i \leq N_{I} .
\label{ch_of_basis}
\end{eqnarray}
It is simple to see from~\ref{ch_of_basis} that the new orthonormal
eigenvectors must share the same chiral properties as the originals
i.e. the coefficient of any negative chirality eigenvector in the
expansion of a positive chirality eigenvector would necessarily be
zero, and {\em vice versa}).

Once we have constructed the matrices $\Sigma$ and $\Omega$, we will
be in a position to write down a true representation of the Dirac
operator:

\begin{equation}
\begin{array}{ccc}
\hspace{1.8cm}\overbrace{\hspace{2.7cm}}^{N_{A}} & \hspace{0.0cm}\overbrace{\hspace{2.7cm}}^{N_{I}} & \\
\di \doteq \widetilde{D} = 
\left(
\begin{array}{l}
\langle\widetilde{\psi}_{i}^{+}|\di|\widetilde{\psi}_{j}^{+}\rangle = 0 \\
\langle\widetilde{\psi}_{i}^{-}|\di|\widetilde{\psi}_{j}^{+}\rangle = \widetilde{M}_{ij}^{\dagger}
\end{array}
\right. & 
\left.
\begin{array}{l}
\langle\widetilde{\psi}_{i}^{+}|\di|\widetilde{\psi}_{j}^{-}\rangle =
 \widetilde{M}_{ij}\\
\langle\widetilde{\psi}_{i}^{-}|\di|\widetilde{\psi}_{j}^{-}\rangle = 0
\end{array}
\right) &
\begin{array}{l}
\Big\}\ N_{A}\\ \Big\}\ N_{I}
\end{array}
\end{array}
\label{prop_rep}
\end{equation}
where
\begin{eqnarray}
\widetilde{M} & = & (\Sigma^{-1})^{*}M(\Omega^{-1})^{T}
\end{eqnarray}
We see that the structure of $\widetilde{D}$ is the same as the
structure of $D$ so that all that has gone before remains
unaffected. Our task now is to calculate the change of basis matrices.

\subsection{Calculating $\Sigma$}
We concentrate on the calculation of the change of basis matrix
$\Sigma$. The construction of $\Omega$ is entirely similar. We see
again that the chiral structure of the Dirac operator allows us to
effectively decompose the change of basis problem into two separate
smaller problems. As all the vectors considered in this section are of
positive chirality, we omit the ``+'' superscript to prevent the
notation becoming too cumbersome. The Gram-Schmidt orthonormalization
can be written as the following recursive process:

\begin{eqnarray}
|\psi_{1}^{\perp}\rangle & = & |\psi_{1}\rangle\nonumber\\
|\widetilde{\psi}_{1}\rangle  & = &
\frac{|\psi_{1}^{\perp}\rangle}{\langle\psi_{1}^{\perp}\rangle^{\frac{1}{2}}}\nonumber\\
& \vdots & \nonumber\\
|\psi_{i}^{\perp}\rangle & = & |\psi_{i}\rangle -
\sum_{j=1}^{i-1}|\widetilde{\psi}_{j}\rangle\langle\widetilde{\psi}_{j}|\psi_{i}\rangle\nonumber\\
|\widetilde{\psi}_{i}\rangle & = &
\frac{|\psi_{i}^{\perp}\rangle}{\langle\psi_{i}^{\perp}\rangle^{\frac{1}{2}}}\qquad
i=2,\ldots,N_{A}
\label{G-S}
\end{eqnarray}
where the intermediate vectors $|\psi_{i}^{\perp}\rangle$ are
orthogonal to all the previous orthonormalized vectors and need only
normalization to generate a new orthogonal vector. The normalization
is given by $\langle\psi_{i}^{\perp}\rangle =
\langle\psi_{i}^{\perp}|\psi_{i}^{\perp}\rangle$ as usual.

Equations~\ref{G-S} show that ${\rm
span}\{|\psi_{1}\rangle,\ldots,|\psi_{k}\rangle\} = {\rm
span}\{|\widetilde{\psi}_{1}\rangle,\ldots,|\widetilde{\psi}_{k}\rangle\}$,
$k = 1,\ldots,N_{A}$, hence $\Sigma$ is lower triangular (all elements
above the diagonal are zero).

\begin{eqnarray}
\Sigma_{ij} & = & 0 \qquad i < j \leq N_{A}\nonumber\\
\Sigma_{ij} & = & \langle\widetilde{\psi}_{j}|\psi_{i}\rangle\nonumber\\
& = & \langle\psi_{i}|\widetilde{\psi}_{j}\rangle\qquad{\rm otherwise}
\label{def_sig}
\end{eqnarray}
where the last equation follows as all matrix elements are real (the
extension to general complex elements are trivial in any case). We
will use equations~\ref{G-S} and~\ref{def_sig} to evaluate the matrix
elements of $\Sigma$. Consider the lower triangular matrix defined by:

\begin{equation}
\begin{array}{lclr}
Z_{jj} & = & \langle\psi_{j}^{\perp}\rangle & 1 \leq j \leq N_{A}\nonumber\\
Z_{ij} & = & \langle\psi_{i}|\psi_{j}^{\perp}\rangle & \qquad 1 \leq j < i
\leq N_{A}
\end{array}
\end{equation}
A simple calculation using~\ref{G-S} shows that:

\begin{eqnarray}
Z_{jj} & = & \sigma_{jj} -
\sum_{l=1}^{j-1}\frac{|Z_{jl}|^{2}}{Z_{ll}}\ \ j = 2,\ldots,N_{A}\nonumber\\
Z_{ij} & = & \sigma_{ij} -
\sum_{l=1}^{j-1}\frac{Z_{il}Z_{jl}^{*}}{Z_{ll}}\ \ 2 \leq j < i \leq N_{A}
\label{recurs}
\end{eqnarray}
The first of these equations is obvious by Pythagoras' Theorem (and is
just an example of the second). The matrix $\sigma$ is as given by
equation~\ref{non-orth}. Equations~\ref{recurs} together with $Z_{i1}
= \sigma_{i1}$ allow us to construct the entirety of the matrix
$Z$. The change of basis matrix elements are given by:

\begin{eqnarray}
\Sigma_{ii} & = & \langle\psi_{i}|\widetilde{\psi}_{i}\rangle = 
\langle\psi_{i}^{\perp}\rangle^{\frac{1}{2}} = Z_{ii}^{\frac{1}{2}} \qquad\ i =
1,\ldots,N_{A}\nonumber\\
\Sigma_{ij} & = & \langle\psi_{i}|\widetilde{\psi}_{j}\rangle = 
\frac{\langle\psi_{i}|\psi_{j}^{\perp}\rangle}{\langle\psi_{j}^{\perp}\rangle^{\frac{1}{2}}}
= \frac{Z_{ij}}{Z_{jj}^{\frac{1}{2}}} \quad 1 \leq j < i \leq N_{A}
\end{eqnarray}
As these matrices are lower triangular, their inversion is simple:

\begin{eqnarray}
(\Sigma^{-1})_{jj} & = & \frac{1}{\Sigma_{jj}} \qquad\qquad\qquad\qquad 1 \leq j \leq
 N_{A}\nonumber\\
(\Sigma^{-1})_{ij} & = & \frac{-1}{\Sigma_{ii}}\sum_{k=j}^{i-1}\Sigma_{ik}(\Sigma^{-1})_{kj}
\quad 1\leq j < i \leq N_{A}
\label{sigfinally}
\end{eqnarray}
It is therefore relatively easy to construct the matrix representation
of the Dirac operator given by equation~\ref{prop_rep}.

One may worry that the recursive nature of this algorithm will amplify
errors and hence render the procedure unstable (for instance a small
norm may become negative and hence the process break down when we take
square roots). In practice one can carry out checks to ensure that
accuracy is maintained (even before the procedure breaks down):

\begin{eqnarray}
\delta_{ij} & = &
\langle\widetilde{\psi}_{i}^{+}|\widetilde{\psi}_{j}^{+}\rangle\nonumber\\
& = & \left((\Sigma^{-1})^{*}\sigma(\Sigma^{-1})^{T}\right)_{ij} .
\end{eqnarray}
We have worked with configurations composed of over a thousand objects
without any problems with maintaining accuracy.

\subsection{Eigenvectors}
We have seen that it is possible to obtain all knowledge of the
eigenvalues of the matrix representation $\widetilde{D}$ by
calculating the eigenvalues of either
$\widetilde{M}\widetilde{M}^{\dagger}$ or
$\widetilde{M}^{\dagger}\widetilde{M}$ (we work with whichever has the
smaller dimensionality). In this section we show how to calculate the
eigenvectors of $\widetilde{D}$ from those of the smaller matrices and
analyse a property of the eigenvectors which we call ``dispersion,'' a
measure of the ``size'' of the eigenvector. In the following we assume
without loss of generality that $N_{A} \leq N_{I}$, so we have the
eigenvalues and eigenvectors associated with
$\widetilde{M}\widetilde{M}^{\dagger}$. We also assume that $\lambda
\neq 0$ so we do not have an eigenvector with definite chirality:

\[
\widetilde{M}\widetilde{M}^{\dagger}\underline{e} =
\lambda^{2}\underline{e} .
\]
Equations~\ref{mm-evec} show that the negative chirality part of the
eigenvector is given by:
\[
\underline{f} = \frac{1}{\lambda}\widetilde{M}^{\dagger}\underline{e}
\]
The ``full'' eigenvector is therefore given by:
\[
\di|e\rangle = \lambda|e\rangle
\]
where
\[
|e\rangle = (\underline{e})_{i}|\widetilde{\psi}_{i}^{+}\rangle +
(\underline{f})_{j}|\widetilde{\psi}_{j}^{-}\rangle
\]
We can use our change of basis matrices to rewrite this eigenvector in
terms of the original would-be zero mode wavefunctions (recall that
the original wavefunctions are localised around a definite centre and
have a definite size associated with them - the orthonormalized
wavefunctions are linear combinations of these and are less useful for
ideas such as ``dispersion'').
\[
|e\rangle = (\underline{c}^{+})_{k}|\psi_{k}^{+}\rangle +
(\underline{c}^{-})_{l}|\psi_{l}^{-}\rangle
\]
where
\begin{eqnarray}
(\underline{c}^{+})_{k} & = &
(\underline{e})_{i}(\Sigma^{-1})_{ik}\nonumber\\
(\underline{c}^{-})_{l} & = & (\underline{f})_{j}(\Omega^{-1})_{jl}
\label{ev_def}
\end{eqnarray}
Consider the function:
\begin{equation}
R(x^{\star}) = \langle e|(x - x^{\star})^{2}|e\rangle
\end{equation}
where $x^{\star}$ is some arbitrary trial point and the eigenvector is
assumed to be normalized $\langle e|e\rangle = 1$. Substituting in
equations~\ref{ev_def} for the eigenvector gives:

\begin{eqnarray}
R(x^{\star}) & = & \sum_{k,m=1}^{N_{A}}\int_{{\mathbb M}}
c_{k}^{+*}c_{m}^{+}\langle\psi_{k}^{+}|x\rangle\langle
x|\psi_{m}^{+}\rangle(x - x_{g}^{\star})^{2} dx\nonumber\\
& & + \sum_{l,n=1}^{N_{I}}\int_{{\mathbb M}}
c_{l}^{-*}c_{n}^{-}\langle\psi_{l}^{-}|x\rangle\langle
x|\psi_{n}^{-}\rangle(x - x_{g}^{\star})^{2} dx
\end{eqnarray}
If our wavefunctions possess spherical symmetry (the cases we deal
with are all of this type) then we can approximate the integrals:

\begin{eqnarray}
R(x^{\star}) & = & \sum_{k,m=1}^{N_{A}} c_{k}^{+*}c_{m}^{+}(x_{km}^{+}
- x^{\star})^{2}\sigma_{km}\nonumber\\
& & + \sum_{l,n=1}^{N_{I}}c_{l}^{-*}c_{n}^{-}(x_{lm}^{-} -
x^{\star})^{2}\omega_{lm}
\end{eqnarray}
where $\sigma,\ \omega$ are the usual overlap matrix
elements~\ref{non-orth} and $x_{ab}$ is determined by symmetry as a
point on the line joining the two centres. If the objects are the same
size then it is simply the mid-point of the line (so with our volume
${\mathbb M} = {\mathbb R}^{4}$ it is given by $x_{ab} =
(x_{a}+x_{b})/2$ - things are of course not so simple on a four
dimensional torus ${\mathbb M} = {\mathbb T}^{4}$ which is the
manifold we will normally use to calculate the overlap matrices). We
define the dispersion $W$ of the eigenvector to be:

\begin{equation}
W = \min_{x^{\star} \in {\mathbb M}} R^{1/2}(x^{\star}) .
\end{equation}
This is a ``natural'' measure of the spread or width of the
eigenvector and we shall use it to investigate how eigenvector spread
is correlated with eigenvalue (e.g. are small eigenvalues related to
isolated objects, hence have small spreads ?).

We have seen that we can construct a representation of the Dirac
operator which has many of the prerequisites of a true
representation. At this stage we still have a few arbitrary functions
to choose, namely those associated with overlaps of zero mode
wavefunctions (used to construct $\sigma$ and $\omega$) and those
associated with these overlaps sandwiching the Dirac operator (used to
construct $M$). Once we have chosen these functions, we will be in a
position to compute eigenvalues of $\widetilde{D}$ for any given
configuration of instantons and anti-instantons. By running through an
ensemble of $N_{c}$ such configurations we will be able to build up a
spectral density:

\begin{equation}
\overline{\nu}(\lambda)\Delta\lambda = \frac{N(\lambda -
\Delta\lambda/2, \lambda + \Delta\lambda/2)}{VN_{c}} ,
\end{equation}
where $N(a,b)$ are the number of eigenvalues in the range $[a,b]$
during the run i.e. the number density of eigenvalues for our
ensemble.

We have stressed already that we do not include the Atiyah-Singer
exact zero eigenvalues in this equation. Chapter~\ref{ch:univ} is
concerned with calculating spectral densities for a variety of
different choices of functions mentioned above. If all properties of
our ostensible Dirac representation are dependent upon the choice of
arbitrary functions then we can have little faith in any results.

%% file: univ.tex
\chapter{Universality}
\label{ch:univ}

\section{Introduction}
The algorithm outlined in the previous chapter has considerable scope
for freedom. Firstly we have freedoms associated with computing matrix
elements. We must decide the form of the would-be zero mode
wavefunction so that we may construct $\sigma_{ij}$ and $\omega_{ij}$
(see equations~\ref{non-orth}). This is necessary if we are to build
the change of basis matrices $\Sigma_{ij}$ and $\Omega_{ij}$ (see
equations~\ref{ch_of_basis}). We must also select a function for the
matrix elements of the Dirac operator between would-be zero modes of
opposite chirality (see equation~\ref{dirm}). This is required to
construct the ``raw'' Dirac matrix $M_{ij}$ (see
equations~\ref{ch_of_basis}). One may wonder why we are free to choose
this function, is it not completely specified once we have an explicit
form for the would-be zero mode wavefunctions and some gauge field
configuration composed of instanton degrees of freedom ? In theory it
is of course; in practice, however, it cannot be computed within a
framework such as our model, and hence we approximate for the presence
of the Dirac operator. These parameters are related to the objects
within a configuration. There are also ``global'' parameters which
determine the configuration as a whole.

Firstly there is the volume of the box $V$ which we are free to
choose. To construct a synthetic configuration we also need to know
how many objects to place within the volume. This will determine the
dimensionality of our matrix representation $\widetilde{D}$. In
reality this is some quantity which is fixed by dynamics (the complex
trade-off between the gauge and fermion parts of the action); in our
model it is a free parameter. We define a packing fraction $f$ as the
average number of topological objects multiplied by their average
volume and divided by the total volume of spacetime:

\begin{equation}
f = \frac{\overline{N}_{T}\overline{V}_{I}}{V}
\label{pack_frac}
\end{equation}
In this chapter, we will deal with objects of a fixed ``size'' i.e all
objects in an ensemble of configurations are such that $\rho_{i}^{\pm}
= \rho$. We define the ``volume'' of an object by $V_{I} =
\pi^{2}\rho^{4}/2$, the volume of a sphere of radius $\rho$ in four
dimensions (as we shall see, we use this formula only when we use
actual hard sphere wavefunctions, we then rely on ``calibration'' to
establish the packing fraction for other wavefunctions). So to
summarize, we have two free parameters. We have some given size
distribution for the instantons, and we are free to choose the volume
of the box and the number of objects to place within the box.

A last freedom is the freedom to choose the manifold over which we are
to compute the overlap integrals. In most cases we choose the manifold
to be a four dimensional box with boundaries identified i.e. the four
torus ${\mathbb T}^{4}$. When this is not possible (for instance the
classical would-be zero mode~\ref{zerom} is for Euclidean spacetime,
not toroidal), or for comparison purposes, we sometimes also do
calculations with overlaps computed over four dimensional Euclidean
spacetime ${\mathbb R}^{4}$.

In this chapter we follow the path of pragmatism. We will use a
variety of wavefunctions, and, ans\"{a}tze for the presence of the
Dirac operator. We will use a range of packing fractions and volumes
and compute overlap integrals on the two different manifolds. We do
this as we do not know the form or values of these quantities in
nature (or even if if makes sense to think of parameterizing nature
using such quantities). We hope that the results are qualitatively
independent of the wavefunction and ansatz used, and that they behave
in a predictable way with relation to packing fraction and volume.

Can we constrain the form of the would-be zero mode wavefunction ? A
minimal requirement is that the function be square integrable,
otherwise our program cannot proceed at all. A second requirement is
that the zero mode wavefunction be localised. In particular, we
require the wavefunction to be localised around the centre of the
instanton. By localised we mean that the wavefunction should decay
rapidly enough so as to occupy a small volume in comparison to the
total volume (we can make the requirement more precise by asking for a
certain percentage of the square integral of the wavefunction to be
within a sphere of some fraction of the box length). We would worry if
the volume of the box was such that the wavefunction was almost a
constant throughout the volume. (This would also cause problems for
the orthonormalization procedure.) Keeping these requirements in mind,
we work with three different wavefunctions namely the hard sphere,
Gaussian and classical zero mode wavefunctions. We use a simple ansatz
for the presence of the Dirac operator, its effect is to introduce a
quantity with the dimension of momentum into the matrix element, swap
the chirality of one of the objects (i.e. it is an operator which
anti-commutes with $\gamma_{5}$) but otherwise acts as the unit
operator. For comparison purposes we will also use another ansatz for
the presence of the Dirac operator based upon a parameterization of
the the operator sandwiched between classical zero mode
wavefunctions. Before we move onto the details of the calculation, we
summarize the main results of this work.

\begin{itemize}
\item{The spectral density is qualitatively independent of the
wavefunction used.}
\item{The spectral density can be parameterized as a power law
$\overline{\nu}(\lambda) = a + b/\lambda^{d}\ \ d \geq 0$ for small
$\lambda$, in particular we find a divergence as $\lambda \rightarrow
0$.}
\item{The power of the divergence $d$ is inversely related to the
density of objects. In particular for high density ``gases'' (high
packing fraction) of instantons, we find $d \approx 0$.}
\item{Finite size effects are small and well under control.}
\item{The power of the divergence is greater if we sample only the $Q
= 0$ sector of net topological charge. The divergence is depleted if
we sample all sectors with some suitable weight. The difference
disappears however, as we take the thermodynamic limit, hence we
minimize finite size effects by concentrating on the $Q=0$ sector.}
\item{The divergence is due to multi-particle interactions. It is not
due to weakly overlapping single instanton--anti-instanton
interactions.}
\item{The dispersion of the corresponding eigenvectors increases
approximately linearly as the eigenvalue decreases. This reinforces
the above result.}
\end{itemize}

Some preliminary results based on these calculations (or on similar
calculations) have been given elsewhere~\cite{Dowrick,Sharan-Proc}. A more
detailed exposition of these results has also been
published~\cite{Sharan-Univ}.

\section{Hard sphere}
\label{hs_sect}
Consider an instanton described by $(x_{j}^{\pm},\rho_{j}^{\pm})$,
using the notation of equation~\ref{coll-co}. One of the simplest
wavefunctions one could try is perhaps the hard sphere:

\begin{equation}
\begin{array}{lcll}
\langle x|\psi_{j}^{\pm}\rangle & = & 1 & \quad |x - x_{j}^{\pm}| \leq
\rho_{j}^{\pm}\\
& = & 0 & \quad {\rm otherwise}
\end{array}
\label{hard_sphere}
\end{equation}
The great advantage with such wavefunctions is that the overlap matrix
elements are given by closed form expressions (see
Appendix~\ref{app:ovlap} for relevant formul\ae). This greatly
increases the speed of the computation and is the main reason why we
use this form for the majority of the calculations in this thesis. It
certainly has some slight drawbacks however, the principal being that
it is ``unnatural'' in that it has a sharp cutoff. This allows for
isolated objects even at finite volume and can invalidate the
Atiyah-Singer theorem by having too many exact zero eigenvalues. (An
object which does not overlap with any object of the opposite
chirality automatically forces an exact ``accidental'' zero
eigenvalue.) These problems are less significant if we work with dense
gases $f \geq 0.5$ which will form the majority of our work.

We make the following set of approximations for the presence for the
Dirac operator.

\begin{eqnarray}
\langle\psi_{k}^{+}|\di[A]|\psi_{l}^{-}\rangle & \approx &
\langle\psi_{k}^{+}|\di[\sum_{i} A_{i}^{+} +
\sum_{j} A_{j}^{-}]|\psi_{l}^{-}\rangle\nonumber\\
& \approx & \langle\psi_{k}^{+}|\di[A_{k}^{+} +
A_{l}^{-}]|\psi_{l}^{-}\rangle\nonumber\\
& = & \langle\psi_{k}^{+}|\di[A_{k}^{+}]|\psi_{l}^{-}\rangle\nonumber\\
& = & \langle\psi_{k}^{+}|-i\!\not\!\partial|\psi_{l}^{-}\rangle\nonumber\\
& \approx & \frac{1}{\sqrt{\rho_{k}^{+}\rho_{l}^{-}}} \langle\psi_{k}^{+}|\psi_{l}^{-}\rangle
\label{seq_approx}
\end{eqnarray}
The first approximation is a consequence of the linear addition
ansatz~\ref{decomp}. The second approximation holds if the objects are
well separated from one another, however it does not hold in general
(especially for the high density gases to which we sometimes apply our
model). It is however necessary if we are to progress beyond the
``dilute'' gas approximation using a model such as ours. The
equalities between the second and third approximations hold due to the
equations of motion~\ref{eq_motion}. The third approximation, which
substitutes for the free Dirac operator, gives the overlap the correct
mass dimension (one should also think of the Dirac operator as
changing the chirality of one of the objects so that the last line is
not identically zero). The quantity $\sqrt{\rho_{k}^{+}\rho_{l}^{-}}$
is the geometric mean of the radii of the two objects. We shall return
to this sequence of approximations shortly.

\begin{figure}[tb]
\begin{center}
\leavevmode
\epsfxsize=100mm
\epsfbox{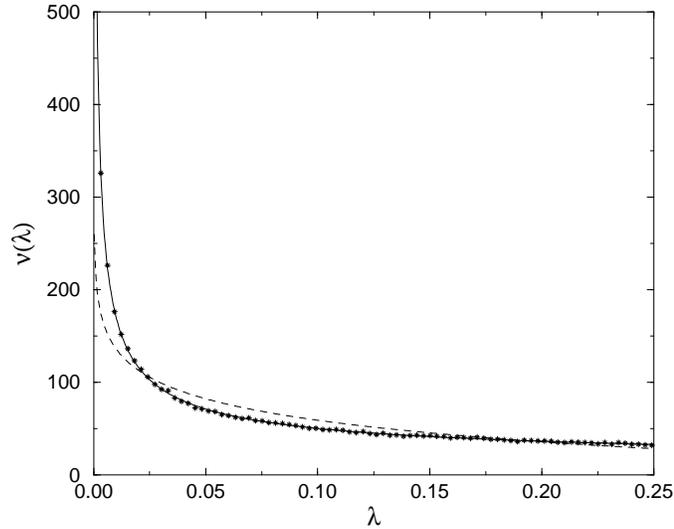}
\end{center}
\caption{The spectral density obtained at $f=1,\ V=1$ for hard
sphere wavefunctions ($\star$). The best power fit (solid line) and
best log fit (dashed line) are also given.}
\label{fig:h_stan}
\end{figure}

\begin{figure}[tb]
\begin{center}
\leavevmode
\epsfxsize=100mm
\epsfbox{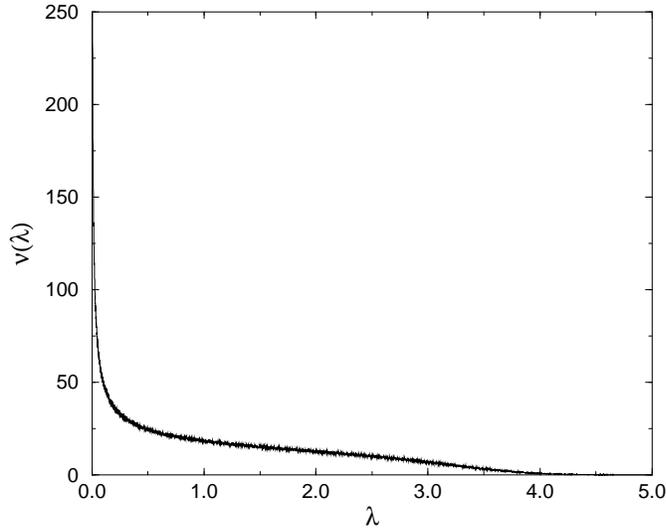}
\end{center}
\caption{The spectral density as in figure~\ref{fig:h_stan} plotted
for a larger range of $\lambda$.}
\label{fig:h_stan_lr}
\end{figure}

We are finally in a position to do some calculations~! We begin with
the simplest case; each configuration consists of a fixed number of
objects of a fixed size (hence every individual configuration has the
same packing fraction as the entire ensemble) with net winding number
zero ($\langle Q\rangle = 0,\ \langle Q^{2}\rangle = 0$). Calculations
are on a four torus with $V=1$. The packing fraction is $f=1$ and we
have chosen the size to be $\rho=0.2$ so that each configuration has
63 instantons and 63 anti-instantons. Figure~\ref{fig:h_stan} shows
the spectral density obtained for an ensemble of 126000 random
configurations (there is no fermion or gauge weighting involved,
hence, each configuration is a random ``snapshot'' of the vacuum). As
we can see, the spectral density appears to diverge as $\lambda
\rightarrow 0$. We try to parameterize the spectral density as a power
divergence:

\begin{equation}
\overline{\nu}(\lambda) = a + \frac{b}{\lambda^{d}} ,
\label{power-law}
\end{equation}
and a log divergence:

\begin{equation}
\overline{\nu}(\lambda) = a + b\ln(\lambda) .
\label{log-law}
\end{equation}
We can see, even by eye, that the power divergence is a far superior
fit to the data than the log divergence. The high statistics involved
in the run (approximately $16000000$ eigenvalues have been computed to
generate figure~\ref{fig:h_stan} - no error bars are shown on the
spectral density for they would be too small to see, and, only one
fifth of the bins are actually plotted to prevent the diagram from
becoming too crowded) imply that the fitting is a highly non-trivial
test. The standard $\chi^{2}$ analysis shows that the power fit (for
the region $\lambda \in [0,0.3]$) has $\chi^{2}/N_{DF} = 1.38$ whereas
the log fit has $\chi^{2}/N_{DF} = 258$. It is fair to say that this
data favours the power fit over the log fit~! A summary of all the
data presented in this chapter is given in
table~\ref{tab:basicdat}.
It is important to note that the fits we propose above should only
hold for small $\lambda$, if at all. We are free, to a large extent,
to decide exactly what we mean by ``small'' $\lambda$. We have chosen
a generous range to fit to, if we were worried by high $\chi^{2}$
figures for the power fit (the log fit is pretty hopeless in any case)
then we could restrict ourselves to $\lambda \in [0,0.2]$ for
instance, and the fit would improve further. To emphasise this point,
we show in diagram~\ref{fig:h_stan_lr}, the same spectrum as
figure~\ref{fig:h_stan}, but plotted for $\lambda \in [0,5.0]$. We do
not claim that a power fit or a log fit could accurately model the
entirety of this spectra. Any universality which is conjectured is
only applicable in the limit as $\lambda \rightarrow 0$.

Returning to the power law fit, the power of the divergence is
calculated to be $d \approx 0.595 \pm 0.002$. All error estimates
given in this work have been carried out using the ``jack-knife''
method. This is summarized in appendix~\ref{app:jk}. Before we make
any conclusions, we should check the behaviour of the spectral density
as we alter the volume of the box (holding the packing fraction
constant), and as we alter the packing fraction of the box (holding
the volume constant).

\begin{figure}[tb]
\begin{center}
\leavevmode
\epsfxsize=100mm
\epsfbox{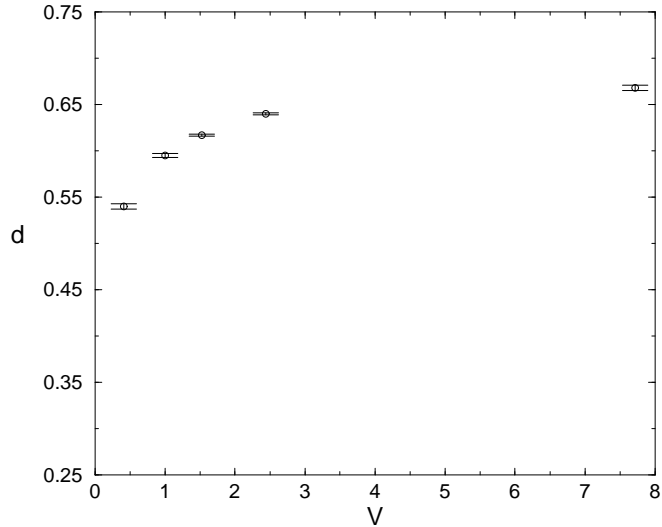}
\end{center}
\caption{The divergence of the power law fit ($d$) as a function of
the volume $V$ of the box. Results are from hard sphere wavefunctions
($\circ$). All calculations have been done at fixed packing fraction
$f=1$.}
\label{fig:h_stan_V}
\end{figure}

\begin{figure}[tb]
\begin{center}
\leavevmode
\epsfxsize=100mm
\epsfbox{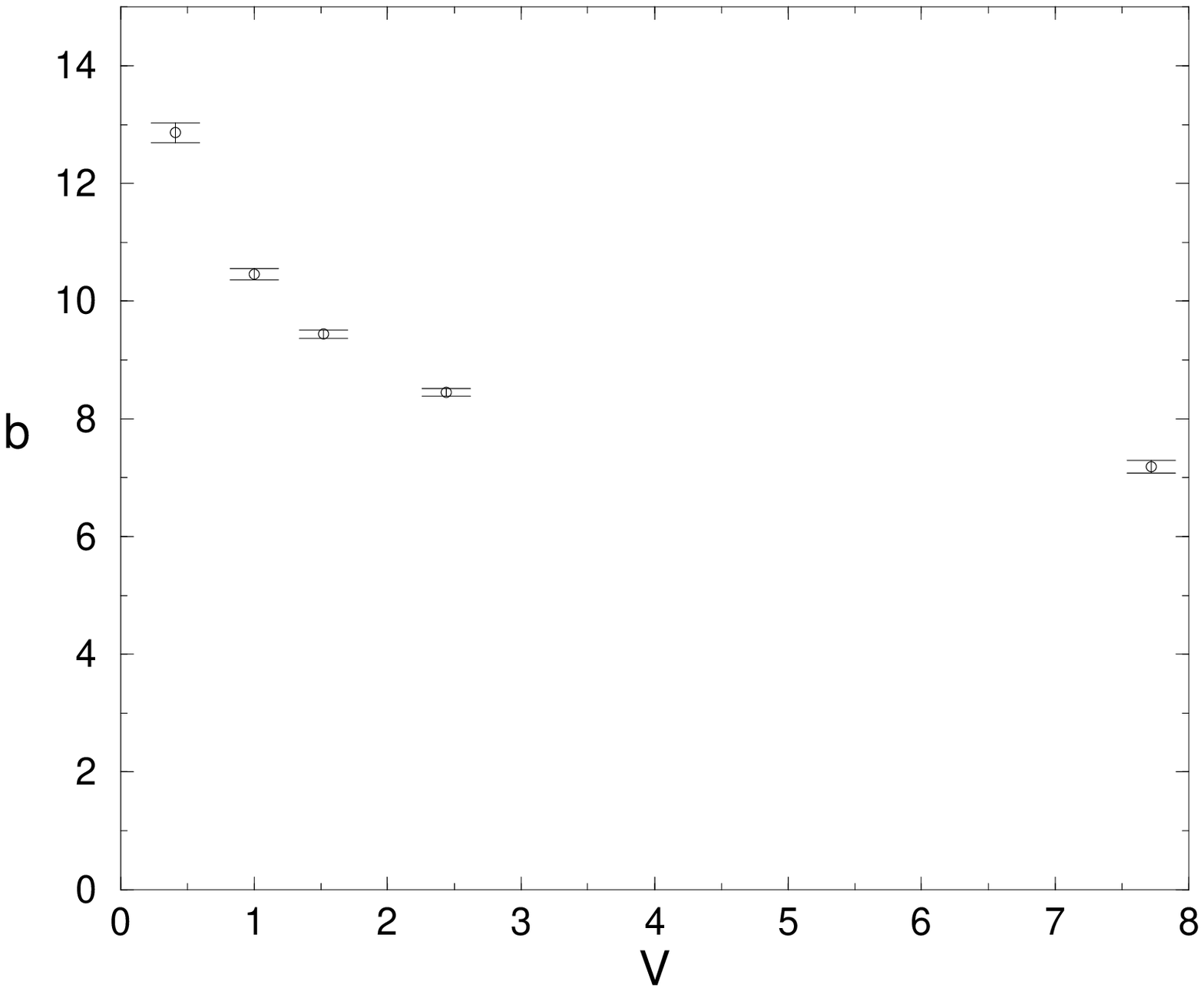}
\end{center}
\caption{The coefficient of the divergence ($b$) from the power law
fit as a function of the volume $V$ of the box. Results are from hard
sphere wavefunctions ($\circ$). All calculations have been done at
fixed packing fraction $f=1$.}
\label{fig:h_stan_Vb}
\end{figure}

Figure~\ref{fig:h_stan_V} plots the divergence (as given by the power
fit) as a function of the volume. We see that the divergence reaches
as plateau at around $d \approx 0.65$ which we can think of as the
power of the divergence in the infinite volume
limit. Figure~\ref{fig:h_stan_Vb} shows the corresponding diagram for
the coefficient $b$ of the divergence as a function of the volume. We
see that this also smoothly approaches a non-zero infinite volume
limit. (We do not plot the different spectral densities as the
variation between them is not easily noticeable on a small graph.)
These results are reassuring, the infinite volume limit is at hand and
finite volume effects are small. We will carry out most of our
calculations at $V = 1$ as this is not too far from the infinite
volume limit and is small enough to allow high statistic calculations
in (relatively) short periods of time.
Greater details of these results are given in
table~\ref{tab:basicdat}. In particular we note that the log fit for
the divergence becomes progressively worse as the volume increases
(though it is always pretty awful).

\begin{figure}[tb]
\begin{center}
\leavevmode
\epsfxsize=100mm
\epsfbox{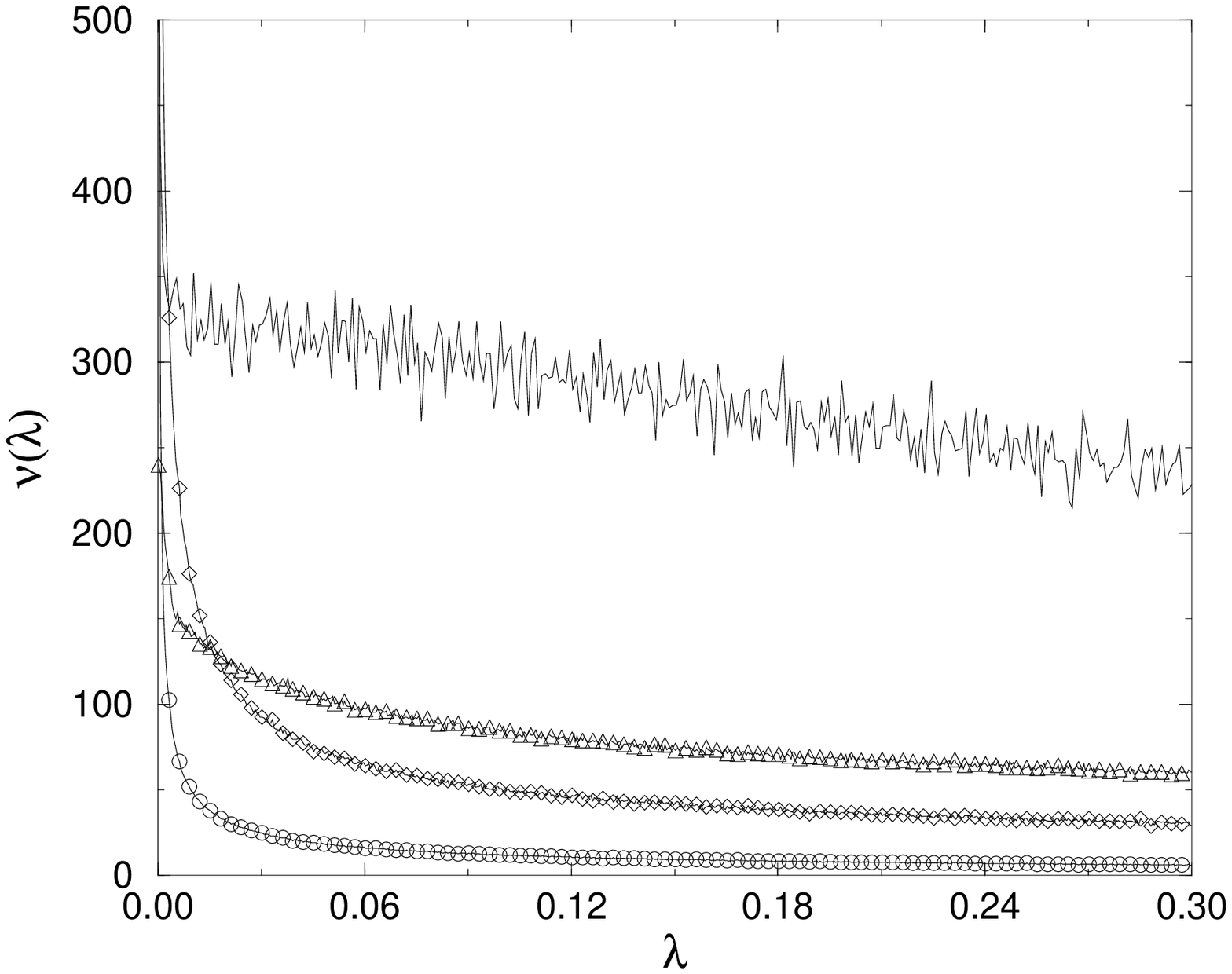}
\end{center}
\caption{Spectral densities from configurations generated by the
random-position model, for various packing fractions, keeping
the volume fixed and varying the number of charges as shown:
$f = 0.2$ ($\circ\ N = 13 + \overline{13}$), 1.0 ($\diamond\ N = 63 +
\overline{63}$), 2.5 ($\triangle\ N = 153 + \overline{153}$) \& 10.0
 (solid $N = 633 + 633$).}
\label{fig:h_xxx_020}
\end{figure}

\begin{figure}[tb]
\begin{center}
\leavevmode
\epsfxsize=100mm
\epsfbox{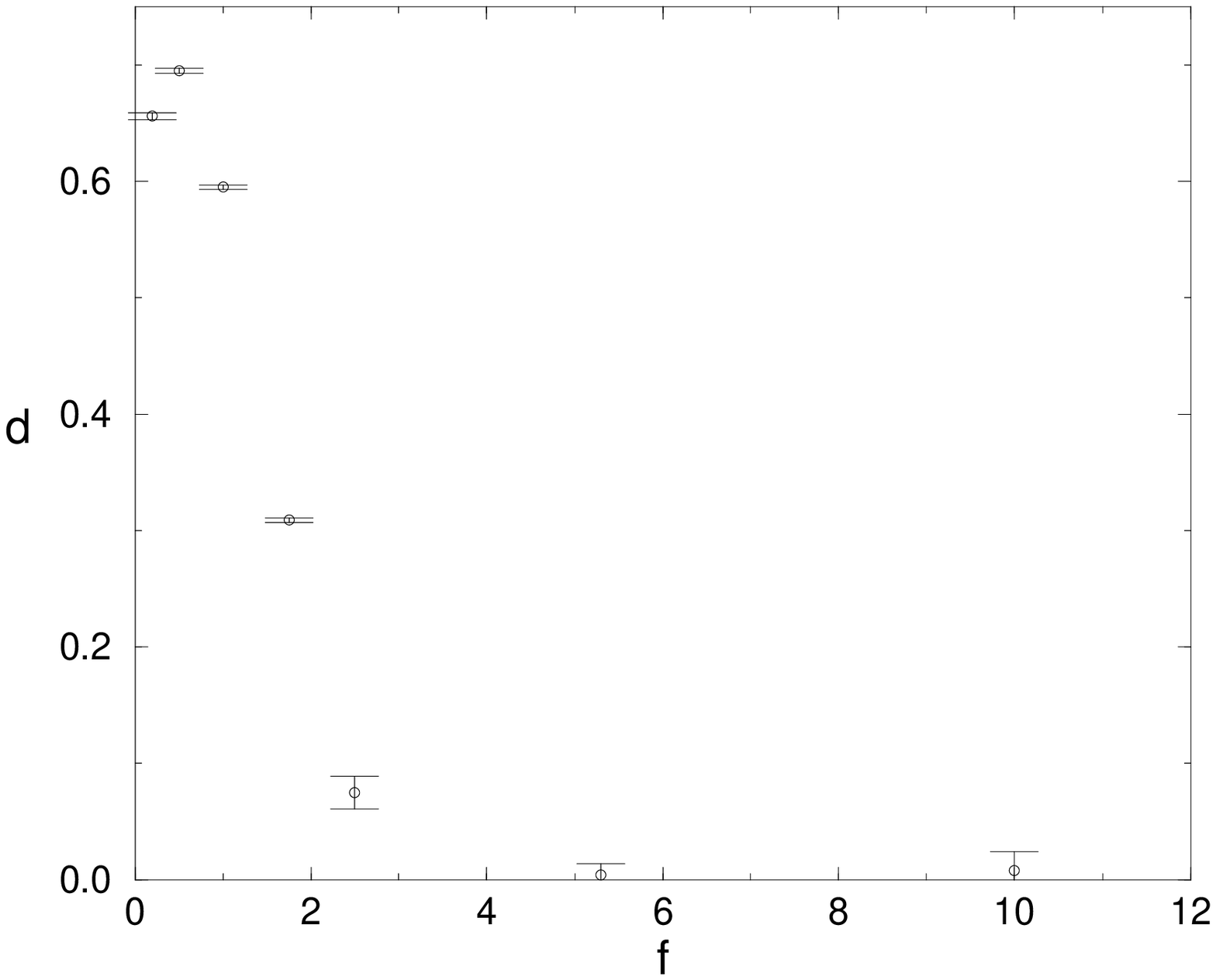}
\end{center}
\caption{The divergence of the power law fit ($d$) as a function of
the packing fraction $f$ of the system. Results for hard sphere
wavefunctions ($\circ$) are shown. All calculations have been done at
fixed volume $V=1$.}
\label{fig:h_stan_f}
\end{figure}

Figure~\ref{fig:h_xxx_020} shows the the spectral densities for a
variety of packing fractions holding the volume constant.
We get a divergence which weakens as the packing fraction in increased
(so that for the highest packing fraction $f = 10$ it is
negligible). These results are borne out by the detailed analysis
summarized in table~\ref{tab:basicdat}. (A greater variety of packing
fractions have been generated than shown in
figure~\ref{fig:h_xxx_020}, this is to prevent the diagram from
becoming too crowded.) The power of the divergence as a function of
the packing fraction is shown explicitly in
figure~\ref{fig:h_stan_f}. We see clearly that the divergence is large
for low packing fractions but becomes negligible for ultra dense
gases.
(The reason why the power of the divergence decreases for the lowest
packing fraction is because we are using hard sphere
wavefunctions. When the gas becomes very dilute, we begin to get a
great number of exact zero eigenvalues - these are not included in the
spectral density.)

\begin{figure}[tb]
\begin{center}
\leavevmode
\epsfxsize=100mm
\epsfbox{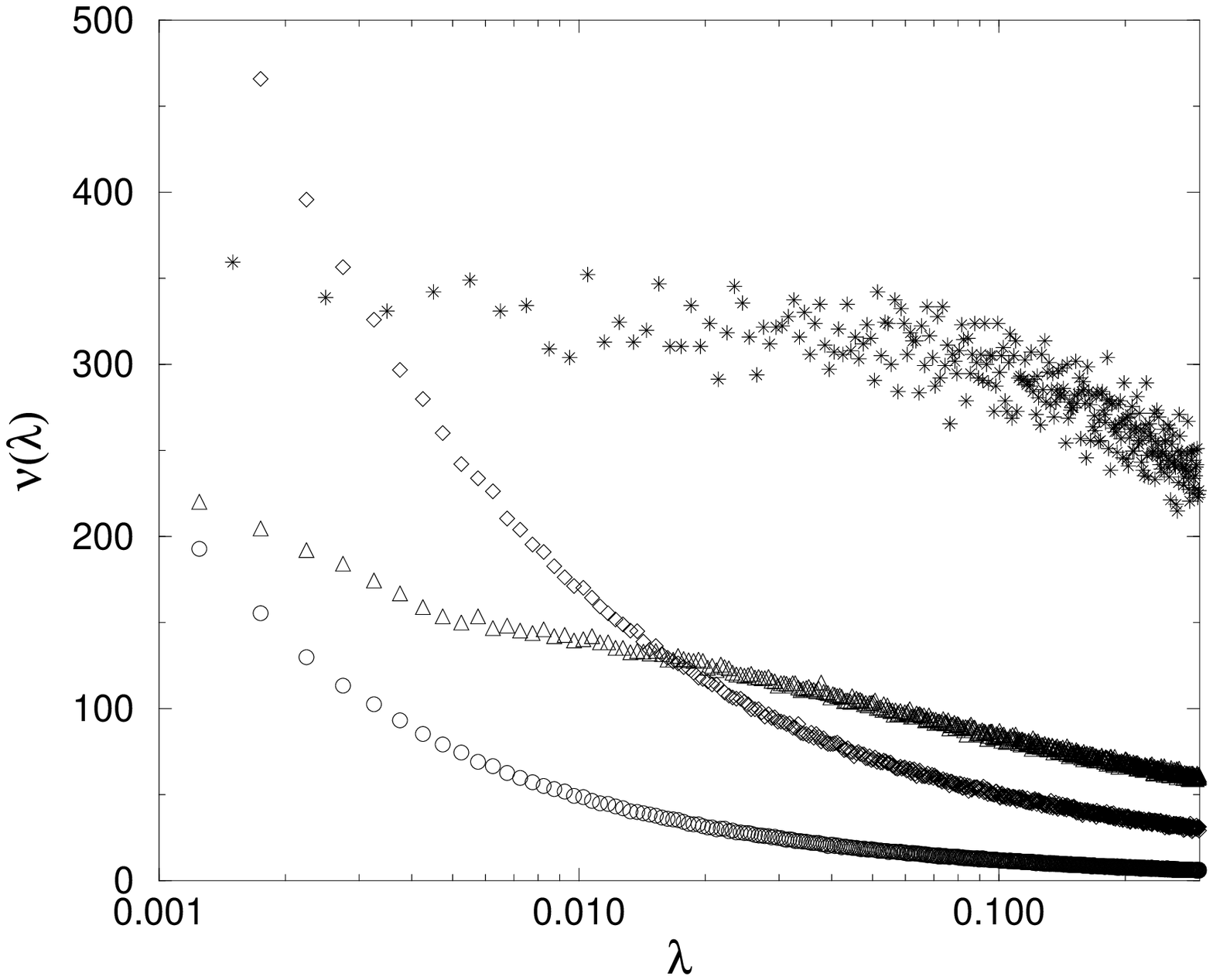}
\end{center}
\caption{Spectral densities in figure (\ref{fig:h_xxx_020}), 
plotted on log-linear axes.}
\label{fig:h_xxx_020_lnlin}
\end{figure}

\begin{figure}[tb]
\begin{center}
\leavevmode
\epsfxsize=100mm
\epsfbox{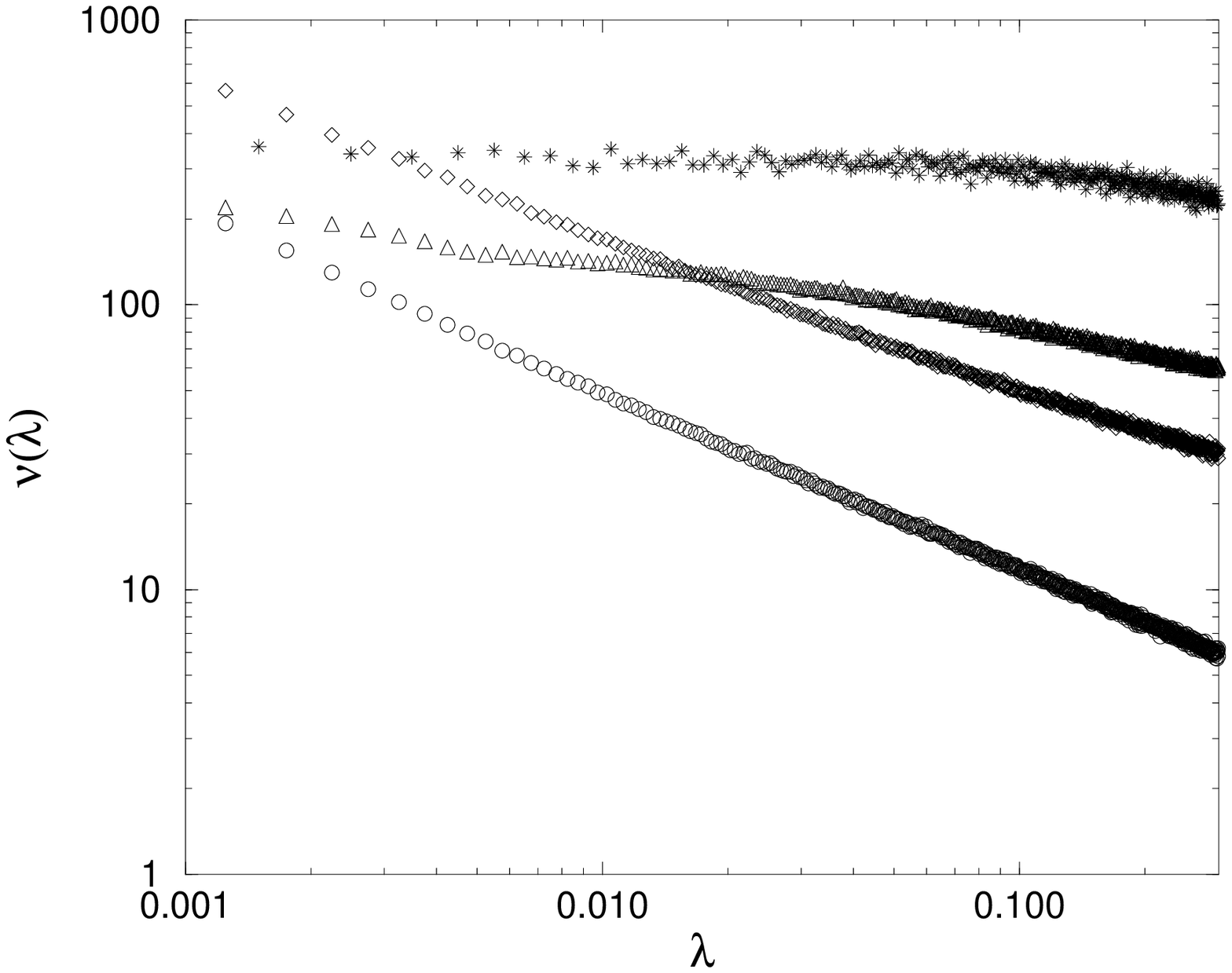}
\end{center}
\caption{Spectral densities in figure (\ref{fig:h_xxx_020}), 
plotted on log-log axes.}
\label{fig:h_xxx_020_lnln}
\end{figure}

We may test the validity of the log and power fits by plotting the
spectral densities on log-linear and log-log plots respectively (the
data should become approximately linear when plotted in such a fashion
if the fits are correct). We show the results of doing so on
figures~\ref{fig:h_xxx_020_lnlin} and~\ref{fig:h_xxx_020_lnln}. This
only reinforces our belief that the data follows a power divergence.
We see from these diagrams and from table~\ref{tab:basicdat} that the
log fit appears to improve somewhat for the high density
data. Presumably this is a trivial consequence of the power of the
divergence being small for such high density configurations and hence:
\begin{equation}
\label{eq:log-power}
b\lambda^{-d} \simeq b - bd\ln(\lambda)
\end{equation}
for small values of the exponent $d$.

These results are reassuring as far as our model is concerned. The
spectral density behaves smoothly as the underlying parameters are
altered. However the results appear pathological. If these results are
true then we are inevitably driven to the conclusion that the spectral
density diverges as $\lambda \rightarrow 0$ in quenched QCD, and
hence, that the chiral condensate diverges in quenched QCD as well
(see equation~\ref{bc}). Is this not a problem ? The answer is, is
that it need not be a problem. We already know that quenched (or even
partially quenched) QCD is not a proper quantum field theory (e.g. it
does not have a Hermitean Hamiltonian), so it should not surprise us
that pathologies lurk hidden within. We would of course be worried if
we were to find such problems when we simulate full QCD (as we will do
in chapter~\ref{ch:unq}). (Recall that a divergent spectral density at
finite quark mass is not necessarily a problem for full QCD, as the
spectral density is itself mass dependent, and such a spectral density
can still lead to a finite quark condensate.) It would be reassuring
though if we could understand how the divergence has originated. A
simple model is perhaps of help in this matter.

In a volume $V$ we expect $\sqrt{V}$ exact zero eigenvalues (we throw
$V$ charges into the box, each with topological charge $\pm 1$ with
equal probability, hence $\langle Q^{2}\rangle \propto V$). If we take
two such volumes and join them together, then by the same argument, we
expect $\sqrt{2V}$ exact zero eigenvalues. The two volumes separately
however, had $2\sqrt{V}$ exact zero eigenvalues. What has happened to
the remaining $(2 - \sqrt{2})\sqrt{V}$ would-be exact zero eigenvalues
? We already know the answer to this, they have split symmetrically
from zero due to mixing. If the two volumes are large however, we
would not expect mixing between objects scattered in the two volumes
to be strong enough to drive these would-be zero modes, far from zero
(just think of ever larger volumes for the plausibility of this
argument). The ever weaker mixing helps to fill in the gap in the
eigenvalue spectrum at $\lambda = 0$ which we get at finite
volume. Hence we expect an accumulation of eigenvalues at zero. This
sort of simple argument makes the idea of a divergent spectral density
a little more palatable. We do not have to accept such an argument
however, we do not actually know where the mixed eigenvalues go to,
perhaps they do not help to fill in the eigenvalue gap ? Perhaps the
pathology we have found is simply a function of the artificial
wavefunction we have used ? After all, the wavefunction is not even
smooth, and furthermore, our ansatz for the Dirac operator is
questionable i.e. shouldn't the derivative in the Dirac operator force
the overlap integrand to be zero everywhere apart from an infinitely
thin ``shell'' at the edge of the hard spheres (where it is
divergent)~? These criticisms of the hard sphere wavefunction are
real, we must check to see if they make a difference.

\section{Gaussian}
A second candidate wavefunction is the Gaussian wavefunction and may
be contructed as follows. For simplicity assume our volume is given by
the unit four torus (the generalization to larger volumes is
immediate). Consider the Gaussian function in ${\mathbb R}^{4}$,

\begin{eqnarray}
G(x;x_{j}^{\pm},\sigma_{j}^{\pm},l) & = &
\frac{1}{\sqrt{2\pi}\sigma_{j}^{\pm}}\exp\left(\frac{-(x-x_{j}^{\pm}-l)^{2}}{2\sigma_{j}^{\pm2}}\right).
\label{gaussa}
\end{eqnarray}
where $x_{j}^{\pm}$ lies in the unit four torus with $l \in {\mathbb
Z}^{4}$. The Gaussian zero mode wavefunction is given by:

\begin{eqnarray}
\langle x|\psi_{j}^{\pm}\rangle & = &
N\sum_{l \in {\mathbb Z}^{4}}G(x;x_{j}^{\pm},\sigma_{j}^{\pm},l),
\label{gaussian}
\end{eqnarray}
where $N$ is a suitable normalization constant. This is, by
construction, a smooth function on ${\mathbb T}^{4}$. The overlap
integral of such functions on a torus is given in
Appendix~\ref{app:ovlap}. The overlap is given as a series with the
terms suppressed exponentially in distance. We may truncate the series
safely after summing over some set of boxes (we normally sum over the
$3^{4}$ boxes around the unit cube).

The only thing left to decide is, what it means for an instanton to be
of a certain ``size''. Consider an object which we model by a hard
sphere of radius $\rho$. What should the corresponding $\sigma$ be, if
we wish to model the object with a Gaussian wavefunction of the form
given above ? Our method of ``calibrating'' the two wavefunctions is
essentially pragmatic. We consider hard sphere configurations, in a
fixed volume $V$, fixed packing fraction $f$, consisting of objects of
a single size $\rho$. For instance the configurations of
figure~\ref{fig:h_stan} had $V=1$ (the unit four torus), $\rho=0.2,\
N_{A}+N_{I}=63+63$ giving $f=1$. We extract the spectral density
from these configurations and fit it using the power law form given in
equation~\ref{power-law}. We then generate configurations using the
Gaussian wavefunction of a size $\sigma$ in the same volume $V$ with
the same number of objects as before. We extract the spectral density
power law fit and compare it to the hard sphere case. We repeat with
different $\sigma$ until we get a good qualitative match in the
spectral densities. If this is possible, then we say the matched
configuration also has packing fraction $f$. This is not a trivial
test; it is by no means obvious that the spectral density will be in
any way similar to the hard sphere case, or that it will be possible
to fit it accurately with the power law fit. Once this ``calibration''
has been made, it should then be possible to test for finite size
effects and variation with packing fraction independently for the two
wavefunctions.

In the concrete example given the calibration required
$\sigma=0.074$. Though this is apparently very different, the
pertinent quantity for calculating the spectral density in our model
is not the ``size'' of the wavefunction, but the distribution of
overlaps. A simple check shows that with this calibration, the mean
overlap is very similar for hard sphere and Gaussian wavefunctions,
which is reassuring (see table~\ref{tab:basicdat}).
\begin{figure}[tb]
\begin{center}
\leavevmode
\epsfxsize=100mm
\epsfbox{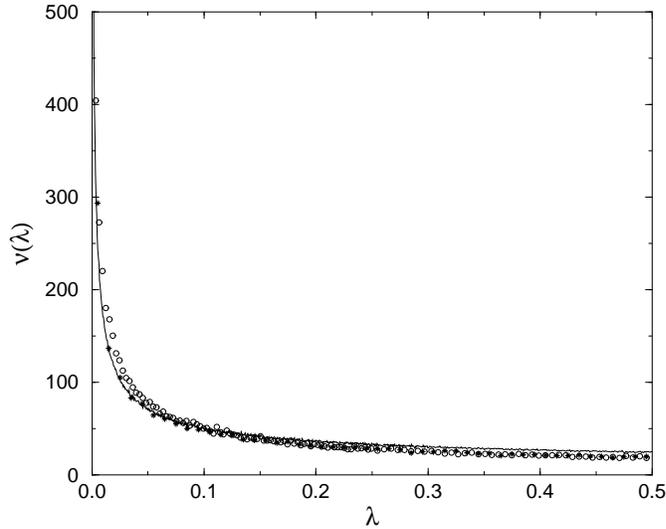}
\end{center}
\caption{The spectral densities obtained at $f=1,\ V=1$ from hard
sphere (solid), Gaussian ($\circ$) and classical zero mode ($\star$)
wavefunctions.}
\label{fig:hgc}
\end{figure}
In figure~\ref{fig:hgc} we plot a comparison of the spectral density
of the hard sphere wavefunction with the calibrated Gaussian
wavefunction for comparison (we will come onto the classical zero mode
wavefunction shown in the diagram later). We find them to be very
similar indeed, a fact borne out by the analysis given in
table~\ref{tab:basicdat}.

\begin{figure}[tb]
\begin{center}
\leavevmode
\epsfxsize=100mm
\epsfbox{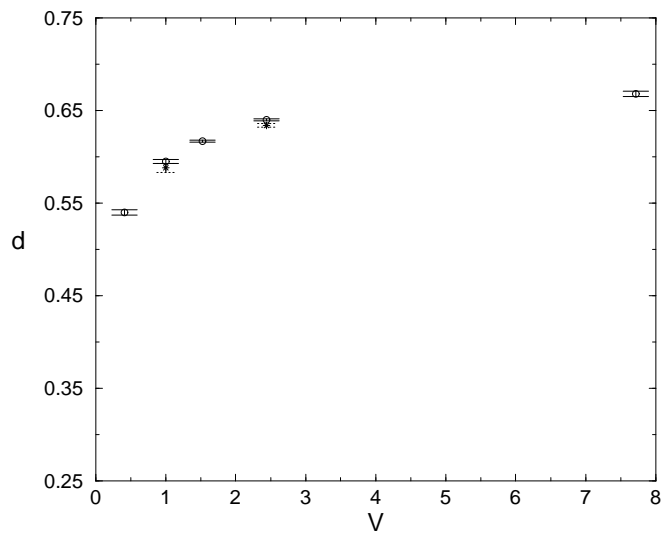}
\end{center}
\caption{The divergence of the power law fit ($d$) as a function of
the volume $V$ of the box. Calibration is done at $V=1$. Results for
hard sphere wavefunctions ($\circ$) and Gaussian wavefunctions
($\star$ and narrow-dashed error bars) are shown. All calculations
have been done at fixed packing fraction $f=1$.}
\label{fig:hg_dvsV}
\end{figure}

\begin{figure}[tb]
\begin{center}
\leavevmode
\epsfxsize=100mm
\epsfbox{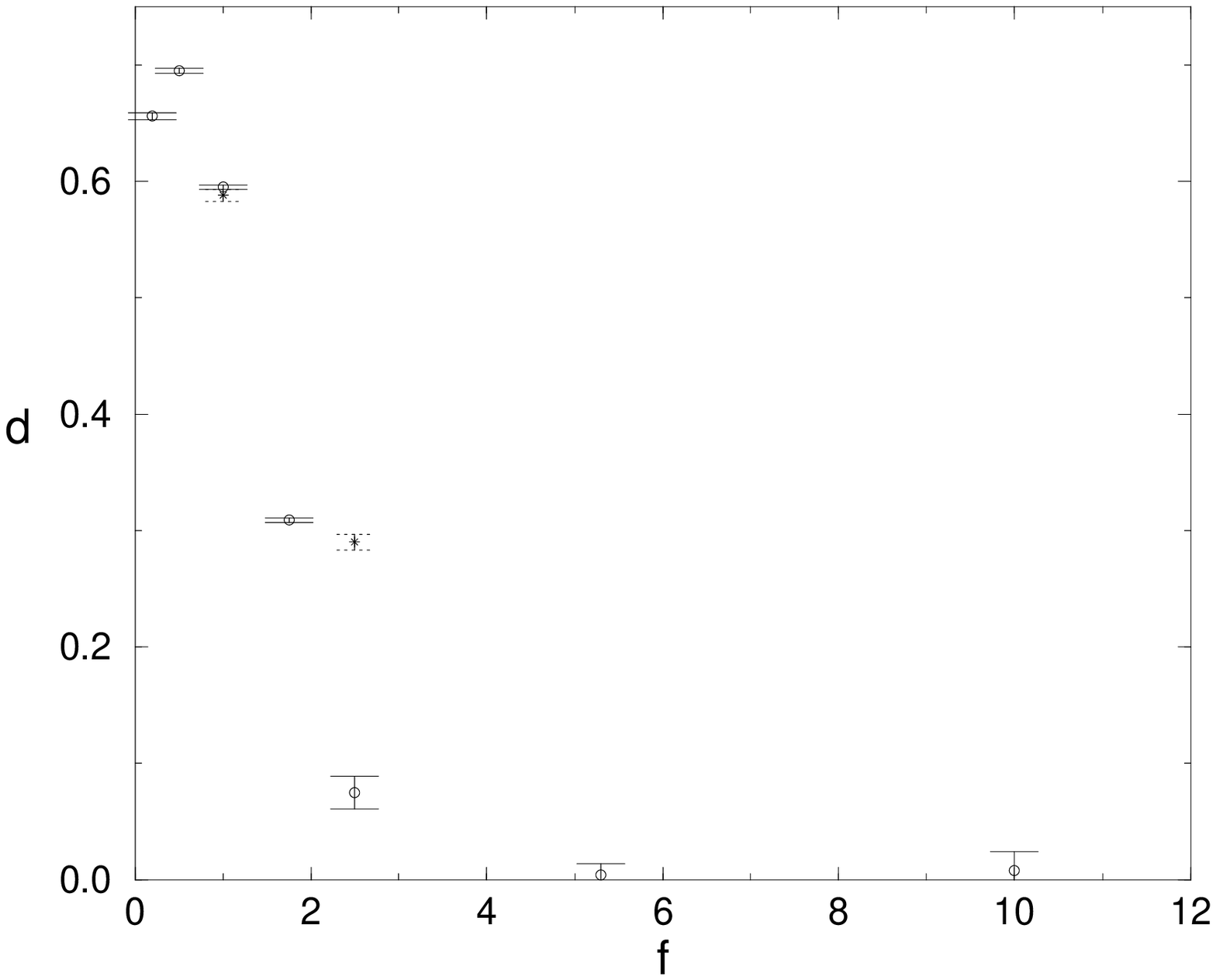}
\end{center}
\caption{The divergence of the power law fit ($d$) as a function of
the packing fraction $f$ of the system. Calibration is done at
$f=1$. Results for hard sphere wavefunctions ($\circ$) and Gaussian
wavefunctions ($\star$ and narrow-dashed error bars) are shown. All
calculations have been done at fixed volume $V=1$.}
\label{fig:hg_dvsf}
\end{figure}

We can check to see if finite volume effects are under control using
the ideas set out above. The result is shown in figure~\ref{fig:hg_dvsV}
which shows that power of divergence behaves as in the case of the
hard sphere. When we increase the packing fraction for Gaussian
wavefunctions (holding the volume constant) we see a decrease in the
power of the divergence (as for the hard sphere) (see
figure~\ref{fig:hg_dvsf} and table~\ref{tab:basicdat}). This shows
that whilst a power divergence is the generic situation for random,
low density gases, it is possible that a dense gas need not suffer
from this pathology. As we shall see, it is also possible to lose the
divergence by constructing a non-random gas i.e. a gas composed of
instanton--anti-instanton dipoles.

Whilst both the hard sphere and the Gaussian wavefunctions are useful
for computational purposes, perhaps we have the greatest faith for the
classical zero mode wavefunction. As we shall see, calculations using
such wavefunctions have their inherent difficulties, but these
notwithstanding, the results are qualitatively ``universal''.

\section{Classical zero mode}
We have seen the form of the classical zero mode previously (see
equation~\ref{zerom}). It is given by:

\begin{eqnarray}
\langle x|\psi_{j}^{\pm}\rangle & = &
\frac{\sqrt{2}}{\pi}\frac{\omega_{j}^{\pm}}{(\omega_{j}^{\pm2} + (x -
x_{j}^{\pm})^{2})^{\frac{3}{2}}}
\end{eqnarray}
In this case the wavefunction is given in ${\mathbb R}^{4}$ and the
overlaps are computed over this space (the objects are contained in
some volume $V$, but the overlaps are computed over the entirety of
${\mathbb R}^{4}$). Whilst this is not ideal, there are both practical
and mathematical difficulties associated with instanton wavefunctions
on tori. Calculating the overlaps on ${\mathbb R}^{4}$ is a compromise
until further work is carried out on the form of the zero mode
wavefunction associated with the approximate instanton on a torus
\cite{Hart}. We ``calibrate'' the size $\omega$ with the
corresponding hard sphere radius $\rho$ as before, and find that
$\omega=0.02$ corresponds to $\rho=0.2$ in the example given above. We
again find the average overlap to be similar for the two cases. We
generically refer to objects of size $\rho$ from now on, it should be
kept in mind that this $\rho$ is dependent upon the wavefunction being
used. We again find impressive agreement between the two
wavefunctions, see figure~\ref{fig:hgc} and table~\ref{tab:basicdat}.

\section{The Dirac ansatz}
We have seen some features which appear universal (requiring only that
the wavefunction be localised around the ostensible ``centre'' of the
instanton), namely the power law form of the spectral density at small
eigenvalues and its behaviour as a function of packing fraction. There
is however the possibility that this is simply the result of the
sequence of approximations we have made (see
equations~\ref{seq_approx}). Whilst we can do little about the first
two approximations, the final approximation, where we substitute the
geometric mean times the identity operator for the classical Dirac
operator, is one approximation whose effects require further close
scrutiny. Consider the following:

\subsection{A subtle fallacy}
\begin{eqnarray}
\frac{1}{\overline{\rho}^{2}}\delta_{ij} & = &
(\langle\widetilde{\psi}_{i}^{+}|\di).(\di|\widetilde{\psi}_{j}^{+}\rangle)\nonumber\\
& = &
\sum_{k}\langle\widetilde{\psi}_{i}^{+}|\di|\widetilde{\psi}_{k}^{-}\rangle\langle\widetilde{\psi}_{k}^{-}|\di|\widetilde{\psi}_{j}^{+}\rangle\nonumber\\
& = & (MM^{\dagger})_{ij}
\end{eqnarray}
This seems to imply that our ansatz for the Dirac operator gives us
the identity matrix (times a constant) and hence all eigenvalues are
either $0$ (from the fact that our matrices obey Atiyah-Singer) or
$1/\overline{\rho}$. What has gone wrong ? The answer is; nothing. We
should expect the identity matrix (times a constant) to appear as we
have sustituted in the identity operator for the Dirac operator. If
this is true then surely the spectra we have seen are wrong~?
Actually, no~! The flaw is in the second line; we have introduced
${\cal I} =
\sum_{k}|\widetilde{\psi}_{k}^{-}\rangle\langle\widetilde{\psi}_{k}^{-}|$
which is false. As we have previously argued, our zero mode
wavefunctions by themselves are not complete and do not span the full
Hilbert space. In fact it would be very unphysical if we could, it
would certainly require an infinite number of objects in a finite
volume, and even this, whilst being necessary to span the space, is
certainly not sufficient. So the spectra we are seeing are a result of
the fact that we are resolving the identity using a ``poor'' set of
basis vectors. What is happening is the following. Ideally, we wish to
use the true Dirac operator without any approximations, and a true
basis. However, we have used approximations for both; we have
approximated the Dirac operator out of necessity (whilst keeping those
properties which we feel to be important) and have approximated the
basis out of choice; we are after all trying to ascertain whether the
mixing of zero modes by themselves can break chiral symmetry. What we
have found is that if we make no approximations to the basis then we
certainly can not approximate the Dirac operator as we have done, or
we would get trivial results. (It would not even help to use the free
Dirac operator with a complete set of eigenvectors for then we would
obtain the free Dirac spectrum~!) Our hope has to be that our
approximations work together; at each stage we have tried to maintain
relevant properties whilst simplifying all we can. This of course may
prove to be not enough.

\begin{figure}[tb]
\begin{center}
\leavevmode
\epsfxsize=100mm
\epsfbox{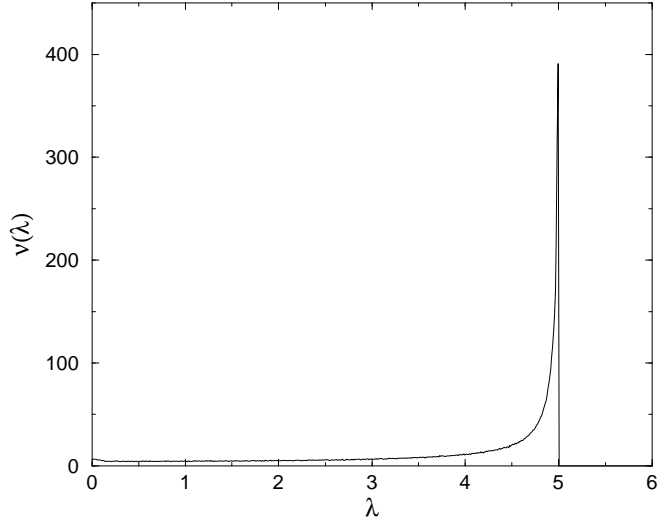}
\end{center}
\caption{Gaussian wavefunctions. Modelling instantons with broad
Gaussian $\sigma = 0.2$, as opposed to calibrated Gaussian $\sigma =
0.074$.}
\label{fig:dspike}
\end{figure}

\begin{figure}[tb]
\begin{center}
\leavevmode
\epsfxsize=100mm
\epsfbox{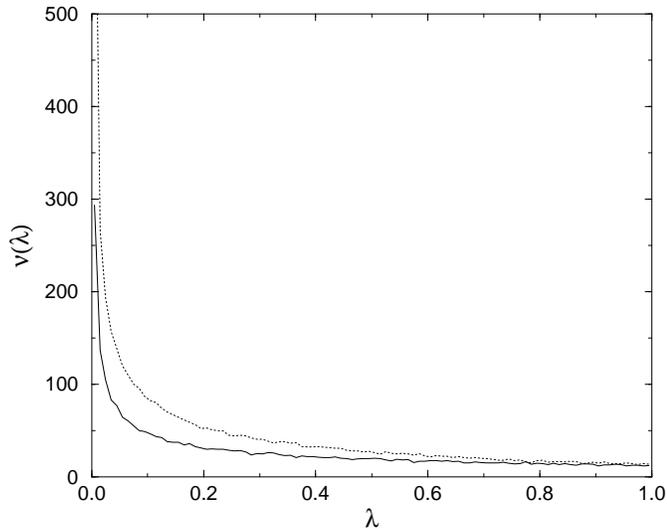}
\end{center}
\caption{Classical zero mode wavefunctions. Replacement of \di[A] with
$1/\sqrt{\rho_{1}\rho_{2}}$ (solid), with linear addition
ansatz (eq.~\ref{linadd}) (dotted).}
\label{fig:cs}
\end{figure}

It is in fact possible to get spectra where most eigenvalues are close
to $1/\overline{\rho}$. This is because we do not require completeness
in the true mathematical sense. Let us take a concrete example for
simplicity. Let us model our zero mode wavefunctions by Gaussian
wavefunctions of the form~\ref{gaussian}. All that is required for the
argument given above to work is that we can write the wavefunction
representing a positive chirality object in terms of the wavefunctions
representing the negative chirality objects in our volume. In
particular we do not require the ability to express an arbitrary
wavefunction as a linear combination (of Gaussian wavefunctions), only
the ability to express certain Gaussian wavefunctions as a linear
combination (of Gaussian wavefunctions). If we were to represent
objects with Gaussian wavefunctions of large $\sigma$ (see
equation~\ref{gaussa}) which are broad objects occupying large
sections of the volume, then this becomes approximately true. A
resultant spectra (with parameters identical to those for
figure~\ref{fig:hgc} except with $\sigma = 0.2$ instead of $\sigma =
0.074$) is given in figure~\ref{fig:dspike} and we can clearly see the
accumulation of eigenvalues at $1/\overline{\rho} = 5$.
These results are certainly worrying. One hope we can have is to use
another ansatz for the presence of the Dirac operator and compare the
spectrum. If we get a qualitatively similar spectrum for a radically
different implementation of the Dirac operator sandwiched between zero
modes of opposite chirality, then we may hope that the results we have
found are robust. There are several parameterizations of the matrix
element $M_{ij}$ for classical zero mode
wavefunctions~\cite{Shuryak-RMP,Shuryak-Param}. We analyzed the
parameteriztion:

\begin{eqnarray}
\langle\psi_{k}^{+}|-i\!\not\!\partial|\psi_{l}^{-}\rangle & \approx &
\frac{16R}{\rho_{k}^{+}\rho_{l}^{-}(2 +
R^{2}/\rho_{k}^{+}\rho_{l}^{-})^{2}}
\label{linadd}
\end{eqnarray}
where $R = |x_{k}^{+} - x_{l}^{-}|$~\cite{Shuryak-RMP}. The results,
shown in figure~\ref{fig:cs} and table~\ref{tab:basicdat} (under
heading ``Class. II'') are encouraging. The identity operator and the
free Dirac operator in the continuum have very different spectra. Yet,
within our model which focuses on just the would-be zero mode degrees
of freedom, they have similar spectra. We can hope that this result
extends to the full Dirac operator.

\section{Further results}
Let us for the sake of argument, accept that the spectral density is
approximately universal (the data is not unhopeful). We now extend the
parameters of the ``minimal'' ensembles used previously to study more
subtle effects and to extend our understanding of the origin of the
divergence.

So far we have concentrated on the $Q = 0$ sector ($\langle
Q^{2}\rangle = 0$). This is because we expect to minimise finite size
effects by doing so; in a large volume $V$ we expect $\langle
Q^{2}\rangle \sim V$ so that $\langle Q^{2}\rangle^{1/2}/V \sim
1/\sqrt{V}$. However, in a finite box we should take each topological
sector into account, with appropriate weighting, to obtain a more
accurate finite volume spectrum. Also, doing so allows us to test the
simple mechanism given previously (at the bottom of
section~\ref{hs_sect}) to explain the divergent peak. We can test this
picture by generating ensembles with realistic $Q$ distributions at
different volumes and seeing where the main difference in the spectral
densities lies. If the picture is correct then we should find that the
split modes lead to a greater difference near zero.
\begin{figure}[tb]
\begin{center}
\leavevmode
\epsfxsize=100mm
\epsfbox{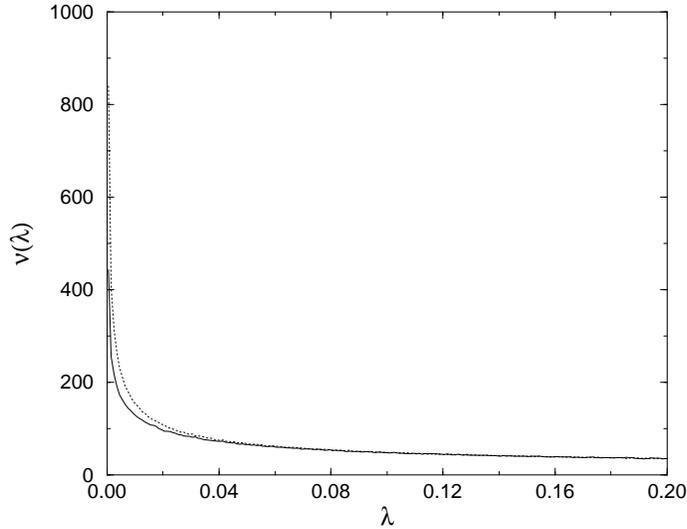}
\end{center}
\caption{Hard Sphere wavefunctions $f=1, \langle Q^{2}\rangle \propto
V$ for volume $V=1$ (solid) and $V\approx 2.44$ (dashed).}
\label{fig:varyq}
\end{figure}
To test these ideas we draw the number of instantons and
anti-instantons in each configuration from a normal distribution:

\begin{equation}
N_{I,A} \sim N(N_{T}/2, N_{T}/4)
\label{eq:varyia}
\end{equation}
where $N_{T} \propto V$ is the mean number of instantons in the volume
$V$ (for instance we have $N_{T}=126$ with $V=1,\ f=1$). We see
therefore that the total number of objects $N_{T} = N_{I} + N_{A} \sim
N(N_{T}, N_{T}/2)$ and that the winding number distribution follows $Q
= N_{I} - N_{A} \sim N(0, N_{T}/2)$. (The actual numerical factors
arise as this is nothing other than the central limit theorem applied
to the binomial distribution $N_{I,A} \sim B(N_{T},1/2)$ for large
$N_{T}$.) The results of a finite volume analysis for such a gas are
shown in figure~\ref{fig:varyq} and table~\ref{tab:varyq}. The results
clearly show a far greater difference at small eigenvalues with the
curves converging for $\lambda \geq 0.04$. This can be seen more
clearly if we simply focus on the difference between the two spectral
densities (figure~\ref{fig:varyq_d}). This seems to fit in with the
intuitive picture given above, if we repeatedly increased the volume,
then eventually we would get the result of the $\langle
Q^{2}\rangle=0$ ensemble. The difference between the the smaller and
larger volumes would be mainly restricted to ever smaller $\lambda$.
\begin{figure}[tb]
\begin{center}
\leavevmode
\epsfxsize=100mm
\epsfbox{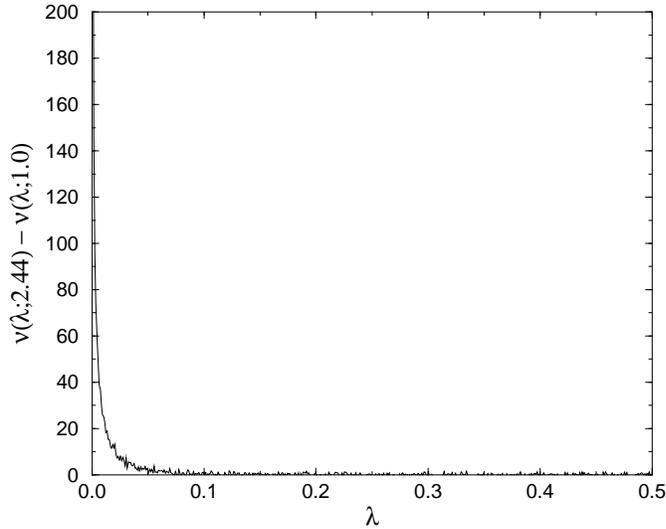}
\end{center}
\caption{The difference of the two spectral densities in
figure~\ref{fig:varyq}.}
\label{fig:varyq_d}
\end{figure}

One concern in our work has been to establish if the divergence we see
is simply the result of isolated dipoles. Is it possible that these
small eigenvalues in the peak originate from objects which overlap
minimally with all objects bar one~? We address this issue in three
ways. Firstly, we generate a ``background spectral density''
consisting of simple pairwise splittings of eigenvalues. Secondly, we
explicitly construct a gas of dipoles and see if we still find a peak
at small eigenvalues. Lastly we study the dispersion of the
eigenvectors. If small modes are associated with isolated dipoles then
the dispersion of these would be correspondingly small, if however, a
complicated multi-particle interaction was responsible for the small
modes then the dispersion need not be small.
\begin{figure}[tb]
\begin{center}
\leavevmode
\epsfxsize=100mm
\epsfbox{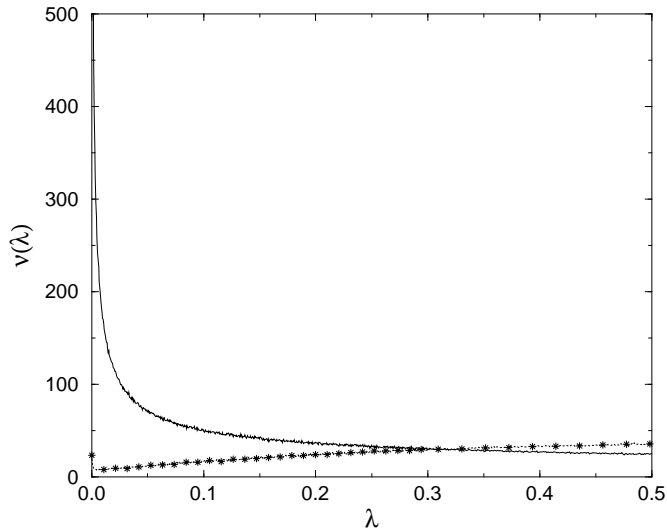}
\end{center}
\caption{Hard Sphere wavefunctions $V=1,\ f=1,\ \langle Q^{2}\rangle
=0$ (solid), background curve (dotted with $\star$).}
\label{fig:hbck}
\end{figure}
Figure~\ref{fig:hbck} shows the spectral density for the hard sphere
wavefunction coming from the actual eigenvalues of our Dirac
matrix. It also shows a ``background curve'' which originates from the
same ensemble. Consider a configuration with $Q=0$. This curve is
generated by simply going through the set of instantons and finding
the largest overlap associated with each object. This is then thought
of as the eigenvalue splitting from zero associated with that object:

\begin{equation}
\lambda_{j} = \max_{i} \{M_{ij}\}\quad j=1,\ldots,N_{I}
\end{equation}
The remaining eigenvalues are given by the $\gamma^{5}$ symmetry as
$\{-\lambda_{j}\}$. If we consider a case with $Q < 0$ then we again
go through the $N_{I}$ instantons associating each with an eigenvalue
as above. The rest of the eigenvalues are given either by the
$\gamma^{5}$ symmetry or the Atiyah-Singer theorem. If we have $Q > 0$
then we go through the $N_{A}$ anti-instantons instead. The background
curve thus generated only takes into account pairwise splitting (the
splitting due to a single object of the opposite charge) and
furthermore ignores totally the effects of other objects of the same
charge in the vicinity. We see that the peak has disappeared from the
background curve; the peak in the Dirac spectrum is not due to
isolated weakly overlapping dipoles.
\begin{figure}[tb]
\begin{center}
\leavevmode
\epsfxsize=100mm
\epsfbox{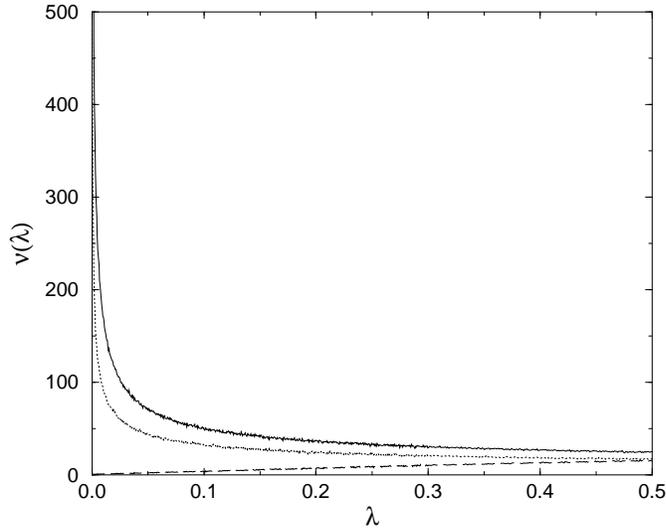}
\end{center}
\caption{Hard Sphere wavefunctions $V=1,\ f=1,\ \langle Q^{2}\rangle
=0$ (solid), spectrum from dipole gas maximum separation $|x^{+} -
x^{-}| \leq \rho^{+} + \rho^{-}$ (dotted), background curve from dipole
gas (long dashed).}
\label{fig:dip1}
\end{figure}

\begin{figure}[tb]
\begin{center}
\leavevmode
\epsfxsize=100mm
\epsfbox{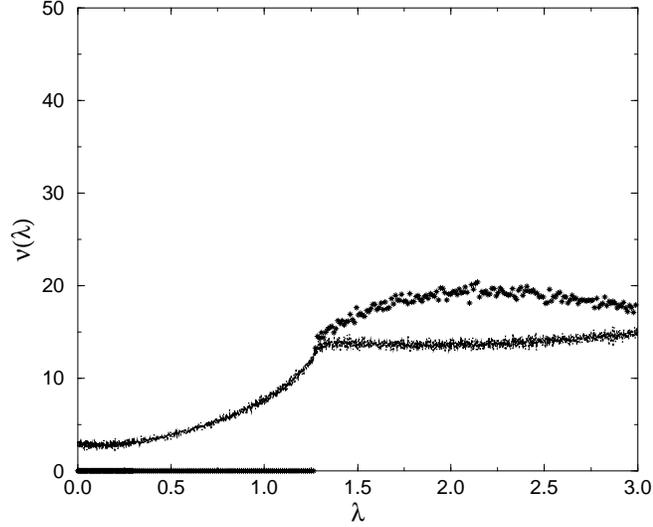}
\end{center}
\caption{Hard Sphere wavefunctions $V=1,\ f=1,\ \langle Q^{2}\rangle
=0$, spectrum from dipole gas maximum separation $|x^{+} -
x^{-}| \leq 1/2(\rho^{+} + \rho^{-})$ (dotted), background curve from dipole
gas ($\star$).}
\label{fig:dip2}
\end{figure}

We can further test this idea by explicitly constructing our gas to be
a gas of dipoles, pairing off each object with (at least) one of
opposite chirality. If we do this then we should find that there are
no small background modes (for instance if we enforce a minimum
overlap then there will be a cut off below which the background curve
will be precisely zero). Any small modes (or excess of small modes)
which arise will be due to multi-particle interactions. We see the
spectrum for such a gas of dipoles in figure~\ref{fig:dip1}. The gas
is constructed using hard sphere wavefunctions and the dipole is
constructed such that the smallest possible overlap is zero (the
spheres are just touching). This ensures that we get a background for
arbitrarily small eigenvalues, though the density should fall to
zero. We see that this is indeed so, but more surprisingly, the full
spectrum still shows a divergent peak. This shows that we can obtain a
divergent peak even for a gas of dipoles. If we enforce a minimum
overlap by ensuring that the separation is less than half the sum of
the radii then we obtain figure~\ref{fig:dip2}. In this case there is
a sharp cut off in the background curve as expected, however we still
find chiral symmetry breaking from the full spectral curve. This shows
that to achieve chiral symmetry restoration requires the dipole gas to
be even more strongly overlapping.

\begin{figure}[tb]
\begin{center}
\leavevmode
\epsfxsize=100mm
\epsfbox{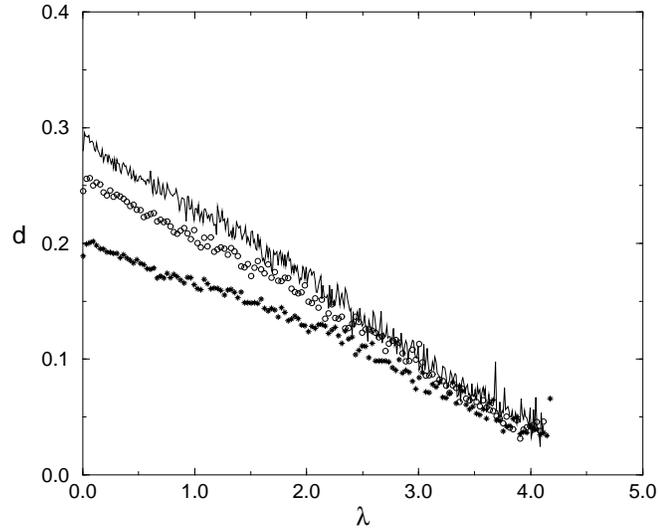}
\end{center}
\caption{Hard Sphere wavefunctions $f=1,\ \langle Q^{2}\rangle
=0$ Dispersion versus eigenvalue, $V=0.4096\ (\star)$, $V=1\ (\circ)$ and $V \approx 1.5$ (solid).}
\label{fig:disp}
\end{figure}

Lastly we analyse the dispersion of the eigenmodes associated with the
spectrum. This is of course analogous to the concept of
``centre-of-mass'' and ``inertia'' for a system of
particles. Figure~\ref{fig:disp} shows a plot of dispersion versus
eigenvalue. We see the dispersion decreases approximately linearly
with increasing eigenvalue. This shows that small eigenvalues are the
product of a linear superposition of many particles (the eigenvector
is spread out) as opposed to originating from isolated dipoles. The
fact that the dispersion curves for the three volumes do not lie on
top of one another is a little worrying. The first thought is that
there is a scale factor missing somewhere (we should be dividing by $L
= V^{1/4}$), but we have been unable to find the culprit. The other
possibility is that the dispersion is subject to finite size effects
for small $\lambda$, and, that the wavefunctions are constricted by
the size of the box. Neither possibility affects the formal result,
the dispersion increases as the eigenvalue decreases: small modes are
not associated with isolated dipoles.

\section{Discussion}
We have shown that within our model, the spectral density is
qualitatively independent of the form of the wavefunction used to
construct the Dirac matrix and the ansatz used for the presence of the
Dirac operator. This is a necessary requirement if we are to have
faith that our results may hold for the true quenched QCD vacuum. We
have found that the universal spectral density follows an inverse
power law and hence diverges as $\lambda \rightarrow 0$. The strength
of the power decreases as the density of instantons increases (other
factors held constant). We have shown that the inclusion of
topologically non-trivial sectors, decreases the strength of the
divergence in general (at finite volume). We have further shown that
the peak is not associated with weakly interacting isolated dipoles of
charges, but is in fact the product of multi-particle interactions.

We should note that others have (independently) found a divergence in
the spectrum of the Dirac operator for quenched QCD using slightly
different
methods~\cite{Osborn-Thou,Osborn-Chi,Verbaarschot,Shuryak-Priv}. Whether
this is related to the appearance, within quenched chiral perturbation
theory, of a logarithmic \cite{Sharpe-Elec,BernardB,Golter-Chip}, or
possibly power divergence \cite{Sharpe-Quen,Golter-Priv}, is also of
interest (we address this in a little more detail in the next
chapter). We now move on and analyze instanton configurations derived
from quenched QCD lattice data.

\begin{table}[htb]
\begin{center}
\begin{tabular}{|c|c|c|c|c|c|c|c|}
\hline
Type & $f$ & $V$ & $N_{c}$ & d & $\chi_{p}^{2}/N_{DF}$ &
$\chi_{l}^{2}/N_{DF}$\\
\hline
Hard Sphere & 0.2 & 1.0 & 650000 & 0.656$\pm$0.003 & 2.40 & 600\\
& 0.5 & 1.0 & 128000 & 0.695$\pm$0.002 &  1.92 & 508\\
& 1.0 & 0.41 & 130000 & 0.540$\pm$0.003 &  1.71 & 123\\
& 1.0 & 1.0 & 126000 & 0.595$\pm$0.002 & 1.38 & 258\\
& 1.0 & 1.52 & 97000 & 0.617$\pm$0.001 & 1.23 & 400\\
& 1.0 & 2.44 & 93000 & 0.640$\pm$0.001 & 1.47 & 702\\
& 1.0 & 7.72 & 9780 & 0.668$\pm$0.003 & 1.73 & 456\\
& 1.75 & 1.0 & 111000 & 0.309$\pm$0.002 & 1.94 & 48\\
& 2.5 & 1.0 & 63200 & 0.075$\pm$0.014 & 1.72 & 2.59\\
& 5.3 & 1.0 & 6720 &  0.004$\pm$0.010 & 2.20 & 2.19\\
& 10.0 & 1.0 & 1266 & 0.008$\pm$0.016 & 1.91 & 1.90\\
Gaussian & 1.0 & 1.0 & 12600 & 0.588$\pm$0.005 & 1.36 & 45\\
& 1.0 & 2.4 & 15500 & 0.634$\pm$0.002 & 1.31 & 201\\
& 2.5 & 1.0 & 6320 & 0.290$\pm$0.007 & 1.46 & 4.87\\
Classical & 1.0 & 1.0 & 2100 & 0.477$\pm$0.010 & 1.32 & 5.33\\
& 1.0 & 2.4 & 450 & 0.538$\pm$0.022 & 1.38 & 3.50\\
Class. (II) & 1.0 & 1.0 & 2520 & 0.555$\pm$0.006 & 1.56 & 34.6\\
\hline
\end{tabular}
\end{center}
\caption{Parameters and results for $\langle Q^{2}\rangle=0$
ensembles. $f$ is the constant packing fraction in each configuration
(a configuration with $Q=0,\ V=1,\ f=1$ contains $63$ instantons and
$63$ anti-instantons). $N_{c}$ is the number of configurations in the
ensemble, $d$ is the degree of divergence, $\chi_{p}^{2}/N_{DF}$ is
the standard chi-square per degree of freedom for the power law fit
and $\chi_{l}^{2}/N_{DF}$ is the chi-square per degree of freedom for
the log fit.}
\label{tab:basicdat}
\end{table}

\begin{table}
\begin{center}
\begin{tabular}{|c|c|c|c|c|c|c|}
\hline
Type & $f$ & $V$ & $N_{c}$ & d & $\chi_{p}^{2}/N_{DF}$ &
$\chi_{l}^{2}/N_{DF}$\\
\hline
Hard Sphere & 1.0 & 1.0 & 126000 & 0.363$\pm$0.006 & 1.90 & 65\\
& 1.0 & 2.4 & 62000 & 0.522$\pm$0.003 & 2.07 & 253\\
\hline
\end{tabular}
\end{center}
\caption{Parameters and results for $\langle Q^{2}\rangle^{1/2}/V
\propto 1/\sqrt{V}$ ensembles. As the number of instantons,
anti-instantons and hence the number of objects in total, varies from
configuration to configuration, $f$ in now the mean packing fraction.}
\label{tab:varyq}
\end{table}

%% file: latt.tex
\chapter{Quenched QCD lattice ensemble}
\label{ch:latt}

So far the ensembles of objects have been artificial. We have imposed
fairly arbitrary values on the fundamental parameters of the ensembles
including the packing fraction $f$ and the size distribution of the
instantons. In fact in all the cases we have considered up to now, the
instantons have all been the same size. Is it possible that the
artificial nature of the ensembles is distorting the resultant
spectral densities ? Ideally we wish to use configurations of
instantons which are derived from gauge field configurations as
opposed to random instanton configurations. In order to do this we
require a representative selection of gauge configurations generated
with the appropriate quenched QCD weighting, and, an algorithm
enabling us to carry out the decomposition given by
equation~\ref{decomp}.

As generating gauge field configurations is relatively simple in
comparison to generating full QCD configurations, the volumes and
lattice spacings available are far superior, and facilitate in depth
analysis of the topological content of the vacuum.  There have been a
number of recent lattice calculations that attempt to determine the
instanton content of the vacuum in
$SU(2)$~\cite{Michael,ForcrandA,DeGrand} and
$SU(3)$~\cite{ForcrandB,Hasenfratz,Smith} gauge theories. To identify
the instanton content of fully fluctuating vacuum gauge fields (to
make the decomposition given by equation~\ref{decomp} but ignoring the
colour orientation of the objects) is far from trivial.  Indeed it is
not entirely clear to what extent it is either meaningful or
possible. Current techniques involve smoothening the lattice gauge
fields on short distances and then using some pattern recognition
algorithm to resolve the topological charge density into an ensemble
of overlapping (anti-)instantons of various sizes and positions. At
present there is only some agreement between the results of the
different approximate methods being
used~\cite{ForcrandB,Hasenfratz,Smith}.  Thus all these calculations
should be regarded as exploratory.

In this thesis we focus on the calculations in~\cite{Smith}.  In that
work the smoothing of the rough gauge fields was achieved by a process
called ``cooling''~\cite{TeperA}.  This is an iterative procedure just
like the Monte Carlo simulation itself, except that the fields are
locally deformed towards the minimum of the action (or some variation
thereof). Although some quantities, such as the topological
susceptibility, are insensitive to the amount of cooling (within
reason) this is not the case for the number, size distribution and
density of the topological charges. Whether this leads to a real
ambiguity for physical observables is an important question. It might
be that the ambiguity is only apparent and that fermionic observables
calculated in these cooled instanton background fields do not show
much variation with cooling. For example it might be the case that the
instantons which disappear with cooling are highly overlapping $Q=\pm
1$ pairs that contribute no small modes to the Dirac operator.  In
that case the spectrum of small modes would be insensitive to cooling,
and so would various fermionic observables such as the chiral
condensate. One of the primary questions addressed in this chapter is
to ascertain how the spectrum for small eigenvalues varies as a
function of the number of cooling sweeps.

We begin with a list of objects (derived via cooling and pattern
recognition from ``hot'' gauge configurations) and construct the
spectral density exactly as in the previous chapter. We have a problem
however. What size should we use for the instanton wavefunctions ?
The lattice decomposition used a classical formula to relate instanton
width $\rho$ to topological peak height $Q_{p}$:

\begin{equation}
Q_{p} = \frac{6}{\pi^{2}\rho^{4}} .
\end{equation}
(We refer the reader to~\cite{Smith} for more details, as this is
simply the initial step in obtaining the size of the objects.) Should
we not simply use classical wavefunctions for our would-be zero modes,
with the width as derived from the lattice data using the above
formula~? This is not so easy and several problems spring to mind. The
greatest is that the lattice data has been calculated using a periodic
box as the volume, if we modelled the data using classical zero mode
wavefunctions then we would be forced to use euclidean spacetime to
compute the overlaps (we could not compromise and calculate
separations on a torus and overlaps in euclidean spacetime for the
orthonormalization procedure would break down). This would be a source
for errors as objects which overlapped heavily ``around the torus''
became objects which overlapped minimally (being at opposite edges of
the non-periodic box). We would have gained by obtaining a (more)
realistic ensemble of instanton configurations and then effectively
thrown away our gain by changing the topology of the manifold. We felt
that we had to maintain as much of the relative positional details as
we could i.e. we had to use a torus. In the end we used hard sphere
wavefunctions for the zero mode wavefunctions, with the size (radius)
as derived from the lattice data. As before we replace the Dirac
operator between two objects with the geometric mean of their radii
times the identity operator. With these replacements we have to rely
upon the universality sketched in chapter~\ref{ch:univ} to give us
hope that our results hold for quenched QCD.

One might think that the reasonable way to approach all these
questions would be to perform calculations directly on the cooled
lattice fields using lattice versions of the Dirac operator. Although
such explicit calculations do show that it is the instanton (would-be)
zero modes that drive chiral symmetry breaking~\cite{Hands}, one also
finds that lattice artefacts spoil the mixing of the instanton
near-zero modes~\cite{Hands} and this makes it difficult to draw
reliable conclusions for the continuum limit. (Although as mentioned
previously, very recent work with domain-wall fermions and related
lattice fermions suggests a promising avenue for progress.)

The results presented in this chapter have been published
elsewhere~\cite{Sharan-Lat}, and, this chapter is, to a large extent,
based upon that paper which was co-authored by Mike Teper.

\section{q-QCD spectra}
In~\cite{Smith} SU(3) lattice gauge field configurations of sizes
$16^3\times48$ at $\beta=6.0$, $24^3\times48$ at $\beta=6.2$ and
$32^3\times64$ at $\beta=6.4$ were cooled and the corresponding
instanton ensembles extracted for various numbers of cooling
sweeps. Over this range of $\beta=6/g^2$ the lattice spacing varies by
a little less than a factor of 2 and these three volumes are
approximately the same in physical units. Comparing the results at the
three values of $\beta$ enables the approach to the continuum limit to
be studied.  Of course, instantons can be large and it is important to
control finite volume effects as well. For this purpose calculations
were also performed at $\beta=6.0$ on a much larger $32^3\times64$
lattice. The conclusion was that finite volume corrections were
negligible and that there was good scaling of, for example, the
instanton size distribution, if one varied the number of cooling
sweeps with $\beta$ so as to keep the average number of instantons
constant. (For an interesting recent analysis of the scaling
properties, see \cite{Ringwald}.)  Some properties of these lattice
configurations are listed in Table~\ref{tab:cdata}.

As we are dealing with decompositions based on real lattice data, the
number of configurations is rather more limited than previously. They
range in number from 20 to 100 depending on the lattice size and the
value of $\beta$. We shall, for simplicity, not employ some of the
rather complicated procedures used in \cite{Smith} for filtering out
possible false instanton assignments. Rather we shall take the raw
instanton ensembles from~\cite{Smith}, corrected for the influence of
the instantons upon each other but without applying any further
filters. (Except that we throw away any charges that are larger than
the volume available.  This usually involves rejecting (much) less
than $1\%$ of the total number.) In addition, we calculate the size
from the (corrected) peak height. We are confident that the results we
obtain from these ensembles differ very little from the results we
would have obtained using the slightly different ensembles obtained by
applying the more complex procedures of~\cite{Smith}.

\begin{figure}[tbh]
\begin{center}
\leavevmode
\epsfxsize=100mm
\epsfbox{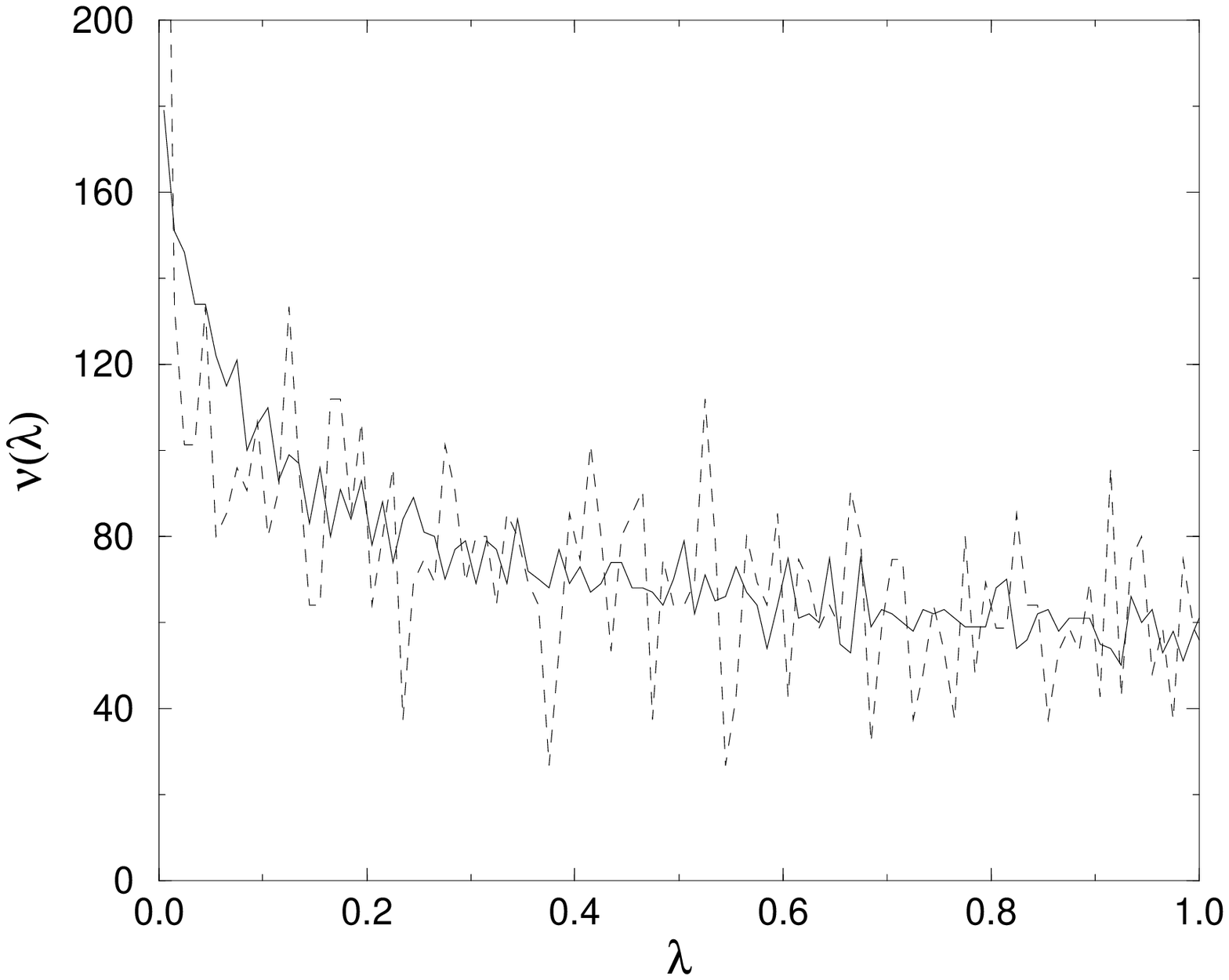}
\end{center}
\caption{Spectral densities from two different volumes
at $\beta=6.0$, after 46 cooling sweeps: $32^3\times64$ (solid) and
$16^3\times48$ (dashed) lattices.}
\label{fig:0.46L.vol}
\end{figure}

\begin{figure}[tbh]
\begin{center}
\leavevmode
\epsfxsize=100mm
\epsfbox{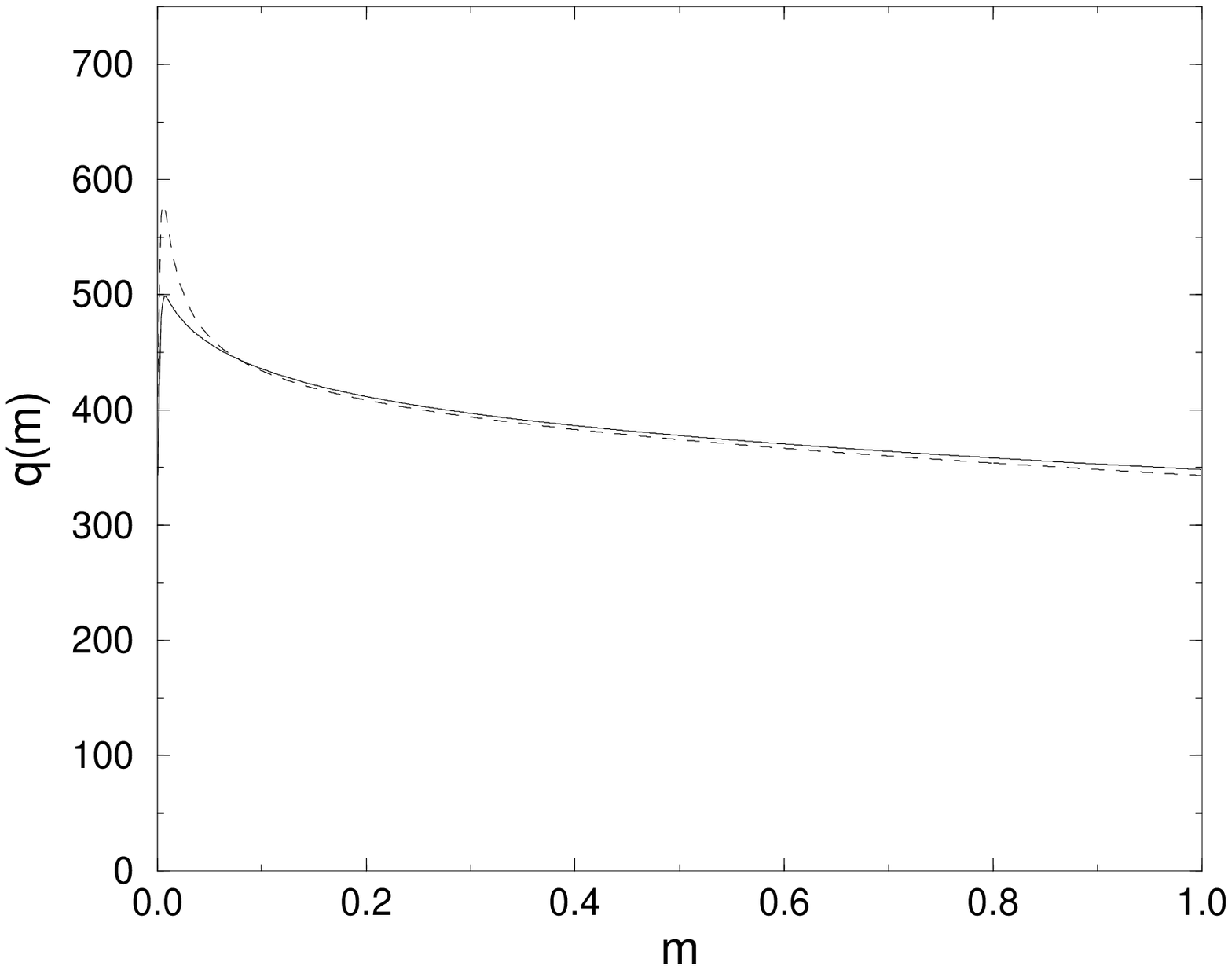}
\end{center}
\caption{\ssi = $(q(m) \rm{MeV})^{3}$ as obtained from the $\beta =
6.0\ 32^{3}\times64$ lattice data (solid), and from the $\beta = 6.0\
16^{3}\times48$ lattice data (dashed).}
\label{fig:0.46x.qc}
\end{figure}

There are several questions we wish to address. These include:
\begin{itemize}
\item{Do fermionic physical observables, such as the spectral density
and the chiral condensate, exhibit a weak variation with cooling,
implying that the rapid variation of the instanton ensemble that one
observes is more apparent than real, or do they exhibit a strong
variation?}
\item{Do these fermionic physical observables also exhibit scaling and
small finite volume corrections?}
\item{Does a realistic ensemble of instantons break chiral symmetry
spontaneously? Lattice calculations find that it does; but the
presence of important lattice artefacts renders the conclusion
suspect. Continuum calculations using model ensembles of instantons
also find that they break chiral symmetry; but it is not clear that
the real world is like the model.}
\item{Is the spectral density of quenched QCD pathological? Some model
calculations (including our own in the previous chapter) have found
that the spectrum appears to diverges at $\lambda=0$. Do we get
similar results for ensembles containing realistic instanton
configurations~?}
\end{itemize}
\begin{figure}[tbh]
\begin{center}
\leavevmode
\epsfxsize=100mm
\epsfbox{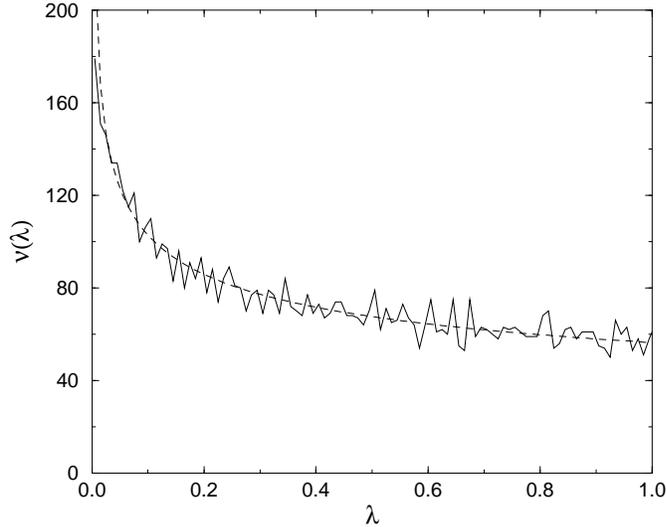}
\end{center}
\caption{The spectral density: $\beta = 6.0,\ 32^{3}64,\ 50$
configurations. Dashed curve is best power law fit $\nu(\lambda) = a
+ b/\lambda^{d}$.}
\label{fig:0.46L}
\end{figure}
Figure~\ref{fig:0.46L} shows the spectral density that results from
the 50 configurations generated after 46 cooling sweeps on the
$32^{3}\times64$ lattice at $\beta = 6.0$. We see that the spectral
density does not smoothly decrease to zero as $\lambda \rightarrow 0$,
so the chiral symmetry will be spontaneously broken.  However we also
see a pronounced peak as $\lambda \rightarrow 0$, just like the
divergence that characterises model instanton ensembles. (Note that as
usual we have removed the $\delta$-function contribution of the exact
zero eigenvalues.) We model the divergence using a power law and a log
law as before (equations~\ref{power-law}
and~\ref{log-law}). Figures~\ref{fig:0.46L.lnlin} and
\ref{fig:0.46L.lnln} show that both models fit the data well, an
observation which is confirmed by the low chi-squared for the fit (see
Table~\ref{tab:cdata}). We should check that this result is not
subject to large finite volume effects and to this end we compare the
spectral density to that obtained from the $\beta = 6.0,
16^{3}\times48$ lattice (a volume approximately ten times smaller). As
shown in figure~\ref{fig:0.46L.vol}, whilst the result on the smaller
volume is noisier, we find the densities are entirely similar, even
down to the details of the forward peak. This shows that at least for
these parameters any finite volume corrections are small. The chiral
condensate as a function of quark mass is given in
figure~\ref{fig:0.46x.qc}. We know that this order parameter must
vanish for very small quark masses because of the gap in the
eigenvalue spectrum (we cannot take the quark mass to zero in a finite
box) and indeed it does.  If we extrapolate to zero quark mass, whilst
ignoring the finite volume dip at very small quark masses and the peak
at small masses, we find that chiral symmetry is broken with an order
parameter $\ssi^\frac{1}{3} \approx 400 {\rm MeV}$. Whilst this is
larger than the phenomenological figure, $\ssi^\frac{1}{3} \sim 200
{\rm MeV}$, it is close considering the qualitative nature of our
calculations. The fact it is larger is presumably a reflection of the
high density of this gas of instantons.
\begin{figure}[tbh]
\begin{center}
\leavevmode
\epsfxsize=100mm
\epsfbox{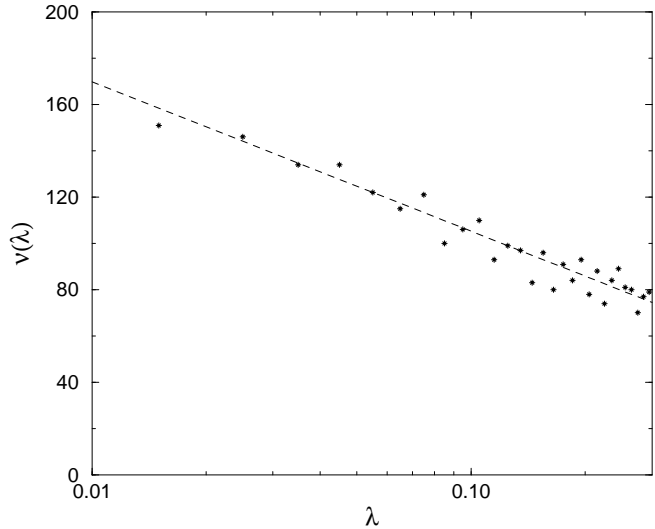}
\end{center}
\caption{The spectral density in figure~\ref{fig:0.46L}, 
plotted on log-linear axes. Dashed curve is best log law fit
$\nu(\lambda) = a + b\ln(\lambda)$.}
\label{fig:0.46L.lnlin}
\end{figure}
\begin{figure}[tbh]
\begin{center}
\leavevmode
\epsfxsize=100mm
\epsfbox{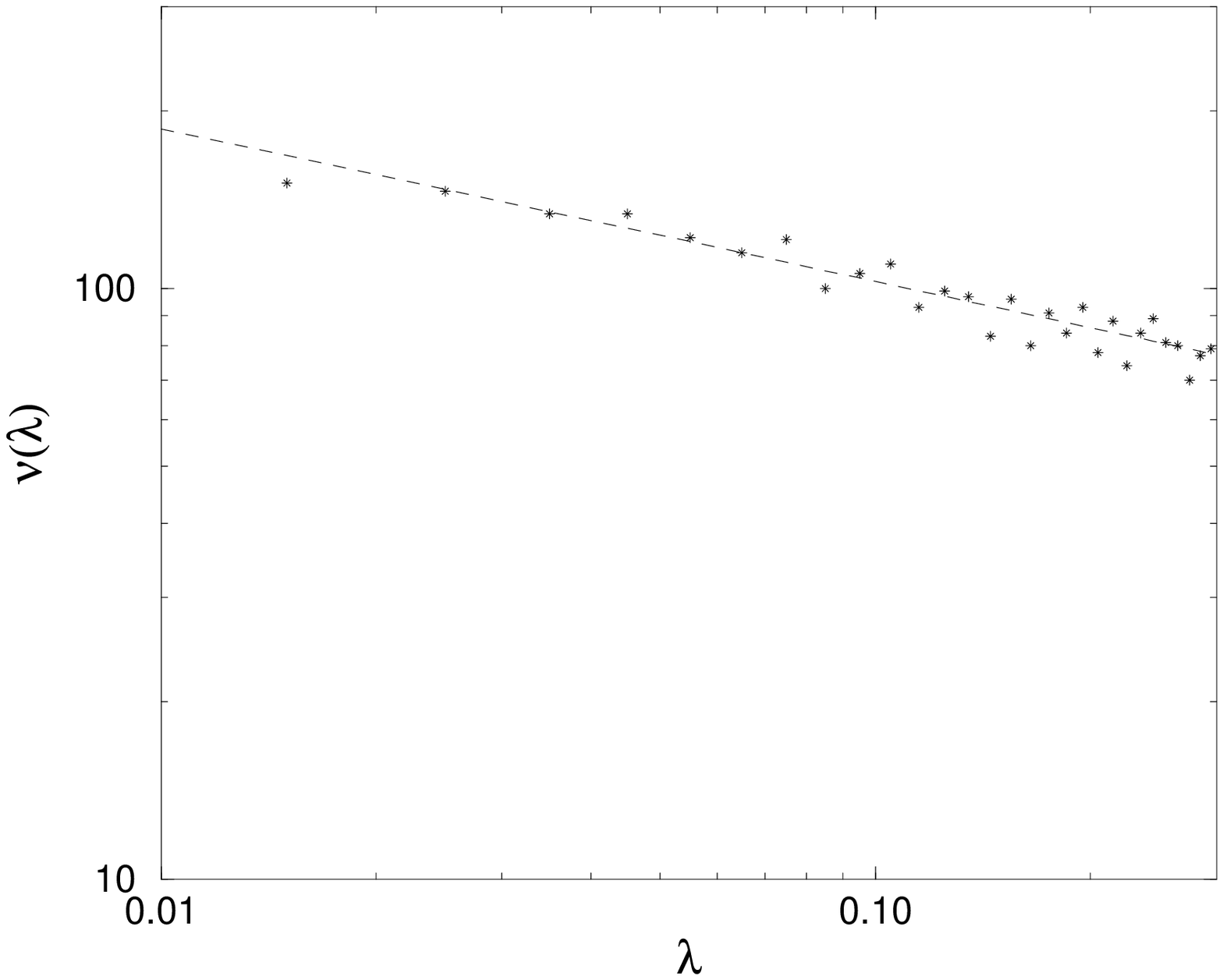}
\end{center}
\caption{The spectral density in figure~\ref{fig:0.46L}, 
plotted on log-log axes. Dashed curve is best power law fit
$\nu(\lambda) = a + b/\lambda^{d}$.}
\label{fig:0.46L.lnln}
\end{figure}
Although the qualitative features of our spectrum do not require a
specification of units, the comparison between different instanton
ensembles does. The units we have chosen are as follows. Our length
unit is chosen to be $32a$ at $\beta=6.0$; so that the $32^3\times64$
and $16^3\times48$ lattices discussed in the previous paragraph have
volumes 2 and 0.1875 respectively.  We see from Table~\ref{tab:cdata}
that this corresponds to taking our length unit as $32a(\beta=6.0)
\simeq 32 \times 0.098fm = 3.136 fm$.  Thus our mass unit is the inverse of
this, $\simeq 64MeV$.  Since $\lambda$ has dimensions $[m]^1$, this
means that the eigenvalues shown in Figure~\ref{fig:0.46L} range from
0 to $\simeq 64MeV$: a reasonable range if what we are interested in
is the spectrum $\lambda < \Lambda_{QCD}$. We maintain this unit
throughout the calculations in this chapter and we use the values of
$a(\beta)$ listed in Table~\ref{tab:cdata} to translate this unit to
other values of $\beta$. Thus if we want to test for scaling all we
need to do is to directly superpose the spectra as shown in our
figures. (We shall do this later on in this section.)

\begin{figure}[tbh]
\begin{center}
\leavevmode
\epsfxsize=100mm
\epsfbox{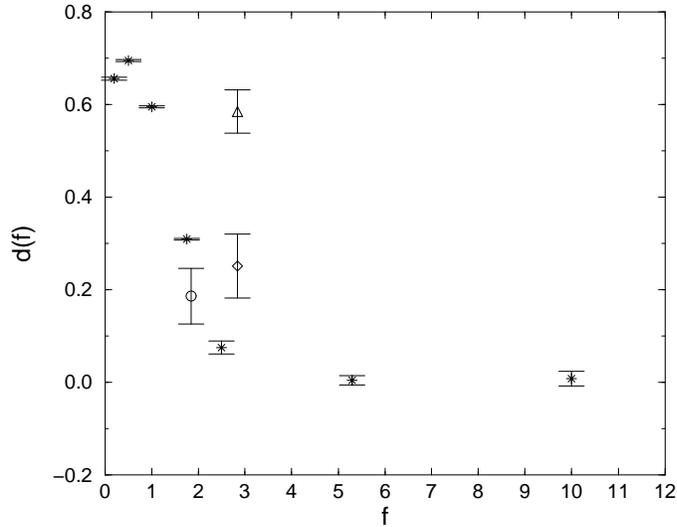}
\end{center}
\caption{Plot of power of divergence $d$ as a function of packing
fraction $f$. $\star$ from random position model. $\diamond$ from
lattice data as in figure (\ref{fig:0.46L}). $\circ$ same, except all
instantons of same size $\overline{\rho}$. $\triangle$ as figure
(\ref{fig:0.46L}) except instantons positioned at random.}
\label{fig:DvsF}
\end{figure}

\begin{figure}[tbh]
\begin{center}
\leavevmode
\epsfxsize=100mm
\epsfbox{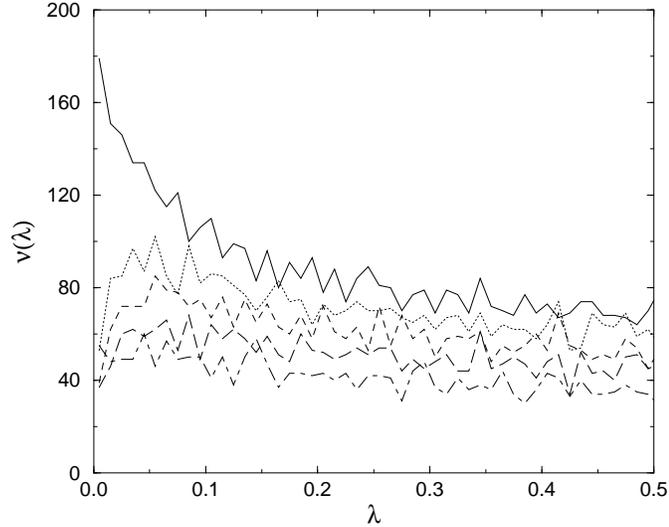}
\end{center}
\caption{Lattice Data: Solid line as figure (\ref{fig:0.46L}). Dotted
line only contains objects with $\rho > 0.12$, dashed line $\rho >
0.14$, long dashed $\rho > 0.16$, dot-dash $\rho > \overline{\rho} \approx
0.182$.}
\label{fig:gt0xx}
\end{figure}

\begin{figure}[tbh]
\begin{center}
\leavevmode
\epsfxsize=100mm
\epsfbox{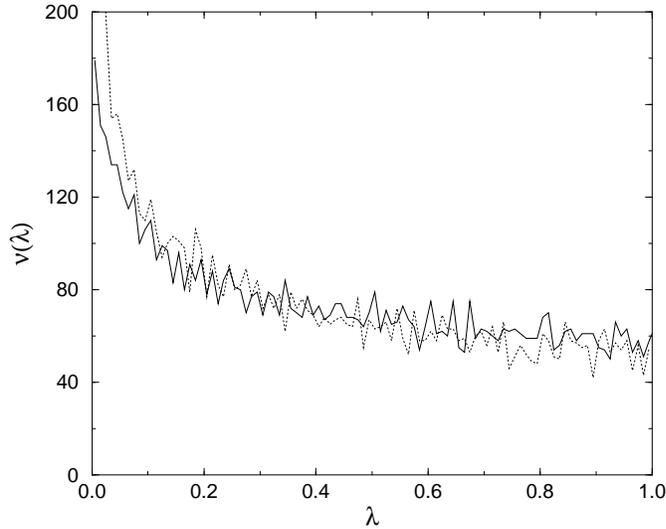}
\end{center}
\caption{The spectral density of figure~\ref{fig:0.46L} (solid), 
compared to the density obtained by positioning the same charges at
random (dotted).}
\label{fig:0.46L.rpos}
\end{figure}

So how do our lattice ensemble spectra compare with those from our
random configuration model ? In figure~\ref{fig:DvsF} we plot the
degree of divergence corresponding to the $\beta = 6.0, 32^{3}64$
lattice data on the graph generated from the random position
model~\ref{fig:h_stan_f}. It would appear that the divergence is too
large for the packing fraction; it lies approximately two sigma above
the synthetic ensemble curve. It is therefore interesting to consider
which particular aspect of the lattice data contributes to the
divergence. The first possible factor is the non-trivial size
distribution of the objects associated with the lattice data. We
therefore set all objects in the lattice data to the same size (the
mean $\overline{\rho}$ of the lattice ensemble). This reduces the
packing fraction of the ensemble, since $\overline{\rho}^4 <
\overline{\rho^4}$, and, as shown in figure~\ref{fig:DvsF}, it also
results in the degree of divergence fitting with that of the synthetic
ensemble.  The fact that a non-trivial instanton size distribution has
a marked impact on the spectrum of small modes leads us to ask whether
it is the small or the large instantons that drive this effect. To
answer this question we systematically cull instantons of ever
increasing size from the lattice instanton ensembles and see how this
affects the spectral density.  The results of this calculation are
shown in figure~\ref{fig:gt0xx}.  In this figure we show the densities
obtained by only including objects with radii above a certain
cut-off. We see that the peak is already significantly reduced if we
exclude the $\sim 10\%$ of instantons with radii $\rho < 0.12$; and it
is eliminated entirely if we exclude all instantons with $\rho <
\bar{\rho} \simeq 0.18$.  (If on the other hand we exclude the largest
instantons, then we find that we strengthen the peaking at
$\lambda=0$.)  This shows that the extra peaking we have observed with
the lattice instanton ensembles is due to the smaller instantons.
More generally, this demonstrates that it is possible to have small
instantons driving a `divergent' spectral density even in a high
density gas. The reason for this unexpected phenomenon is actually
quite simple. The large packing fraction of such a gas is driven by
the larger instantons (since the volume $V_{I} \propto \rho^4$). The
smaller instantons are rather dilute and are not likely to overlap
significantly with each other. Instead they typically overlap
completely with some of the much larger instantons.  However this
overlap is small: if we have a small instanton of radius $\rho_{s}$
sitting on a large one of size $\rho_{l}$ (of the opposite charge)
then this will contribute $\propto
\rho_{s}^{4}/\rho_{s}^{2}\rho_{l}^{2} = (\rho_{s}/\rho_{l})^{2}$ to
the overlap matrix.  The larger instantons, on the other hand, will
have large overlaps with other large instantons (of the opposite
charge) in addition to their small overlaps with small instantons.  So
they are less likely candidates for producing small eigenvalues. Thus
the smaller instantons in an apparently dense gas can behave as a
dilute gas with a corresponding peak at small eigenvalues.
It is also interesting to ask whether the non-random positioning of
the instantons in the lattice ensembles makes a difference to the
small-$\lambda$ peak in the spectral density. We see from
figure~\ref{fig:DvsF} and figure~\ref{fig:0.46L.rpos} that it does;
positioning the objects at random (but incorporating other information
such as the size distibution) increases the degree of
divergence. Whilst there are systematic uncertainties (due to deciding
which region of the data to fit the power divergence to), this result
is seen in all the lattice data that we have analysed. The simplest
explanation for this is that we are seeing an effect of the
topological charge screening that was observed in~\cite{Smith}. This
tendency for opposite charges to `pair up', will lead to an increased
eigenvalue splitting and a weaker divergence. When we position the
objects at random, this screening is lost and the degree of divergence
is increased.

\begin{figure}[tbh]
\begin{center}
\leavevmode
\epsfxsize=100mm
\epsfbox{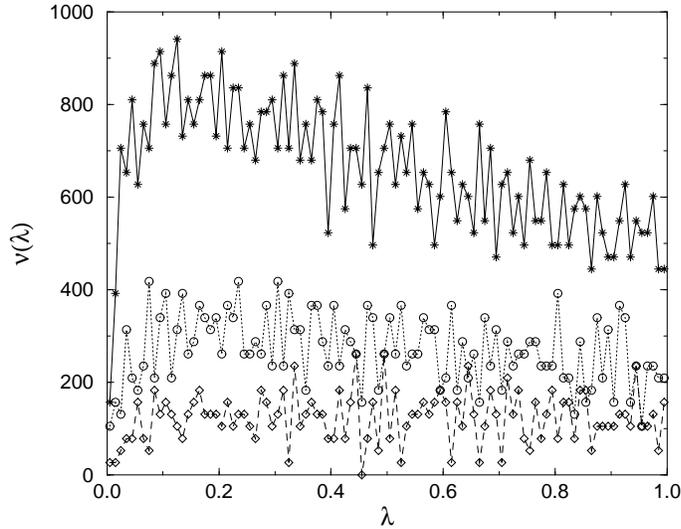}
\end{center}
\caption{The spectral densities obtained from the 
$\beta = 6.4,\ 32^{3}64$ configurations for various numbers of 
cooling sweeps: ($\star$) 30 cools, ($\circ$) 50 cools, ($\diamond$)
80 cools.}
\label{fig:4.xx.sp}
\end{figure}

\begin{figure}[tbh]
\begin{center}
\leavevmode
\epsfxsize=100mm
\epsfbox{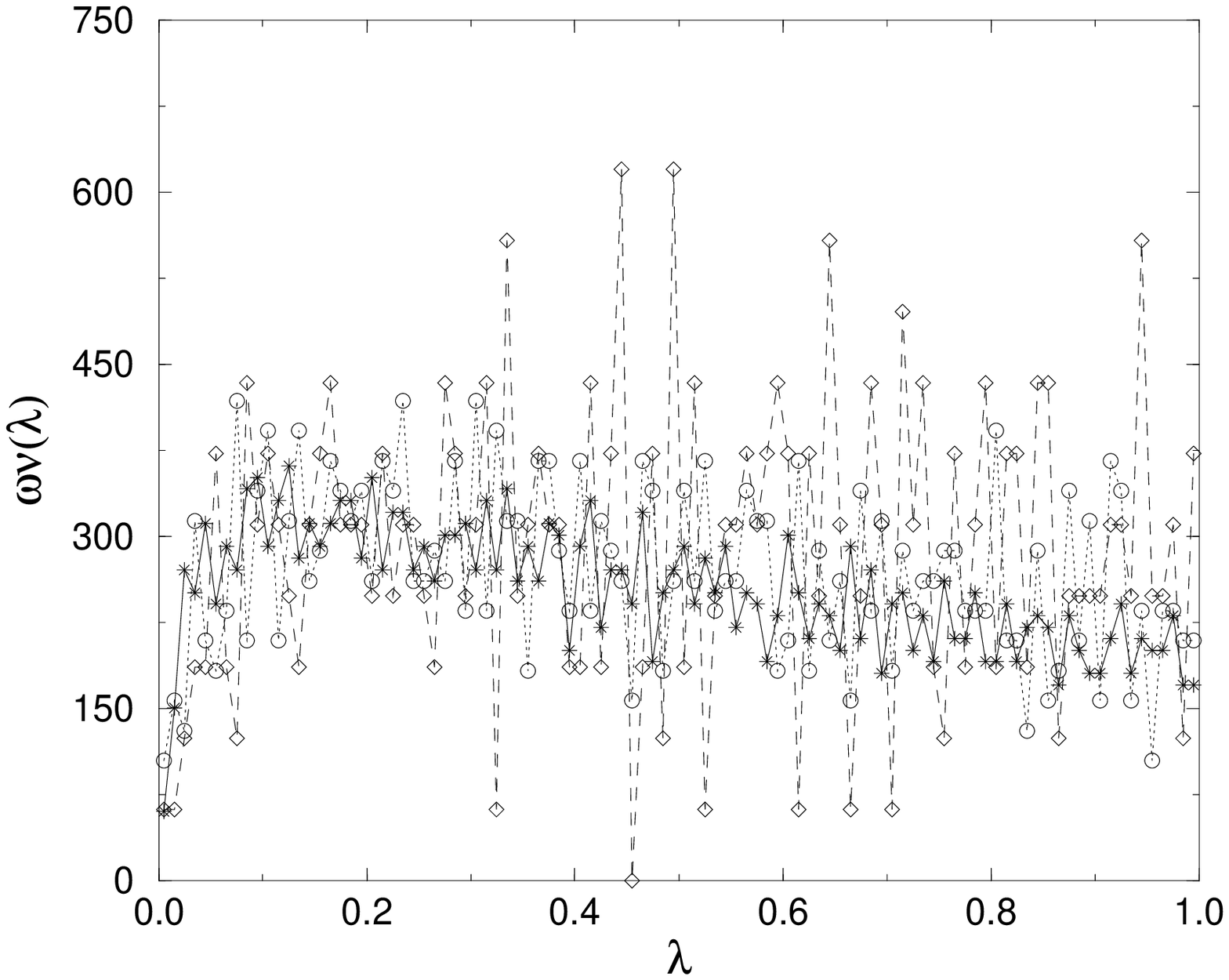}
\end{center}
\caption{The same densities as in figure~\ref{fig:4.xx.sp} but 
rescaled by $\omega = n_{50}/N_{I}$, where $N_{I}$ is number of 
topological charges after $i$ cooling sweeps.}
\label{fig:4.xx.sp.res}
\end{figure}

\begin{figure}[tbh]
\begin{center}
\leavevmode
\epsfxsize=100mm
\epsfbox{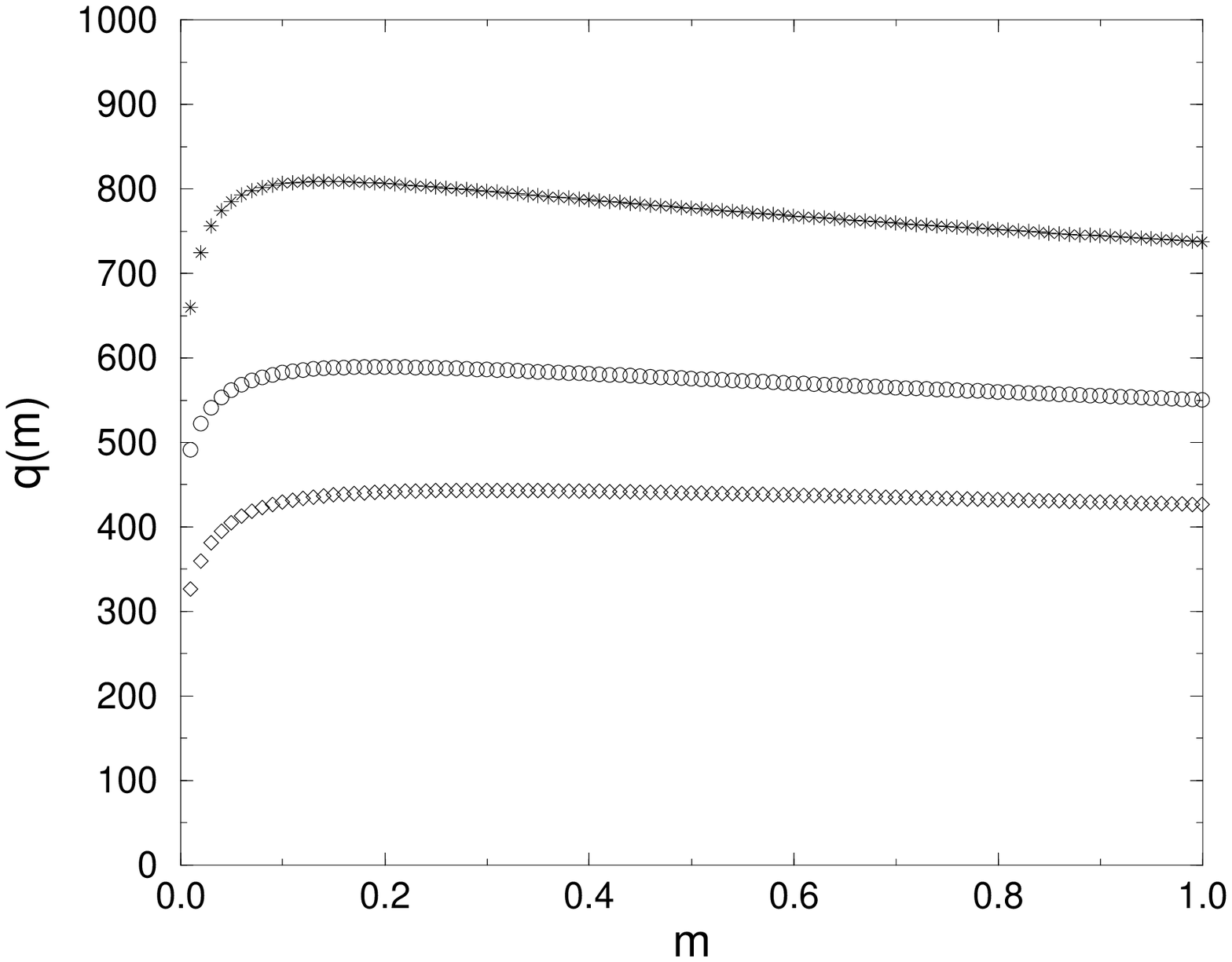}
\end{center}
\caption{\ssi = $(q(m) \rm{MeV})^{3}$
 obtained from the spectral densities in
figure~\ref{fig:4.xx.sp}.} 
\label{fig:4.xx.qc}
\end{figure}

\begin{figure}[tbh]
\begin{center}
\leavevmode
\epsfxsize=100mm
\epsfbox{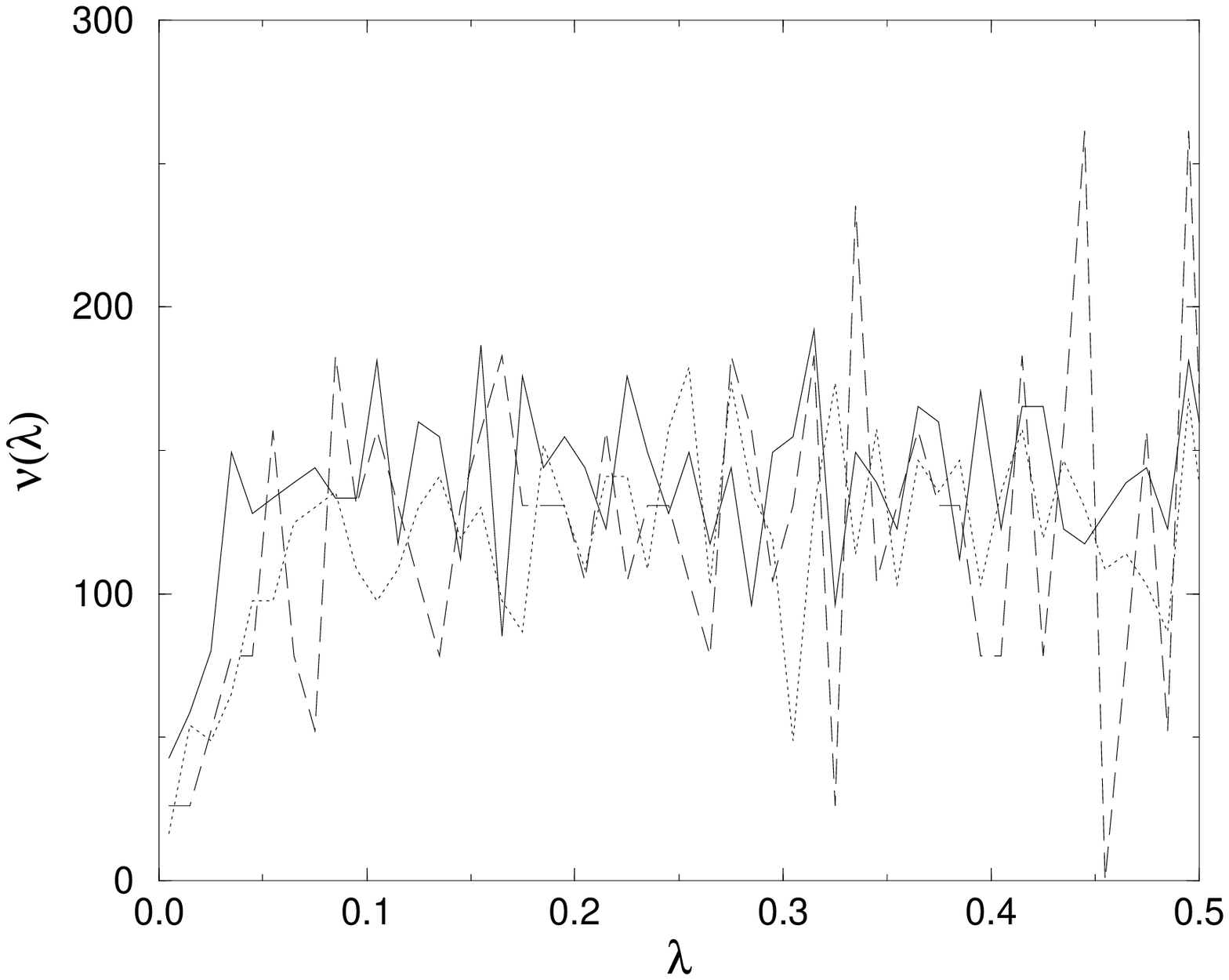}
\end{center}
\caption{The spectral densities obtained at $\beta=6.0, \ 6.2, \ 6.4$
after 23 (solid), 46 (dotted) and 80 (dashed) cooling sweeps respectively.}
\label{fig:x.xx.scl}
\end{figure}

We turn now to analyzing the effects of cooling. As a lattice field
configuration is cooled, one finds~\cite{Smith} that the average size
of the instantons increases and opposite charges annihilate. The
former leads to fewer smaller objects whilst the latter leads simply
to fewer objects in total per unit volume. Figure~\ref{fig:4.xx.sp}
shows the spectral density for the 20 configurations generated at
$\beta = 6.4$ (corresponding to the smallest lattice spacing) on a
$32^{3}\times64$ lattice for 30, 50 and 80 cooling sweeps
respectively. The configurations after 80 sweeps are thought to
correspond to configurations after 23 sweeps at $\beta = 6.0$ (see
\cite{Smith} and figure~\ref{fig:x.xx.scl}). Hence all these
configurations are denser than those analysed previously. We might
therefore expect the peaking at $\lambda=0$ to be weaker, or even
non-existent. This is indeed what we see in
figure~\ref{fig:4.xx.sp}. We also see something rather striking; as we
cool more, and as we find fewer objects in the same volume, the entire
spectral density shifts downwards in a way that is roughly
proportional to the change in instanton number (see
figure~\ref{fig:4.xx.sp.res}). This is in contradiction with the
optimistic expectation that cooling, being a local smoothing, should
have less effect at small eigenvalues (`infrared physics') and more
effect at large eigenvalues (`ultraviolet physics') -- as would occur
if the main reason for the decrease of the number of charges with
cooling was that heavily overlapping objects which produce large
eigenvalues were annihilating. This na\"{\i}ve hope is seen to be
unrealized.  Cooling will also therefore alter the quark condensate,
as we see in figure~\ref{fig:4.xx.qc}. (As usual this plot excludes
the exact zero modes which would give a finite-volume peaking of the
condensate at small quark masses.)  Whilst we should not pay too much
attention to the absolute normalisation of the quark condensate (given
the qualitative nature of the calculation), our observation that
cooling rapidly alters the quark condensate should be reliable.  This
creates an ambiguity that is particularly acute in the context of the
small-$\lambda$ divergence: depending on the amount of cooling, the
instanton ensemble produces a divergence in quenched QCD that ranges
from being very strong to being negligibly weak. The clear message is
that these instanton ensembles differ strongly in the long-distance
fermionic physics that they encode and that this is a problem that
needs to be resolved before one can be confident that one understands
the instanton content of the quenched QCD vacuum.

\begin{figure}[tbh]
\begin{center}
\leavevmode
\epsfxsize=100mm
\epsfbox{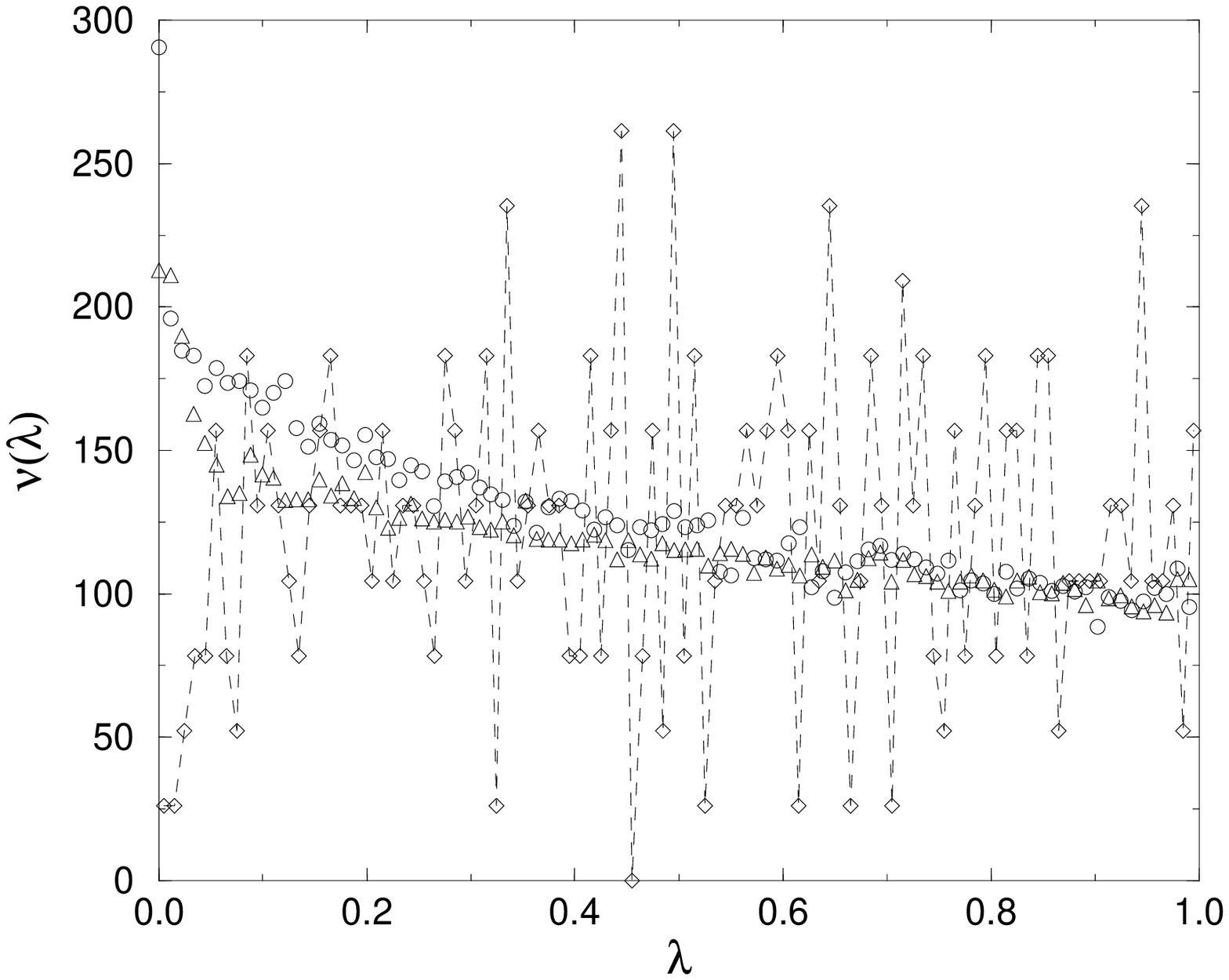}
\end{center}
\caption{Extending the calculations from actual lattice
data. ($\diamond$) $\beta = 6.4\ 32^{3}64$ data. $(\circ),\
(\triangle)$ from synthetic ensembles with approximately four times
the volume, and half the volume respectively.}
\label{fig:h_latxtend}
\end{figure}

It would be nice to have a study of the large volume limit at $\beta =
6.4$, similar to the one at $\beta= 6.0$. Unfortunately that would
require lattices much larger than $32^3\times64$ and this is clearly
impractical (at the moment). By contrast, a nice feature of using our
model is that it is easy to increase the volume and number of
configurations and so test whether one has reached the infinite volume
limit (and to obtain some idea of what a high statistics spectrum
would look like).  We show the results of such a calculation in
figure~\ref{fig:h_latxtend}. We compare the spectral density generated
from the instanton ensembles obtained after 80 cooling sweeps at
$\beta = 6.4$ to that from high statistics synthetic ensembles with
approximately half the volume and four times the volume
respectively. The packing fraction has been chosen to equal that of
the lattice ensemble. The lattice and model ensembles differ in that
the latter contain objects of a single size positioned at random and
with a total charge that is always zero, $Q=0$. We observe however
that the model spectra compare quite well with the lattice
spectrum. One difference is that the lattice spectrum lacks a forward
peak but this is in part due to the fact that the model ensemble
always has $Q=0$ while the lattice configurations do not. We note from
the figure that the two volumes produce essentially identical (model)
spectra.  Thus the $V\to\infty$ limit appears to be under control.
Finally we address the question of scaling. In \cite{Smith} it was
shown that if we vary the number of cooling sweeps with $\beta$
appropriately, then many properties of the instanton ensemble become
independent of $\beta$ once they are expressed in physical units. Is
this also true of the more subtle features that are embodied in
physical observables such as the chiral condensate?  To investigate
this we plot in figure~\ref{fig:x.xx.scl} the spectral densities
obtained after 23, 46 and 80 cooling sweeps on the $16^3\times48$,
$24^3\times48$ and $32^3\times64$ lattices at $\beta\ = 6.0,\ 6.2,\
6.4$ respectively. These lattices have nearly equal volumes in
physical units and the variation with $\beta$ of the number of cooling
sweeps is as prescribed in~\cite{Smith}.  As we see, the corresponding
spectral densities are very similar showing that the important
fermionic physical observables do indeed scale.

\section{Discussion}
One qualitative feature that is common to all the instanton ensembles
that we have investigated, is that they lead to spontaneous chiral
symmetry breaking.  A second, and striking, qualitative feature is
that the spectral density diverges as $\lambda \to 0$. The divergence
follows an approximate power law, $\propto \lambda^{-d}$, where $d$
decreases as the density of the instantons increases. Moreover we have
seen that it is possible to have a stronger divergence for denser
gases if one has a sufficient range of instanton sizes, of the kind
that one finds in the lattice instanton ensembles.  We have also tried
to fit the divergence with a logarithmic form, since this is what one
expects in leading-order quenched chiral perturbation theory
\cite{Sharpe-Elec,BernardB,Golter-Chip}. (It has also been seen by
unfolding the microscopic spectral density obtained via Random Matrix
Theory \cite{Osborn-Thou,Osborn-Chi,Verbaarschot,Shuryak-Priv}.) However
such logarithmic fits are usually unacceptable, and where they are not
it is a trivial consequence of the power exponent $d$ being small, as
in equation~(\ref{eq:log-power}). It is interesting to note that if
one attempts to sum the leading-logs of quenched chiral perturbation
theory, one can obtain \cite{Sharpe-Quen,Golter-Priv} a power
divergence. The exponent of this divergence is $d=\delta/(1+\delta)$
where the parameter $\delta$ is simply related to the elementary
pseudoscalar flavour singlet annihilation diagram (whose iteration
provides an estimate of the mass of the $\eta^{\prime}$ in full
QCD). The strength of this diagram is related, in turn, to the
topological structure of the quenched
vacuum~\cite{'tHooft,Witten,Veneziano}, and so this suggests an
approach to constructing a detailed link between our approach and that
of chiral perturbation theory. It is amusing to note that the most
recent quenched QCD estimates~\cite{Aoki,Sharpe-Prog} of $\delta$,
obtained from chiral extrapolations where this parameter multiplies
the quenched chiral log term, suggest a value $\delta \sim 0.1$ which
is consistent with the kind of weak divergence we typically observe on
the cooled instanton ensembles (see Table~\ref{tab:cdata}).

We have furthermore found evidence, from a comparison of the Dirac
spectral densities, for some of the claims in \cite{Smith}: in
particular for the screening of topological charges in the quenched
QCD vacuum, for the smallness of finite volume corrections and for the
claim that if the number of cooling sweeps is varied with $\beta$ so
that the number of topological charges per unit physical volume is
constant, then the physical observables show scaling.

However we have also found that fermionic physical observables, such
as the chiral condensate, vary strongly with the number of cooling
sweeps.  This contradicts the expectation that a moderate amount of
cooling should only eliminate short distance fluctuations and so
should not alter the physically important small-$\lambda$ end of the
spectral density. Whether this is a problem with cooling {\it per se}
or whether, as one would expect, it indicates the increasing
unreliability of the instanton ``pattern recognition'' algorithms
of~\cite{Smith} as one decreases the number of cooling sweeps, is a
question we are not able to address. The resulting uncertainty is
particularly important for the significance of the small-$\lambda$
divergence. We have seen that this divergence ranges from being strong
to being negligible depending on which of the lattice instanton
ensembles is used. It is strong for the larger number of cooling
sweeps, which is where the instanton pattern recognition should be
more reliable. On the other hand, a recent analysis~\cite{Ringwald}
suggests that it is the instanton ensembles obtained with less
cooling, where the low-$\lambda$ peaking is negligible, that are the
more physical.  So although we do find that instantons generically
produce a divergence in the chiral condensate of quenched QCD, it is
not clear whether it is strong enough to have any impact on the
predictions for physical quark masses.  One lesson is unambiguous:
there is more that needs to be done before one can claim to have
completely understood the true instanton structure of the gauge theory
vacuum.

\begin{table}[tbh]
\begin{center}
\begin{tabular}{|c|c|c|c|c|c|c|c|}
\hline
Set&$\beta$&$L^{3}T$&a(fm)&Cools &$N_{c}$&$\overline{N}_{T}/V$
($fm^{-4}$)&f\\ \hline
A&6.0&$16^{3}48$&0.098&23&100&9.1&4.2\\
B&6.0&$16^{3}48$&0.098&46&100&3.2&1.9\\
C&6.0&$32^{3}64$&0.098&46&50&3.5&2.84\\
D&$6.0_{RP}$&$32^{3}64$&0.098&46&50&3.5&2.84\\
E&$6.0_{SS}$&$32^{3}64$&0.098&46&50&3.5&1.85\\
F&$6.0_{\rho>0.12}$&$32^{3}64$&0.098&46&50&3.1&2.82\\
G&6.2&$24^{3}48$&0.072&46&100&8.9&4.9\\
H&6.4&$32^{3}64$&0.0545&30&20&56.6&12.4\\
I&6.4&$32^{3}64$&0.0545&50&20&21.7&8.5\\
J&6.4&$32^{3}64$&0.0545&80&20&9.2&5.3\\
\hline
\end{tabular}
\end{center}
\caption{Some information about the data analysed in this chapter. $V$
is the total spacetime volume, $\overline{V}_{I}$ is the average
volume of an object in the ensemble, $\overline{N}_{T}$ is the average
number of topological charges per configuration and $N_{c}$ is the
number of configurations in the ensemble. The subscript RP stands for
random positioning of objects whilst SS stands for all the objects
being set to the same size. Power law fits of the data were possible
(had reasonable statistical and systematic errors) for sets C, D and
E, and had degree of divergence $d = 0.251\pm0.069,\ 0.585\pm0.047$
and $0.186\pm0.060$ with $\chi^{2}/N_{DF} = 1.6,\ 2.2$ and 2.4
respectively.}
\label{tab:cdata}
\end{table}

%% file: unq.tex
\chapter{Unquenched QCD ensemble}
\label{ch:unq}

In chapter~\ref{ch:univ} we carried out simulations on random
configurations. We then proceeded in chapter~\ref{ch:latt} to study
instanton ensembles derived from gauge configurations generated with
the correct gauge weighting. In the spirit of D{\ae}dalus (whilst
recalling the misfortune of his child) we turn now to simulations
carried out on ensembles of instantons which incorporate both a gauge
weighting and a fermion weighting. The results presented in this
chapter have been published elsewhere~\cite{Sharan-Univ}.

One possible method for doing so would be to follow the equivalent
path to the previous chapter, but, based on dynamical fermion
ensembles. This is however not possible currently as there is only
limited data available on the topological content of the full QCD
vacuum (that is to say, the decomposition~\ref{decomp} has yet to be
carried out in a detailed and systematic way for a variety of volumes,
lattice spacings, pattern recognition algorithms etc.). We therefore
revert to the method of chapter~\ref{ch:univ} and generate synthetic
configurations. How do we incorporate the correct weightings into our
model~?

\section{Ensemble generation}
\subsection{The gauge weighting}
We use a Poisson distribution for the gauge weighting for the number
of instantons and anti-instantons in any configuration. The two
distributions are independent so the joint distribution is simply the
product of Poisson distributions:

\begin{equation}
P(N_{A}=s, N_{I}=t) =
\exp(-2\mu)\frac{\mu^{s+t}}{s!t!}
\end{equation}
where $\mu$ is the mean number of instantons (as well as
anti-instantons) in the gas if we used the gauge weighting alone. This
is analogous to equation~\ref{eq:varyia} with $\mu = N_{T}/2$. (The
variance is a factor of 2 greater using the Poisson weighting but such
numerical factors should be unimportant.)

\subsection{The fermion weighting}
We claim that we are generating the low lying eigenvalue spectrum of
the Dirac operator for any given instanton configuration. It is
therefore easy to give the fermion weighting as the determinant for
the configuration:

\begin{equation}
\det(\di[A] - im) \doteq (\overline{\lambda}_{NZ}^{2} +
m^{2})^{(T-N_{A}-N_{I})/2}m^{|N_{A}-N_{I}|}\prod_{i=1}^{\min(N_{I},N_{A})}(\lambda_{i}^{2}+m^{2})
,
\label{mod-det}
\end{equation}
where $\overline{\lambda}_{NZ}$ is a representative eigenvalue from
mixing with/of non-zero modes, $T$ is the total number of modes of the
Dirac operator and $\lambda_{i}$ are the eigenvalues generated for the
configuration. The logic is as follows. Different configurations will
contain different numbers of objects, and hence, direct comparison of
determinants will not produce the correct weighting. The total number
of eigenvalues is fixed, however, for a system with a fixed volume and
ultra-violet cutoff. (Recall the lattice operator of dimension 786432
mentioned previously.) We think of the eigenvalues we generate as
forming the crucial low lying spectrum; the remaining eigenvalues
should be larger. We use a constant $\overline{\lambda}_{NZ}$ to
represent these higher modes. It should be apparent that whilst the
total number of modes $T$ appears in equation~\ref{mod-det}, it should
not appear at all during the Monte Carlo simulation. This is because
the Monte Carlo simulation requires only the ratio of determinants
between old (accepted) configurations and new (trial) configurations,
and hence, the constant $(\overline{\lambda}_{NZ}^{2} + m^{2})^{T/2}$
drops out. (It should also be noted that whilst we have written a
factor of $(\overline{\lambda}_{NZ}^{2} + m^{2})^{1/2}$ for each
non-zero mode, in the numerical code we used a factor
$(\overline{\lambda}_{NZ} + m)$ instead. Given the qualitative nature
of our definition of $\overline{\lambda}_{NZ}$ this is hopefully not a
serious flaw.)

\subsection{Monte Carlo simulation}
To summarize, we can incorporate a reasonable gauge and fermion
weighting into our model using only two paramters $\mu$ and
$\overline{\lambda}_{NZ}$. The parameter $\mu$ is of course related to
the packing fraction we desire for our configurations but does not
determine it wholly. This is because the fermion determinant will also
play a part in finding the equilibrium number of objects in the gas.

The Monte Carlo simulation begins with a random configuration. We then
move a single object to generate a new trial configuration. This
process is repeated as we ``sweep'' through the gas, moving each
object in turn, accepting moves according to the standard Metropolis
algorithm:

\begin{equation}
\begin{array}{lcll}
P({\rm accept}) & = & 1 & \det({\rm new}) > \det({\rm old})\\
& = & \frac{\det({\rm new})}{\det({\rm old})} & {\rm otherwise} .
\end{array}
\end{equation}
We incorporate different numbers of flavours of fermions by raising
the ratio to the power $N_{f}$. Periodically we attempt to either
increase or decrease the number of instantons or anti-instantons by
one (the period being every 10 attempted moves). We expect the system
to come into equilibrium after some number of sweeps. As the change
between successive configurations is small (differing only in the
position of a single object or in having one extra or one fewer
object), we normally require long separations to obtain independent
configurations. Whilst we use all the configurations we generate
(after leaving enough for the system to equilibriate), we ensure we
have long runs to maximize the number of independent configurations.

\subsection{Correlation functions}
As we have a Monte Carlo simulation which moves objects around trying
to generate configurations according to their overall weighting (both
gauge and fermion), we can calculate correlation functions and use
these to estimate particle masses~! We have two gluonic operators
which we can use within our model, namely we have the topological
charge density $Q$ and the number density of objects $N$ (which is our
equivalent of the action density). How can we use these to calculate
particle masses~?

\subsubsection{$\langle Q(0)Q(t)\rangle$}
We divide up our volume into a number of ``strips'' each of width
$\delta t$. We can calculate the total charge in any strip simply by
adding up the charges of all the objects contained within the
strip. We call this $Q(t)$ for the strip $[t,t+\delta t)$. That is to
say, $Q(t) = N_{I}(t) - N_{A}(t)$ where $N_{I/A}(t)$ are the number of
instantons/anti-instantons contained with the strip. We can then work
out the correlation function $\langle Q(0)Q(t)\rangle$ as a function
of separation $t$.

\begin{equation}
\langle Q(0)Q(t)\rangle = \sum_{n}c_{n}\exp(-M_{n}t) .
\end{equation}
We can fit the resultant correlation function and attempt to extract
the lowest mass (this will be the exponential which dies away
slowest). Examination of the quantum numbers of the operator
($0^{-+}$, flavour singlet - as the operator is purely gluonic)
reveals that this will be the mass within our model corresponding to
the $\eta^{'}$.

\subsubsection{$\langle N(0)N(t)\rangle - \langle N(0)\rangle^{2}$}
We can do the same as above but using our operator equivalent of the
action. In this case $N(t) = N_{I}(t) + N_{A}(t)$. The only difference
is that this time we have to subtract the constant vacuum expectation
value, before fitting to an exponential. The mass we extract (if we
can) will be the mass within our model corresponding to the $\sigma$
($0^{++}$, flavour singlet).

\subsubsection{Effective mass plots}
As well as fitting exponentials to the correlation functions, we can
plot the following quantity known as the effective mass:

\begin{equation}
m_{eff}(t) = -\ln\left(\frac{\langle
{\cal O}(0){\cal O}(t)\rangle}{\langle{\cal O}(0){\cal O}(t-1)\rangle}\right)
,
\end{equation}
where we have written our correlator with a generic operator ${\cal
O}$ which can represent either the charge or the number density. It
should be apparent that if the correlation function is in fact given
by an exponential then our effective mass $m_{eff}(t)$ will be a
straight line.

\section{A few questions}
There are a surprisingly large number of non-trivial questions we can
ask within the framework of our model. Quantities of interest include:

\begin{itemize}
\item{Spectral density. How does the spectral density behave with
dynamical quarks~? Do we still get a power divergence as seen
previously, $b(m)\lambda^{-d(m)}$ where $b(m),\ d(m)$ are now
dependent upon the quark mass~? What is the behaviour of the spectral
density as a function of the number of quark flavours $N_{f}$~?}
\item{Chiral condensate. What is the behaviour of \ssi(m)~? What is
its behaviour as a function of the number of quark flavours~?}
\item{Topological susceptibility. General arguments give the behaviour
of this quantity as:

\begin{equation}
\begin{array}{lcll}
\langle Q^{2}\rangle(m) & \propto & mV & \chi SB\\
& \propto & m^{N_{f}}V & {\rm symmetric\ phase} .
\end{array}
\label{q2_behav}
\end{equation}
Can our model reproduce such behaviour~?}
\item{Particle masses. How do the masses of our versions of the
$\sigma$ and $\eta^{'}$ behave as functions of quark mass~? In
reality, neither of these particles belongs to the octet of Goldstone
bosons (in the chiral limit), hence, we wish for both particles to
remain massive as we take the quark mass to zero. On a simpler level,
can we even extract masses for such particles~?}
\end{itemize}

We carry out simulations at $\overline{\lambda}_{NZ} = 2$. This value
is chosen because the mean eigenvalue for our ``standard'' quenched
simulation with $f=1,\ V=1$ is approximately $1.12$. If we compare
with figure~\ref{fig:h_stan} then it is apparent that this value is
also far larger than the median eigenvalue (of the quenched
ensemble). Whilst we do not know the ``correct'' value to use for
$\overline{\lambda}_{NZ}$ (there is almost certainly no single correct
value), $\approx 200\%$ of the quenched figure seems a reasonable
place to start. As in the standard quenched simulation, we use $V=1$
with all objects of a fixed size $\rho=0.2$ (any objects which are
inserted into the gas also have this size). The gauge weighting is
given with $\mu=63$. Initially we have $N_{I}=63,\ N_{A}=63$, hence
initially, the parameters are exactly those of the standard quenched
simulation. We do not expect however for the system to find
equilibrium with these values; once the fermion weighting is taken
into account, the mean number of objects will undoubtedly be something
else.

A difficulty with this approach lies with the concept of a replacement
eigenvalue $\overline{\lambda}_{NZ}$. Our entire model is based around
would-be zero modes and a replacement eigenvalue is admittedly, a very
ad hoc method to incorporate other modes. We therefore also carry out
unquenched simulations with a fixed total number of objects. The
advantage of this is that we can have a fermion weighting given by the
determinant without having to introduce a replacement eigenvalue at
all. We still ``sweep'' through the configuration, moving objects as
before, we still introduce and remove objects as before. The only
difference is that every time we introduce an object, we remove an
object of the opposite chirality, and, {\em vice versa}. The winding
number can therefore only take even values but this is not a great
loss considering that we have one fewer arbitrary parameter in our
simulation.

We use a range of quark masses $m = 0.15,\ 0.3$ and $0.5$ with $N_{f}
= 1,\ 2$ for both the case when we have a replacement eigenvalue and
when we have the total number of objects fixed. In the latter case we
also carry out simulations at some larger quark masses $m=1.0,\ 2.0$
and $3.0$. (It will turn out that there are technical and conceptual
difficulties associated with running simulations for ``large'' quark
masses when we have a replacement eigenvalue.) An ensemble consists of
630000 configurations. It should be noted however that the number of
independent configurations is far smaller (perhaps two orders of
magnitude smaller). We omit the first 63000 configurations (from data
gathering) to allow the system to achieve equilibrium. (In truth a
dynamic form of checking during the run would be preferable.)

\section{Results}
\subsection{$N_{f} = 1,\ \overline{\lambda}_{NZ} = 2.0$.}
\begin{figure}[tb]
\begin{center}
\leavevmode
\epsfxsize=100mm
\epsfbox{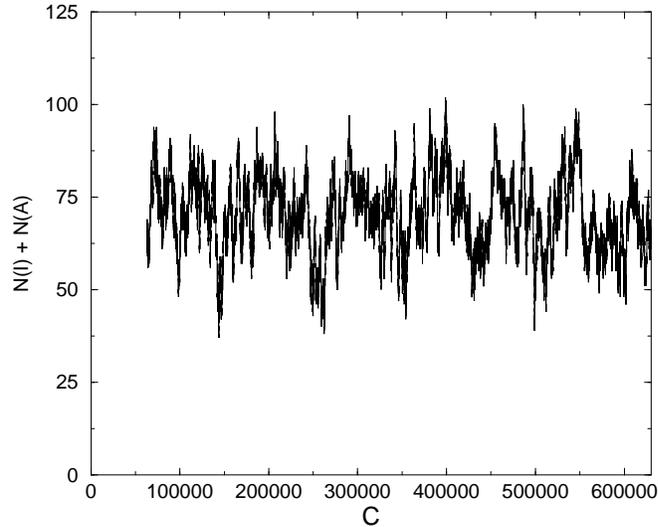}
\end{center}
\caption{$N_{f}=1,\ \overline{\lambda}_{NZ}=2.0,\ V=1,\ m = 0.5$. The
total number of objects as a function of the configuration number.}
\label{fig:050_eqt_rm1}
\end{figure}

We begin with one flavour of fermion with mass $m = 0.5$. As a test of
whether the system has achieved equilibrium, we plot the total number
of objects in the gas $(N_{I} + N_{A})(c)$ as a function of the
configuration number $c$ in figure~\ref{fig:050_eqt_rm1}.

\begin{figure}[tb]
\begin{center}
\leavevmode
\epsfxsize=100mm
\epsfbox{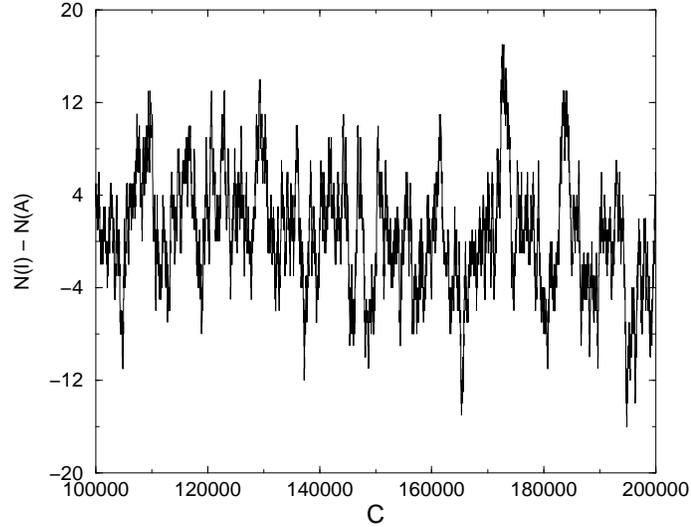}
\end{center}
\caption{$N_{f}=1,\ \overline{\lambda}_{NZ}=2.0,\ V=1,\ m = 0.5$. The
net winding number as a function of the configuration number. We show
only a section of the run for clarity.}
\label{fig:050_eqw_rm1}
\end{figure}

\begin{figure}[tb]
\begin{center}
\leavevmode
\epsfxsize=100mm
\epsfbox{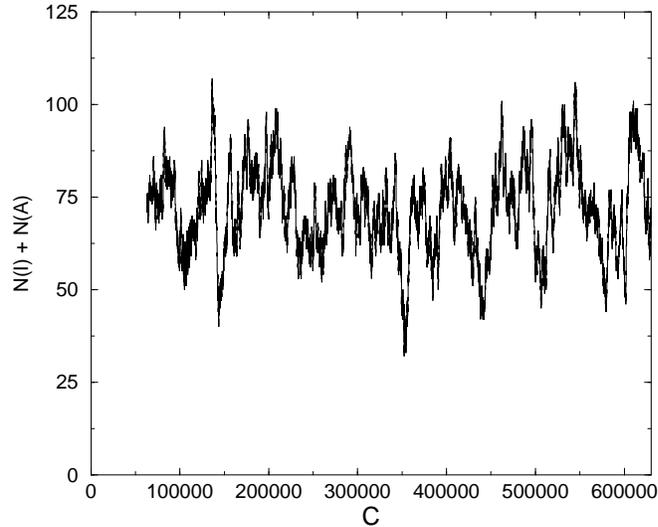}
\end{center}
\caption{$N_{f}=1,\ \overline{\lambda}_{NZ}=2.0,\ V=1,\ m = 0.15$. The
total number of objects as a function of the configuration number.}
\label{fig:015_eqt_rm1}
\end{figure}

\begin{figure}[tb]
\begin{center}
\leavevmode
\epsfxsize=100mm
\epsfbox{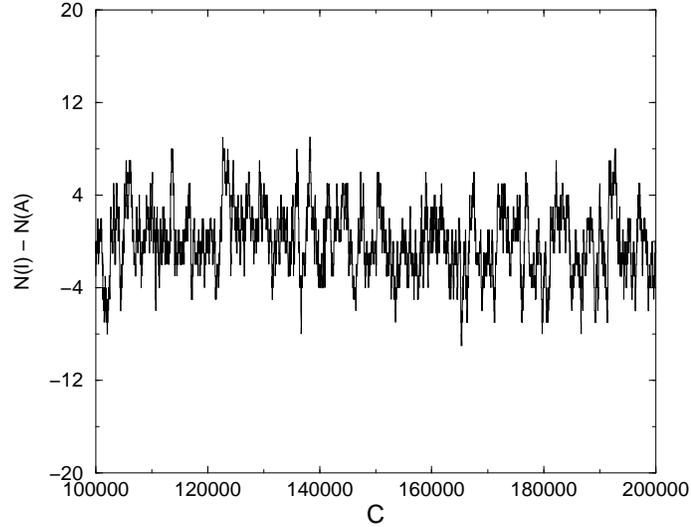}
\end{center}
\caption{$N_{f}=1,\ \overline{\lambda}_{NZ}=2.0,\ V=1,\ m = 0.15$. The
net winding number as a function of the configuration number. We show
only a section of the run for clarity.}
\label{fig:015_eqw_rm1}
\end{figure}

We see that the system oscillates around 70 objects in total (precise
figures are given in table~\ref{tab:unq}), radically different from
the initial 126. We shouldn't pay too much attention to the actual
figure, for this will depend upon the value of
$\overline{\lambda}_{NZ}$ we choose to use, the main importance is as
a check of the algorithm and as a test of whether the system has come
into equilibrium. We should also ensure that we are scanning net
non-trivial topological sectors and are not stuck in the $Q = 0$
sector. This can be seen in figure~\ref{fig:050_eqw_rm1}, which, is
reassuring.
What is the behaviour for lighter quark masses~? The lighter the
quark, the greater the difficulty for simulation. This is because it
becomes progressively more difficult to accept a trial configuration
(e.g. consider a change from 63 objects of each chirality to 62
objects of each chirality; this requires us to go through a stage with
63 of one and 62 of the other - if we have massless objects then this
will pose an insurmountable barrier as the determinant for the trial
configuration will be precisely zero). The algorithm thus begins to
``slow down'' and its ergodicity gradually breaks down. We expect it
to take longer to achieve equilibrium and longer to scan different
sectors. Figures~\ref{fig:015_eqt_rm1} and~\ref{fig:015_eqw_rm1} which
are the corresponding plots (to figures~\ref{fig:050_eqt_rm1}
and~\ref{fig:050_eqw_rm1} respectively) are therefore pleasing; whilst
there is undoubtedly some slowing down, there appears to be little
difficulty with this range of quark masses.

\begin{figure}[tb]
\begin{center}
\leavevmode
\epsfxsize=100mm
\epsfbox{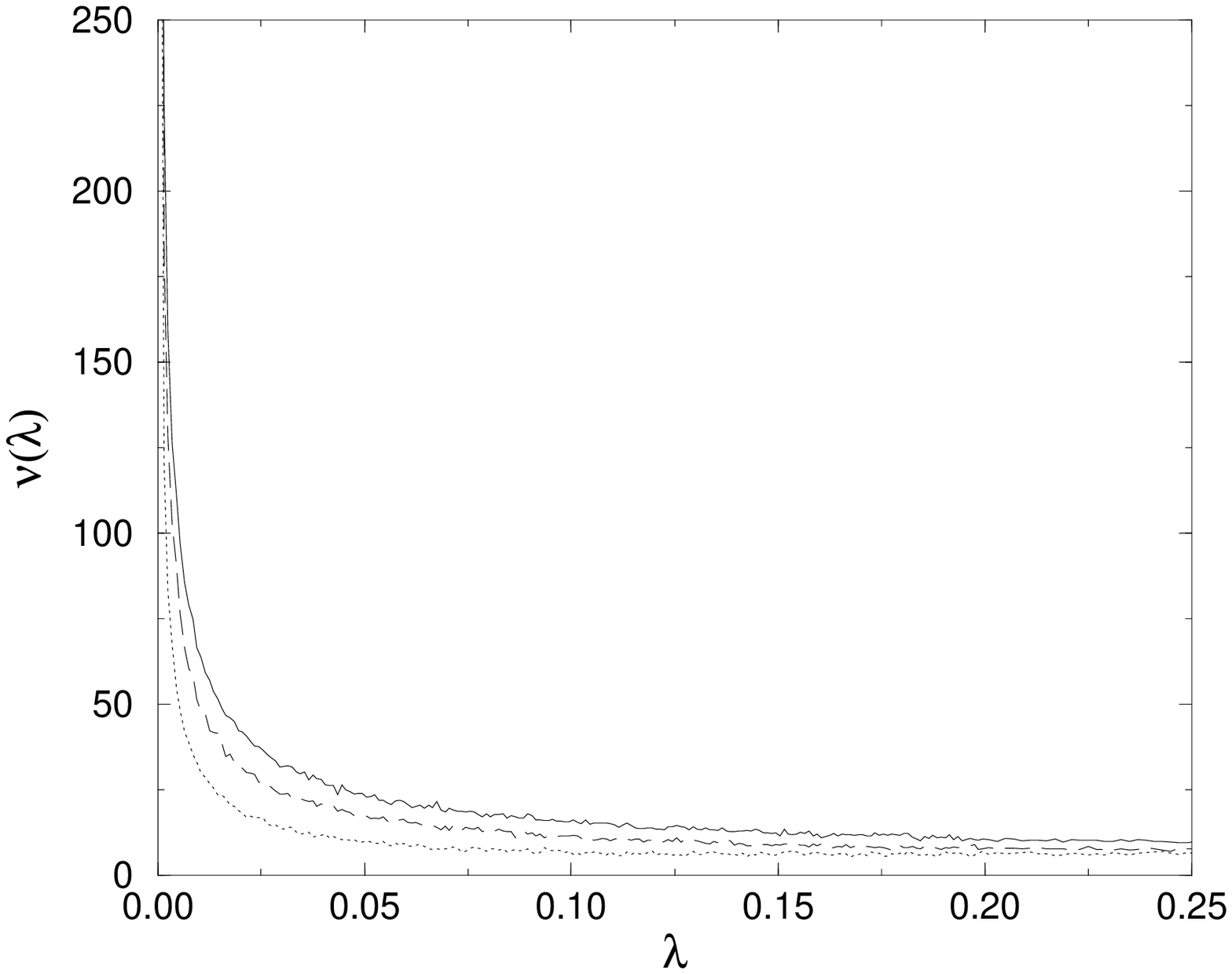}
\end{center}
\caption{$N_{f}=1,\ \overline{\lambda}_{NZ}=2.0,\ V=1$. The spectral
density for quark mass $m=0.5$ (solid), 0.3 (long dashed) and 0.15
(dotted).}
\label{fig:sp_un_rm1}
\end{figure}

\begin{figure}[tb]
\begin{center}
\leavevmode
\epsfxsize=100mm
\epsfbox{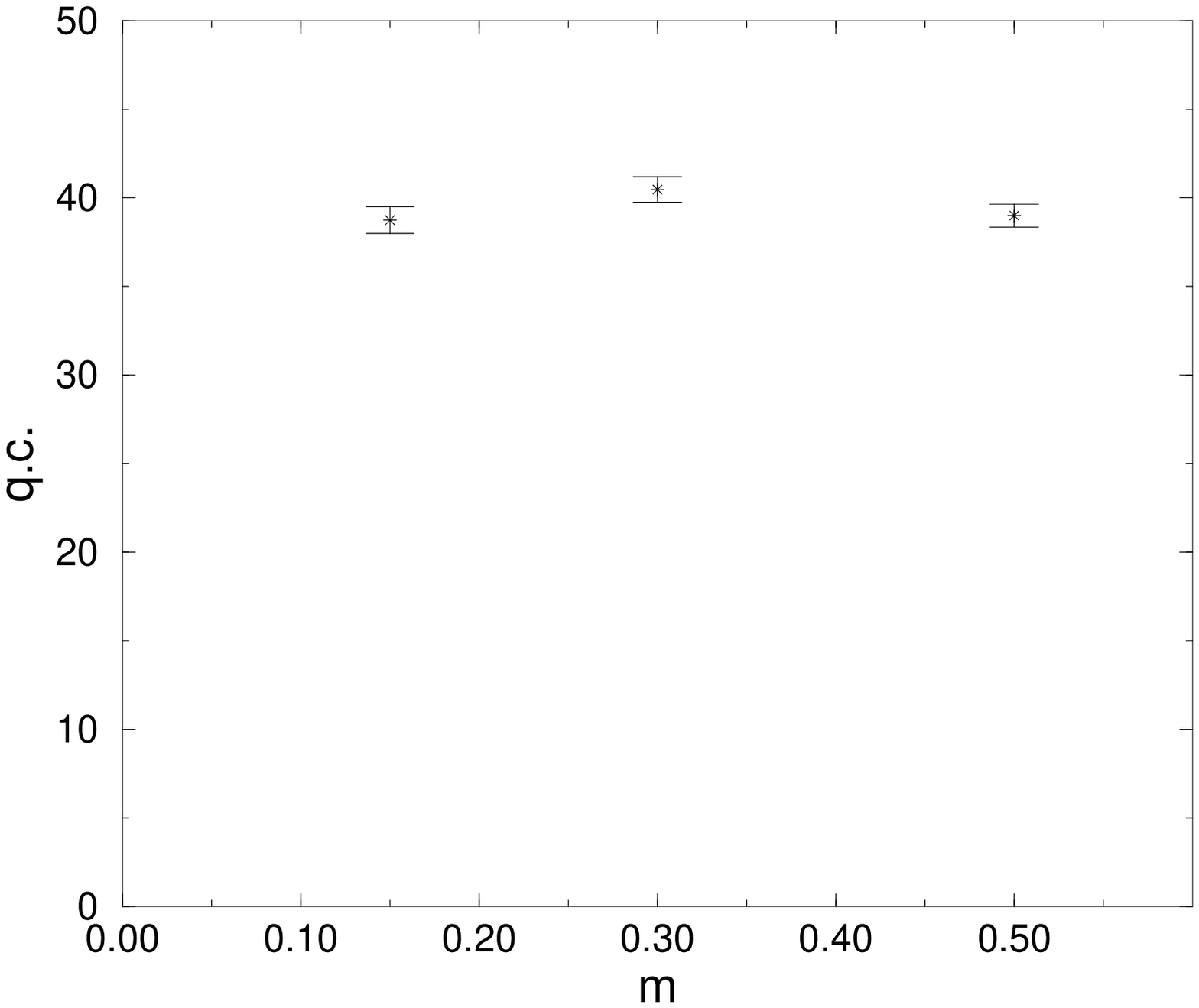}
\end{center}
\caption{The quark condensate \ssi\ as a function of the quark mass
$m$ from the spectra plotted in figure~\ref{fig:sp_un_rm1}.}
\label{fig:qc_rm1}
\end{figure}

\begin{figure}[tb]
\begin{center}
\leavevmode
\epsfxsize=100mm
\epsfbox{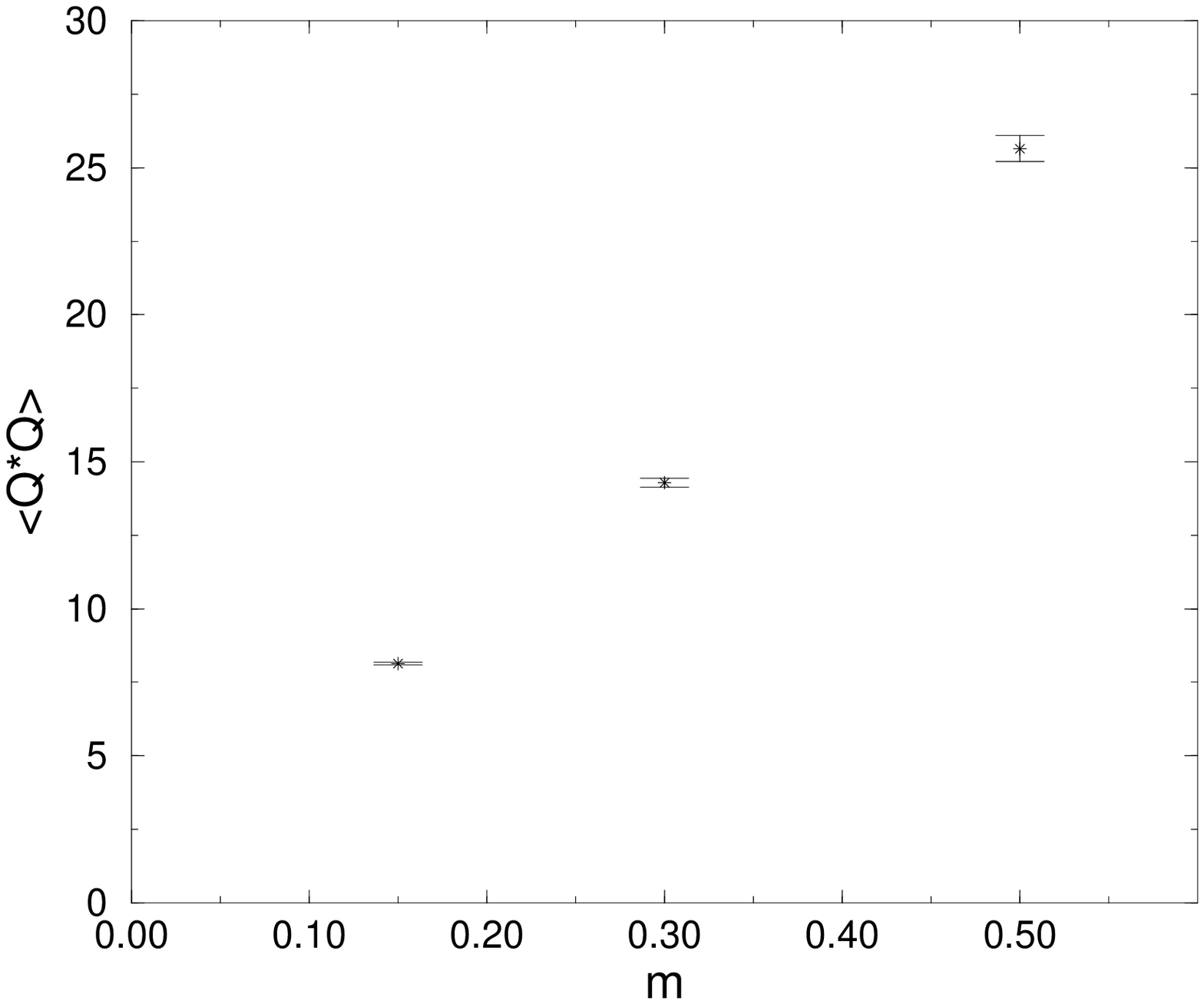}
\end{center}
\caption{$N_{f}=1,\ \overline{\lambda}_{NZ}=2.0,\ V=1$. The second
moment of the winding number distribution $\langle Q^{2}\rangle$ as a
function of the quark mass $m$.}
\label{fig:qq_rm1}
\end{figure}

Reassured that the algorithm is behaving, we move onto the spectral
density. Figure~\ref{fig:sp_un_rm1} shows the spectral density for the
three quark masses. The first thing to note is that the spectral
density appears to diverge as $\lambda \rightarrow 0$ for all three
masses.
It is however, equally clear that the spectral density is not
independent of the quark mass. This gives us hope that we may yet
evade the Banks-Casher relation and achieve a finite quark
condensate. Table~\ref{tab:sp} clearly shows that the coefficient of
the divergence $b \rightarrow 0$ as $m \rightarrow 0$. Whilst this is
not in itself enough to show that we will achieve a finite quark
condensate, it is at least promising. In figure~\ref{fig:qc_rm1} we
integrate these spectra and plot the corresponding quark condensate, a
linear extrapolation shows that we obtain chiral symmetry breaking
with a finite condensate.
We can also verify that the ensembles we have generated possess the
correct distribution of winding numbers (at least the first two
moments have the correct behaviour~!). This can be seen in
table~\ref{tab:unq} and in figure~\ref{fig:qq_rm1}. Whilst we
expect to obtain a linear relationship for $\langle Q^{2}\rangle(m)$
even in the symmetric phase (for $N_{f}=1$), this is not a trivial
test for our model.

\begin{figure}[tb]
\begin{center}
\leavevmode
\epsfxsize=100mm
\epsfbox{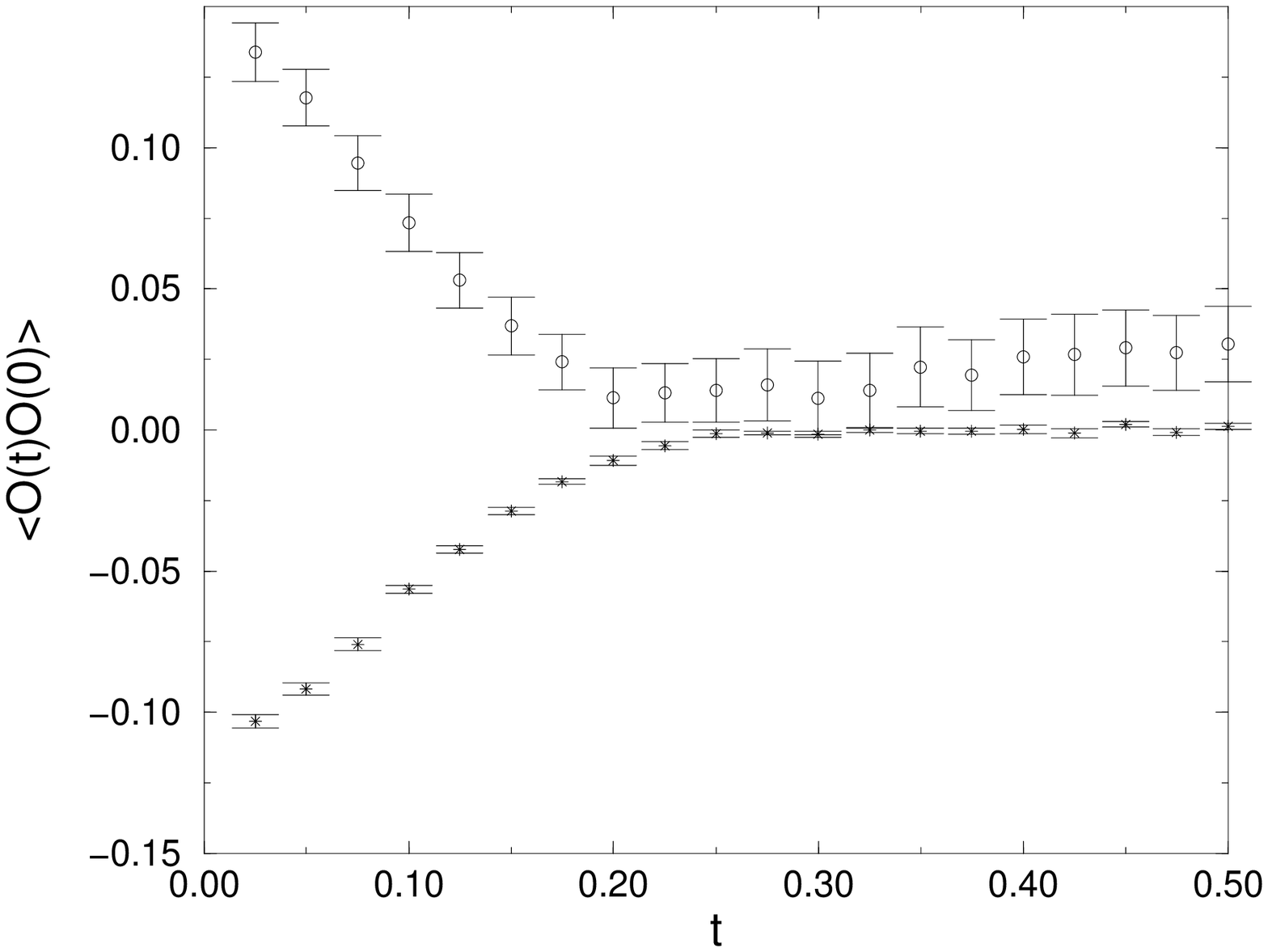}
\end{center}
\caption{$N_{f}=1,\ \overline{\lambda}_{NZ}=2.0,\ V=1$. Correlation
functions for $\eta^{'}\ (\star)$, and for $\sigma\ (\circ)$.}
\label{eta_sig_corr}
\end{figure}

\begin{figure}[tb]
\begin{center}
\leavevmode
\epsfxsize=100mm
\epsfbox{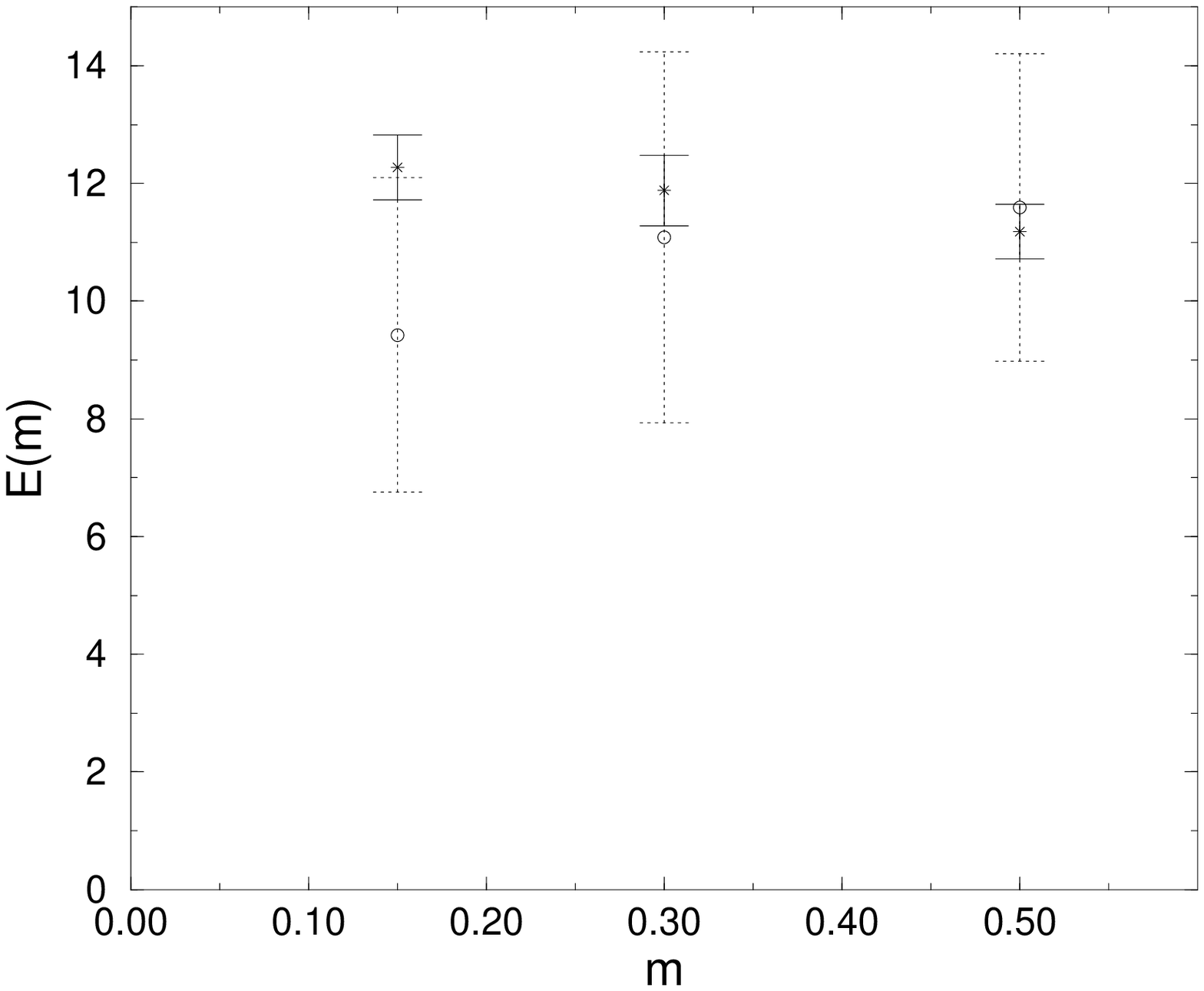}
\end{center}
\caption{$N_{f}=1,\ \overline{\lambda}_{NZ}=2.0,\ V=1$. Particle
masses. $\eta^{'}$ ($\star$) and $\sigma$ ($\circ$) as a function of
the quark mass.}
\label{eta_sig_mass}
\end{figure}

\begin{figure}[tb]
\begin{center}
\leavevmode
\epsfxsize=100mm
\epsfbox{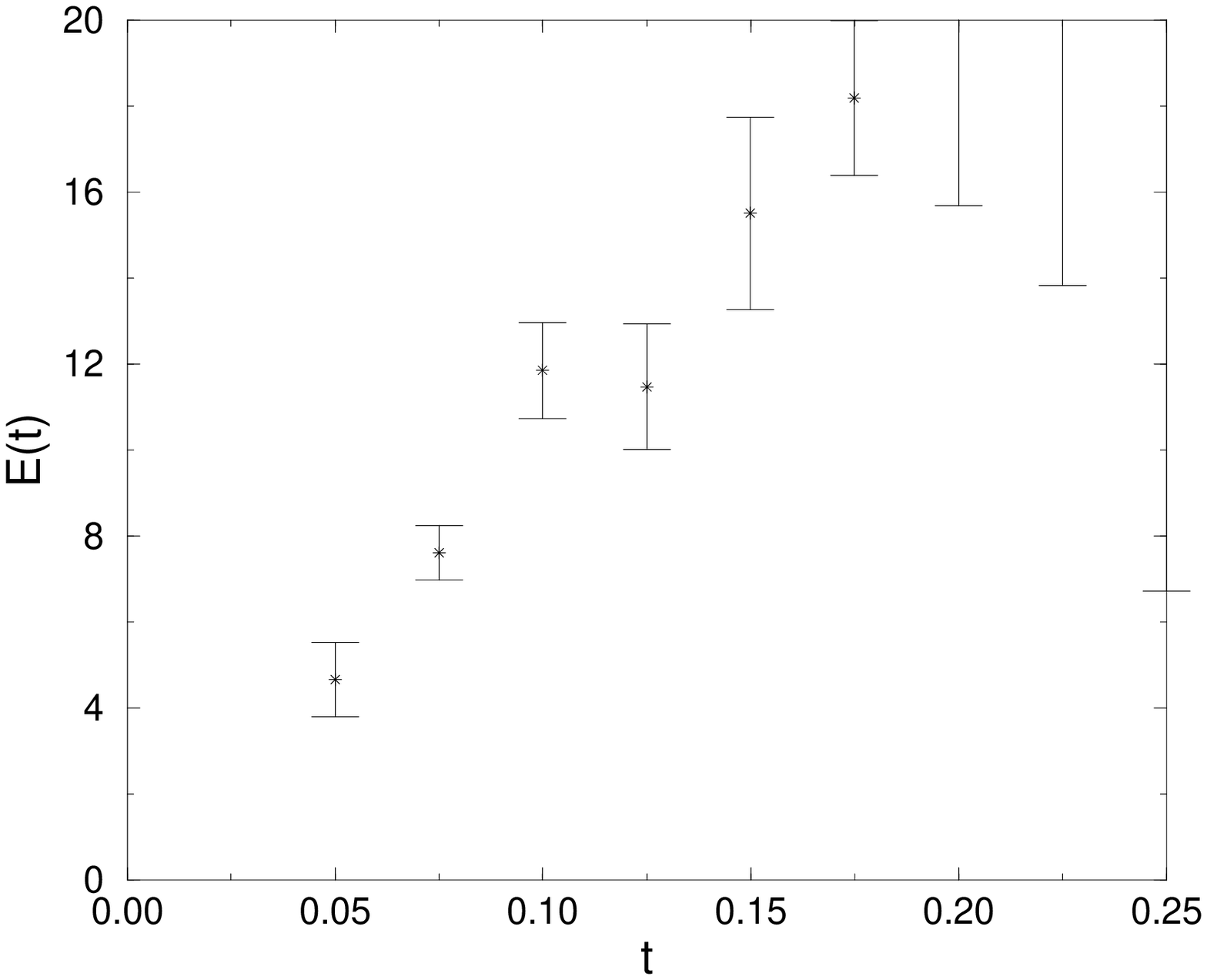}
\end{center}
\caption{$N_{f}=1,\ \overline{\lambda}_{NZ}=2.0,\ V=1,\ m=0.15$. The
effective mass for the $\eta^{'}$ data shown in
figure~\ref{eta_sig_corr}.}
\label{meff_eta}
\end{figure}

Our results seem to be suprisingly good considering the simplicity of
the underlying model. We move onto an extreme test of the model, that
of obtaining particle masses. Figure~\ref{eta_sig_corr} shows the
correlation function for the $\eta^{'}$ and the $\sigma$ for $m =
0.15$. We can see that the fermion weighting has set up highly
non-trivial correlations amongst the objects. We consistently find the
$\eta^{'}$ channel to give a far cleaner signal than the $\sigma$
channel. A consequence of this is that we get far higher values for
$\chi^{2}/N_{DF}$ for the exponential fit to the $\eta^{'}$ data (see
table~\ref{tab:partmass}). It should be stressed however that we have
fitted the exponential to the entire range shown in
figure~\ref{eta_sig_corr}. The maximum of this range $t=0.5$
corresponds to a separation half way around the torus i.e. the maximum
possible separation involving only one co-ordinate. If we were
concerned with the large $\chi^{2}/N_{DF}$ for this fit then we could
certainly reduce the range and excise the region which is in theory
the most useful, but in practice, mostly noise (e.g. $t \geq
0.3$). Coversely, the low $\chi^{2}/N_{DF}$ for the $\sigma$ is a
consequence of the weakness of the signal, not the goodness of
fit. Regardless of these reservations, this is still a remarkable
result for our model, a fact underlined by the mass plot given in
figure~\ref{eta_sig_mass} which shows that both particles remain
massive in the chiral limit.
Lastly, in figure~\ref{meff_eta}, we give the effective mass plot
corresponding to the $\eta^{'}$ with $m=0.15$. This shows that whilst
the particle masses we have given are plausible, we could easily pick
a slightly different mass due to the noise in the data (even for the
$\eta^{'}$).

\subsection{$N_{f} = 2,\ \overline{\lambda}_{NZ} = 2.0$.}
\begin{figure}[tb]
\begin{center}
\leavevmode
\epsfxsize=100mm
\epsfbox{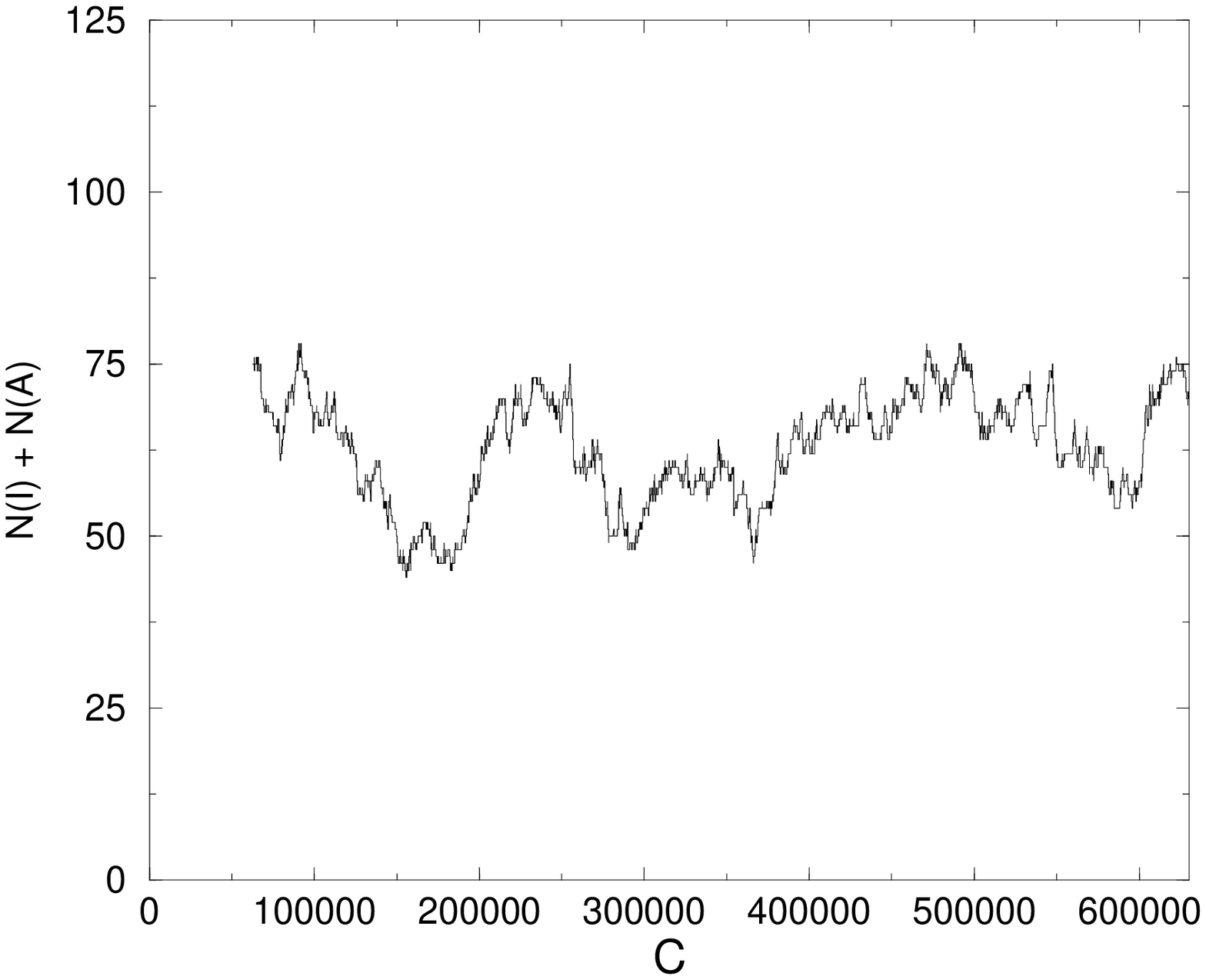}
\end{center}
\caption{$N_{f}=2,\ \overline{\lambda}_{NZ}=2.0,\ V=1,\ m = 0.15$. The
total number of objects as a function of the configuration number.}
\label{fig:015_eqt_rm2}
\end{figure}

\begin{figure}[tb]
\begin{center}
\leavevmode
\epsfxsize=100mm
\epsfbox{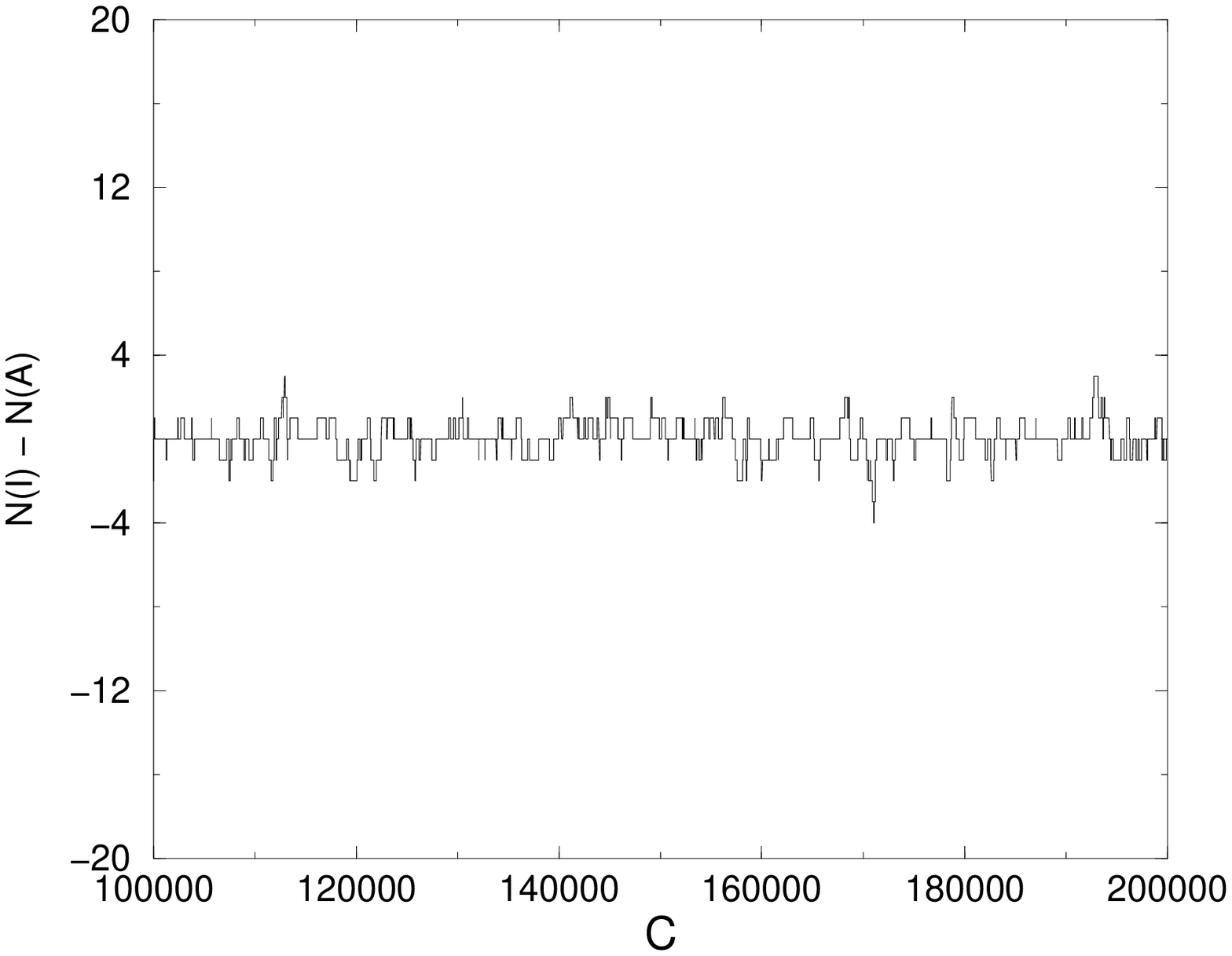}
\end{center}
\caption{$N_{f}=2,\ \overline{\lambda}_{NZ}=2.0,\ V=1,\ m = 0.15$. The
net winding number as a function of the configuration number. We show
only a section of the run for clarity.}
\label{fig:015_eqw_rm2}
\end{figure}

The results we have found seem to indicate that we can obtain a large
amount of (qualitative) information about QCD by focusing on simply
instanton degrees of freedom. It is of course trivial to vary the
number of fermion flavours, all that is required is to take the ratio
of the determinants to the power $N_{f}$ in the Metropolis step. It
should be apparent however that the algorithm will struggle even more
as we reduce the quark mass, effectively our suppression is enhanced
and it is even more difficult to accept trial configurations. This can
be clearly seen in figure~\ref{fig:015_eqt_rm2} where we show the
total number of objects as function of the configuration number for a
quark mass of $m=0.15$. This plot should be compared with
figure~\ref{fig:015_eqt_rm1} to see the degree of difficulty
encountered by the algorithm. We are effectively becoming trapped in
certain sectors, in order to sample the phase space as effectively as
for the $N_{f}=1$ case, we would require even larger runs.
This is only emphasised by figure~\ref{fig:015_eqw_rm2} which shows
the difficulty in moving between sectors of different net topological
charge.
We should therefore be a little sceptical of these results, it is not
a question of being wrong, it is more a question of being incomplete -
we have not sampled as much of the phase space as we would wish. It is
however not all bad, we can see from these plots that we appear to
have achieved equilibrium; we are oscillating around $N_{T}=60$, not
far from the $N_{f}=1$ figure, as opposed to the initial $N_{T}=126$.

\begin{figure}[tb]
\begin{center}
\leavevmode
\epsfxsize=100mm
\epsfbox{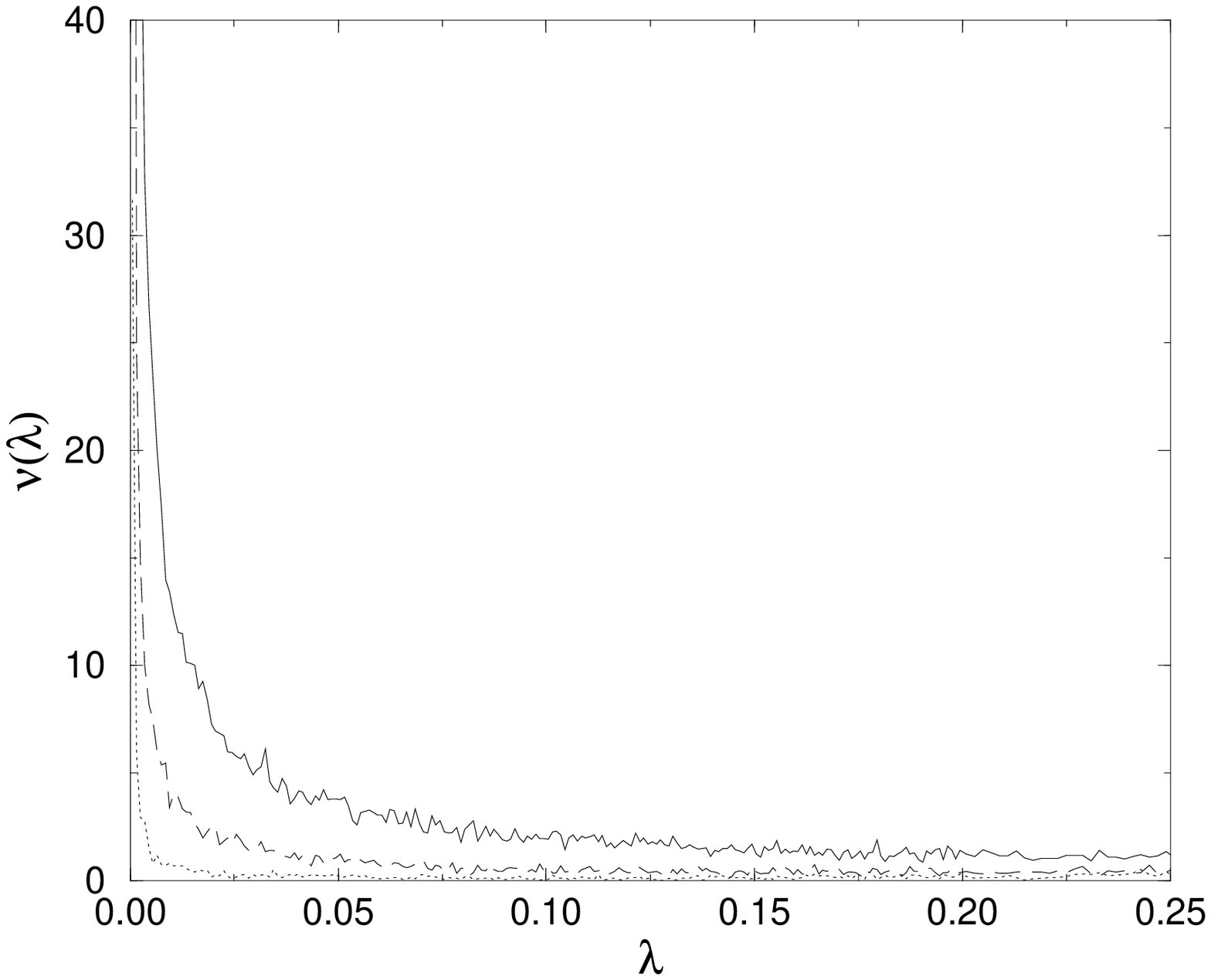}
\end{center}
\caption{$N_{f}=2,\ \overline{\lambda}_{NZ}=2.0,\ V=1$. The spectral
density for quark mass $m=0.5$ (solid), 0.3 (long dashed) and 0.15
(dotted).}
\label{fig:sp_un_rm2}
\end{figure}

\begin{figure}[tb]
\begin{center}
\leavevmode
\epsfxsize=100mm
\epsfbox{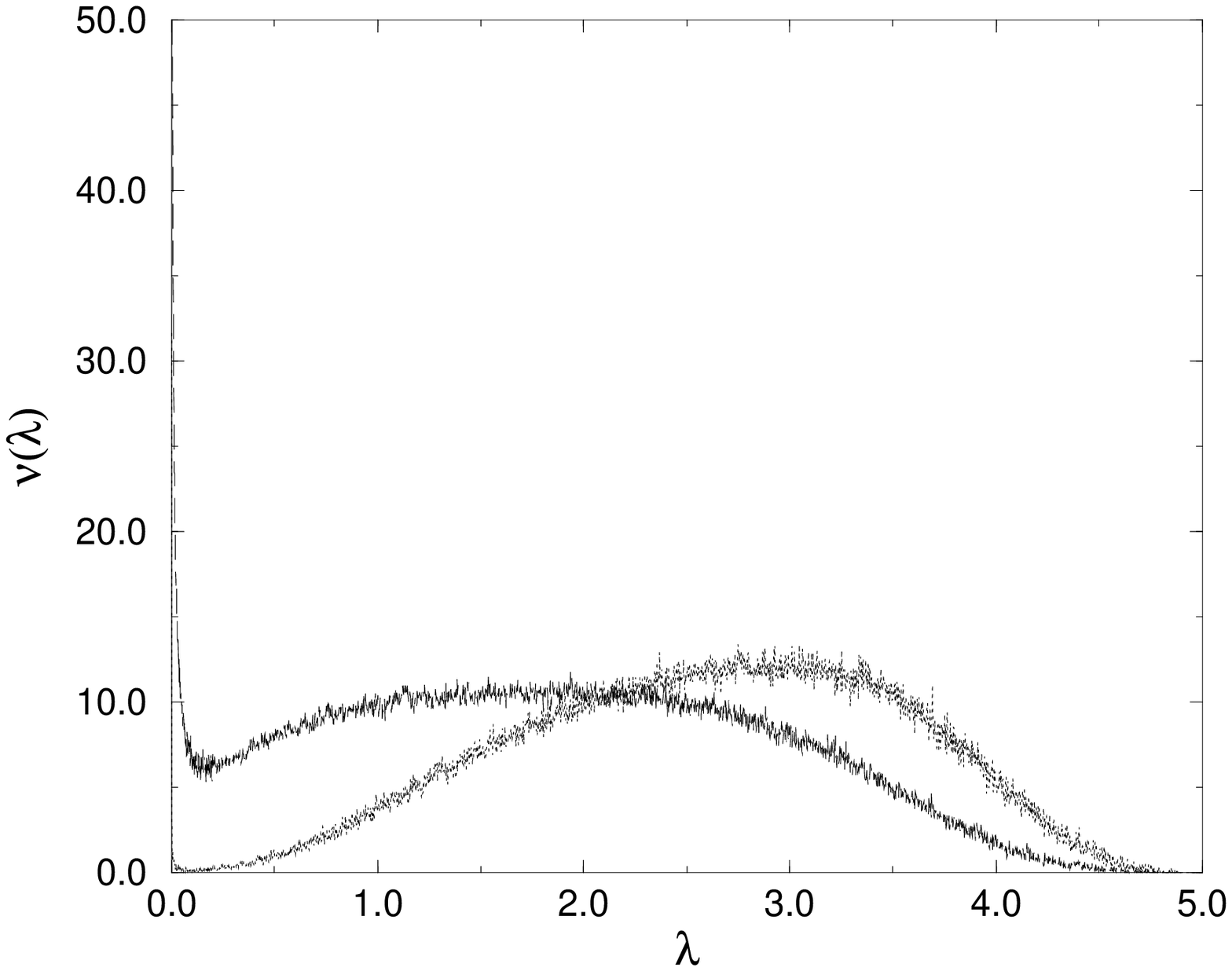}
\end{center}
\caption{$\overline{\lambda}_{NZ}=2.0,\ V=1,\ m=0.15$. A comparison of
the spectral densities obtained for $N_{f}=1$ (dark-solid) and
$N_{f}=2$ (light-dotted).}
\label{fig:sp_un_rm12_cmp}
\end{figure}

\begin{figure}[tb]
\begin{center}
\leavevmode
\epsfxsize=100mm
\epsfbox{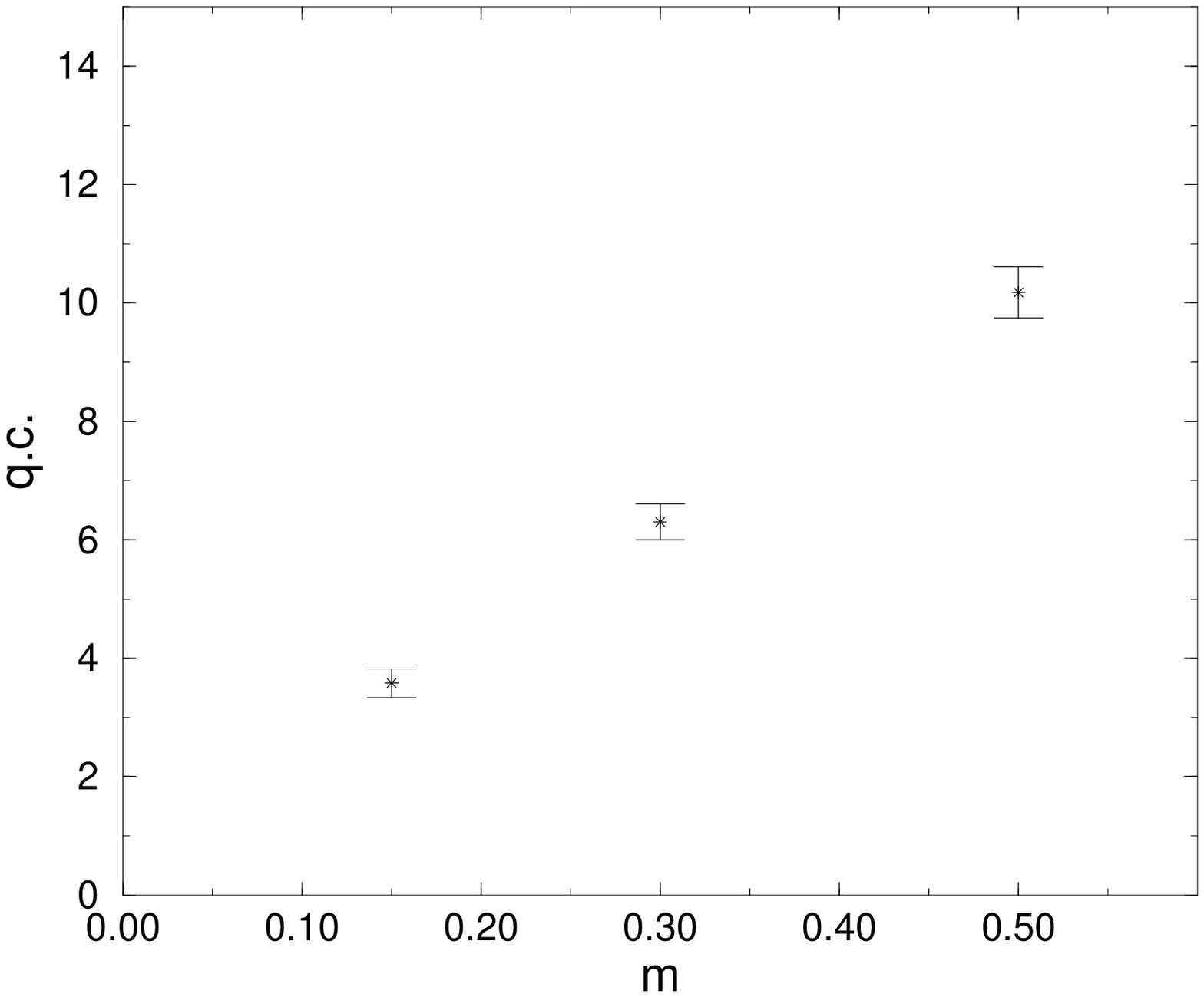}
\end{center}
\caption{The quark condensate \ssi\ as a function of the quark mass
$m$ from the spectra plotted in figure~\ref{fig:sp_un_rm2}.}
\label{fig:qc_rm2}
\end{figure}

\begin{figure}[tb]
\begin{center}
\leavevmode
\epsfxsize=100mm
\epsfbox{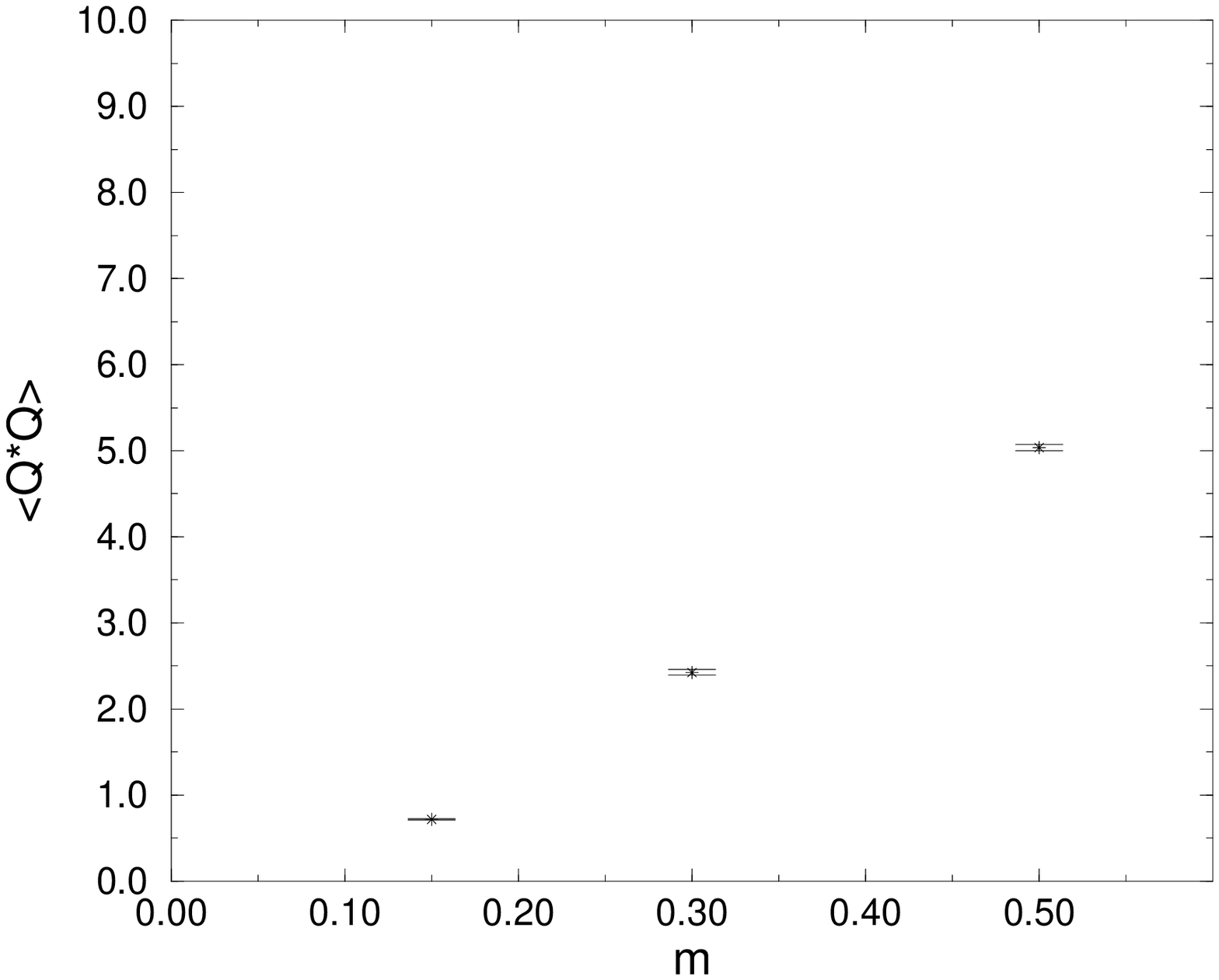}
\end{center}
\caption{$N_{f}=2,\ \overline{\lambda}_{NZ}=2.0,\ V=1$. The second
moment of the winding number distribution $\langle Q^{2}\rangle$ as a
function of the quark mass $m$.}
\label{fig:qq_rm2}
\end{figure}

The spectral density for the three masses is depicted in
figure~\ref{fig:sp_un_rm2}. A quick comparison with
figure~\ref{fig:sp_un_rm1} indicates that something very different is
occuring for two flavours of fermions. Why is the spectral density so
small in magnitude in comparison to the one flavour case~? It is
evident that we are seeing chiral symmetry restoration; the
eigenvalues have been pushed to larger values, and hence, the spectrum
is greatly depleted at the crucial low eigenvalues. This can be seen
explicitly by comparing the spectrums obtained at $m=0.15$ for
$N_{f}=1$ and $N_{f}=2$, as shown in figure~\ref{fig:sp_un_rm12_cmp}.
Whilst we would expect to get chiral symmetry restoration for a
sufficiently high number of fermion flavours, we most certainly do not
expect chiral symmetry restoration for $N_{f}=2$, this is, after all,
the physical world to a good approximation~! We can see the problem
graphically in figure~\ref{fig:qc_rm2} which shows the chiral
condensate decreasing to zero with the quark mass. The fact that we
get the correct quadratic behaviour of the topological susceptibility
for chiral symmetry restoration, is, at best, small comfort (see
figure~\ref{fig:qq_rm2}).

There are a number of possibilities as to where the difficulties
lie. The first is that we are seeing a facet of the critical slowing
down of our algorithm. In other words, it is possible that if we were
to undertake longer runs which sample more of the phase space, or
design a more efficient algorithm, then our results would be
different. Whilst this is a possibility, it is one which we have not
been able to test, mainly due to time constraints. A second (more
likely) possibility, is that we are seeing results of finite size
effects. Recall, we get chiral symmetry restoration if we take the
quark mass to zero in a finite volume. But this is precisely what we
have been doing~! (The tacit assumption being that we would have
needed to go to smaller masses yet, if we were to see a similar effect
for $N_{f}=1$.) If this is so, then instead of being a sign of the
breakdown of our model, it will reaffirm that it seems to capture some
essential properties of the underlying field theory. How can we test
this possibility~? The easiest way would be to work with larger quark
masses; if we are seeing finite size effects, then these should become
less important for larger quark masses and we should recover the
signal of symmetry breakdown.

There are a few problems with implementing this idea however. The
first concerns the concept of a replacement eigenvalue. This should be
larger than the eigenvalues from the would-be zero modes, preferably
far larger than the median such eigenvalue. This ensures that in the
chiral limit, the main contribution to the spectral density is from
the would-be zero modes (hence we concentrate upon them). When we are
working with larger quark masses, our model is still based around the
would-be zero modes. However, the contribution to the spectral density
is no longer given mainly by the would-be zero modes (as all the
eigenvalues are in fact shifted $\lambda \rightarrow (\lambda^{2} +
m^{2})^{1/2}$), and the non-zero modes also contribute. A second
problem is of a more practical nature. It lies with the fact that the
number of objects is a dynamical quantity, furthermore, it is a
dynamical quantity which depends upon the quark
mass. Table~\ref{tab:unq} shows that we get fewer objects as we
increase the quark mass, we have only $\approx 21$ objects of each
chirality per configuration for $N_{f}=2,\ m=0.5$. What is happening
is that the gas is becoming increasing dilute, for larger quark masses
we get increasingly trivial configurations containing very few
objects. (This is especially a problem as we are using hard sphere
wavefunctions so it is easy to get accidental zero eigenvalues and,
hence, zero determinants.) We therefore move to the second type of
ensemble. In this case we fix the total number of objects as described
previously. This alleviates both the conceptual, and, the practical
problem, and, allows us to probe whether the restoration of chiral
symmetry which we are seeing is a finite size effect, or, whether we
face a breakdown of our model.

\subsection{$N_{f}=1,2$. Fixed $N_{T}$.}
\begin{figure}[tb]
\begin{center}
\leavevmode
\epsfxsize=80mm
\epsfbox{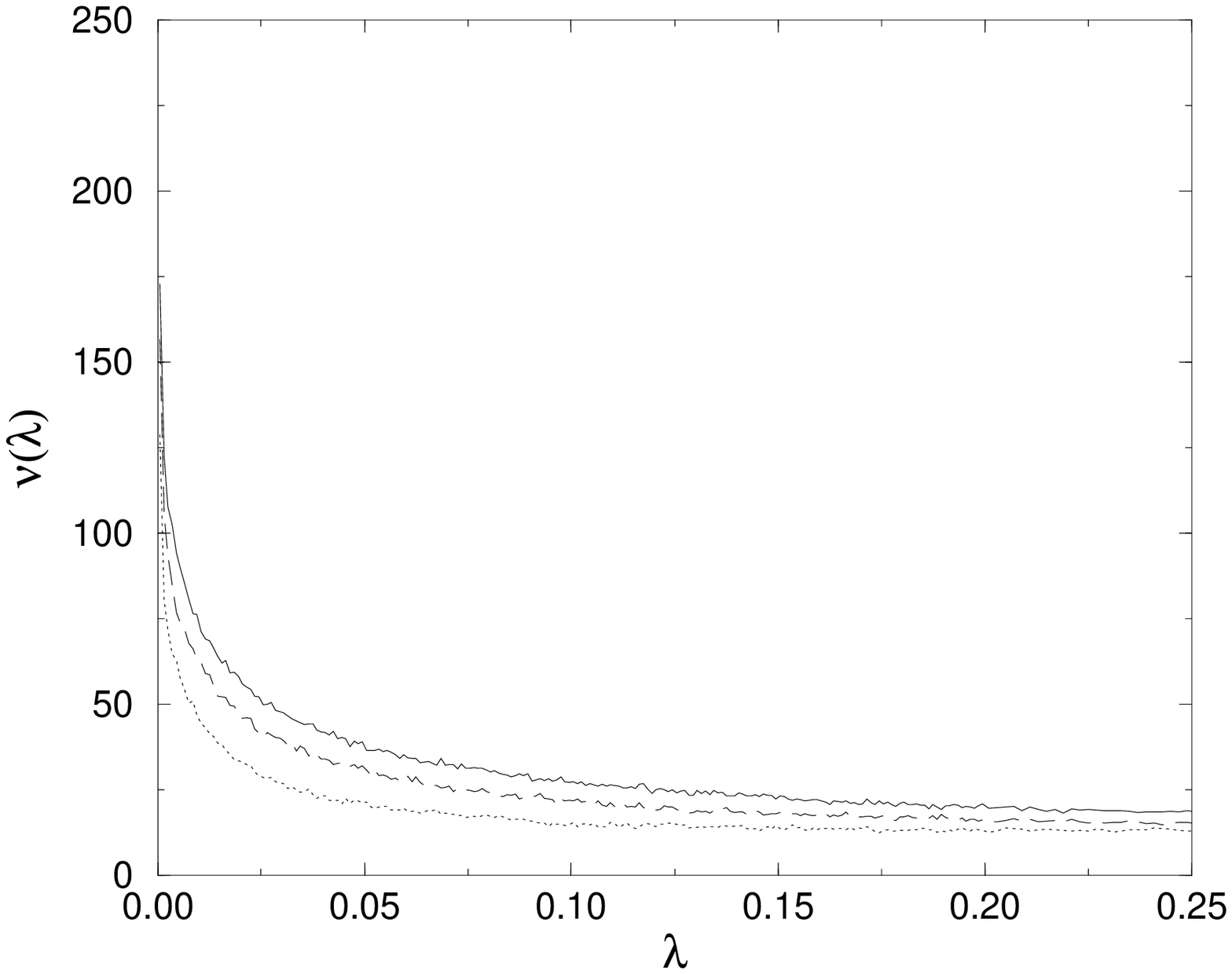}
\end{center}
\caption{$N_{f}=1$, Fixed $N_{T},\ V=1$. The spectral
density for quark mass $m=0.5$ (solid), 0.3 (long dashed) and 0.15
(dotted).}
\label{fig:sp_un_rx1}
\end{figure}

\begin{figure}[tb]
\begin{center}
\leavevmode
\epsfxsize=80mm
\epsfbox{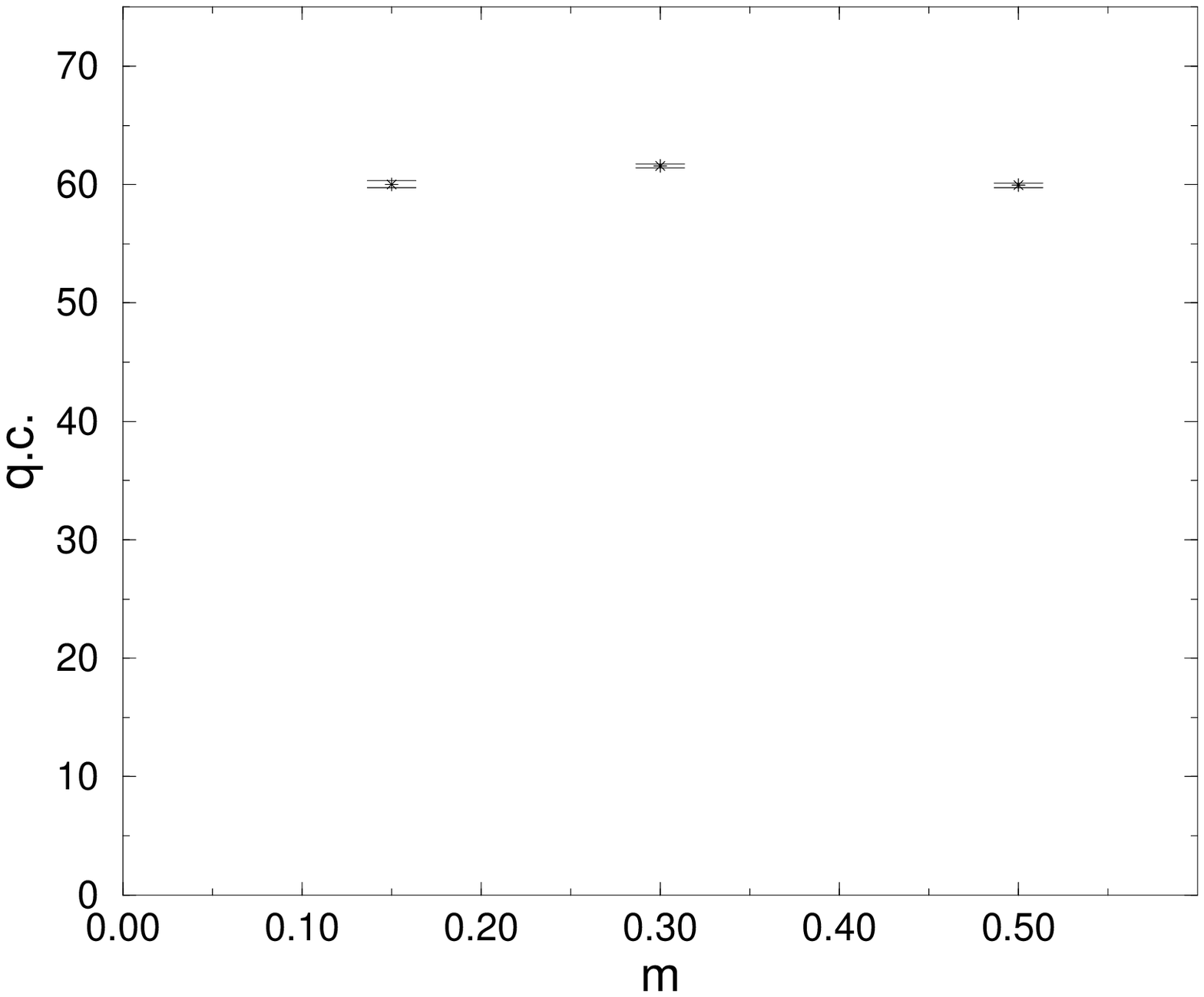}
\end{center}
\caption{$N_{f}=1$, Fixed $N_{T},\ V=1$. The quark condensate \ssi\ as
a function of the quark mass $m$.}
\label{fig:qc_rx1}
\end{figure}

Tables~\ref{tab:unq},~\ref{tab:sp} and~\ref{tab:partmass} indicate
that we get the same qualitative behaviour as before for $m=0.15,\
0.3$ and $m=0.5$ for both $N_{f}=1$ and $N_{f}=2$. In
figure~\ref{fig:sp_un_rx1} we plot the spectral density for $N_{f}=1$
for these quark masses. We see as in the cases where we varied
$N_{T}$, a divergent spectral density (see
figure~\ref{fig:sp_un_rm1}). We plot the quark condensate
corresponding to these spectra in figure~\ref{fig:qc_rx1}, and we see,
as in figure~\ref{fig:qc_rm1}, chiral symmetry breakdown with a finite
quark condensate.
So, for $N_{f}=1$ we get the behaviour we would expect. We study the
more interesting $N_{f}=2$ case in more detail, and, in particular we
simulate at higher quark masses. 
\begin{figure}[tb]
\begin{center}
\leavevmode
\epsfxsize=80mm
\epsfbox{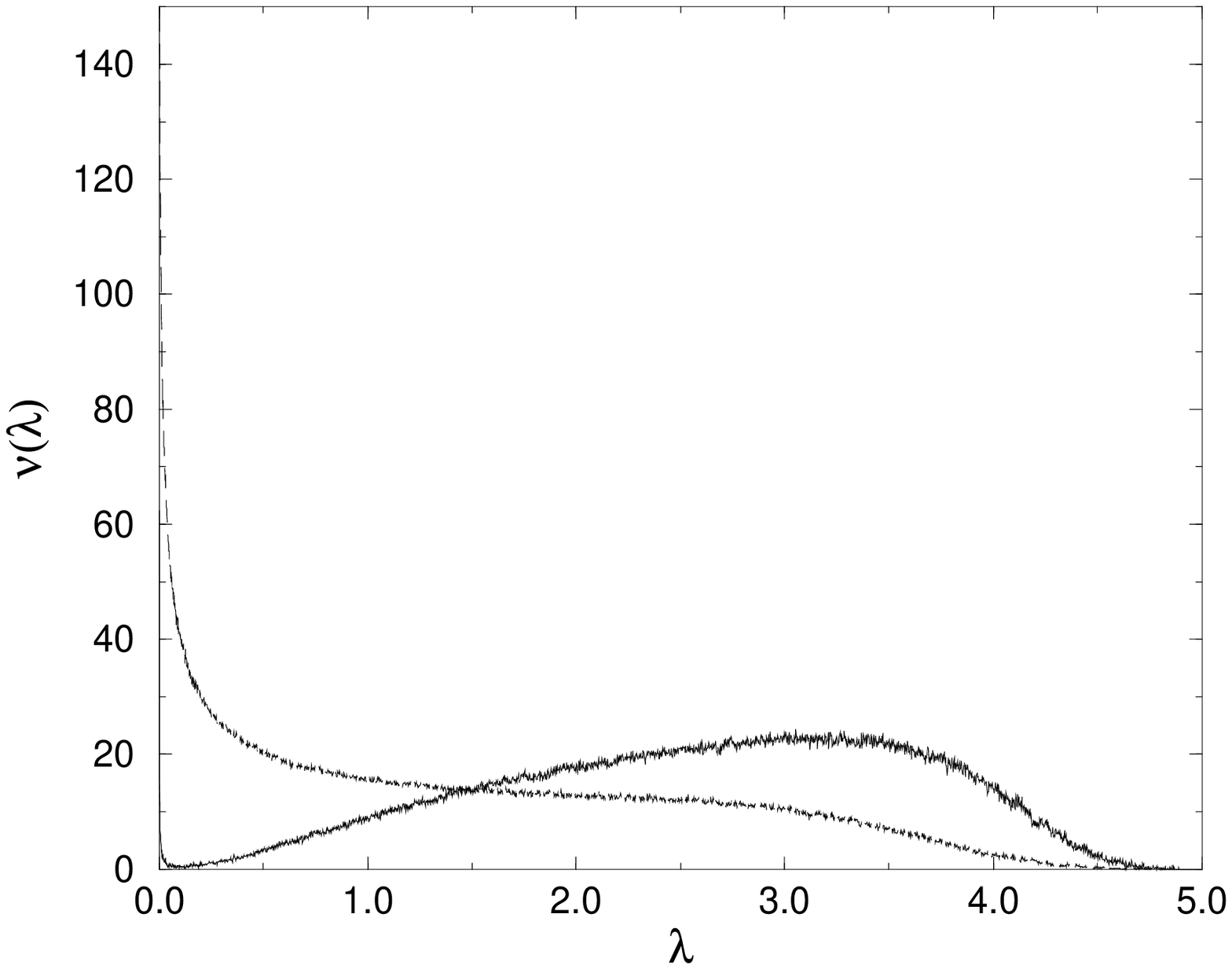}
\end{center}
\caption{$N_{f}=2$, Fixed $N_{T},\ V=1$. The spectral density for
quark mass $m=0.15$ (solid) and $m=3.0$ (long dashed).}
\label{fig:sp_un_rx2}
\end{figure}
Figure~\ref{fig:sp_un_rx2} show the spectra we obtain for $m=0.15$ and
$m=3.0$ respectively. We can immediately see the difference between
the two; the spectra corresponding to the smaller mass shows the
depletion of eigenvalues at small $\lambda$ that we expect to see if
chiral symmetry is to be restored (compare with
figure~\ref{fig:sp_un_rm12_cmp}); the spectra corresponding to $m=3.0$
shows no such depletion.

\begin{figure}[tb]
\begin{center}
\leavevmode
\epsfxsize=80mm
\epsfbox{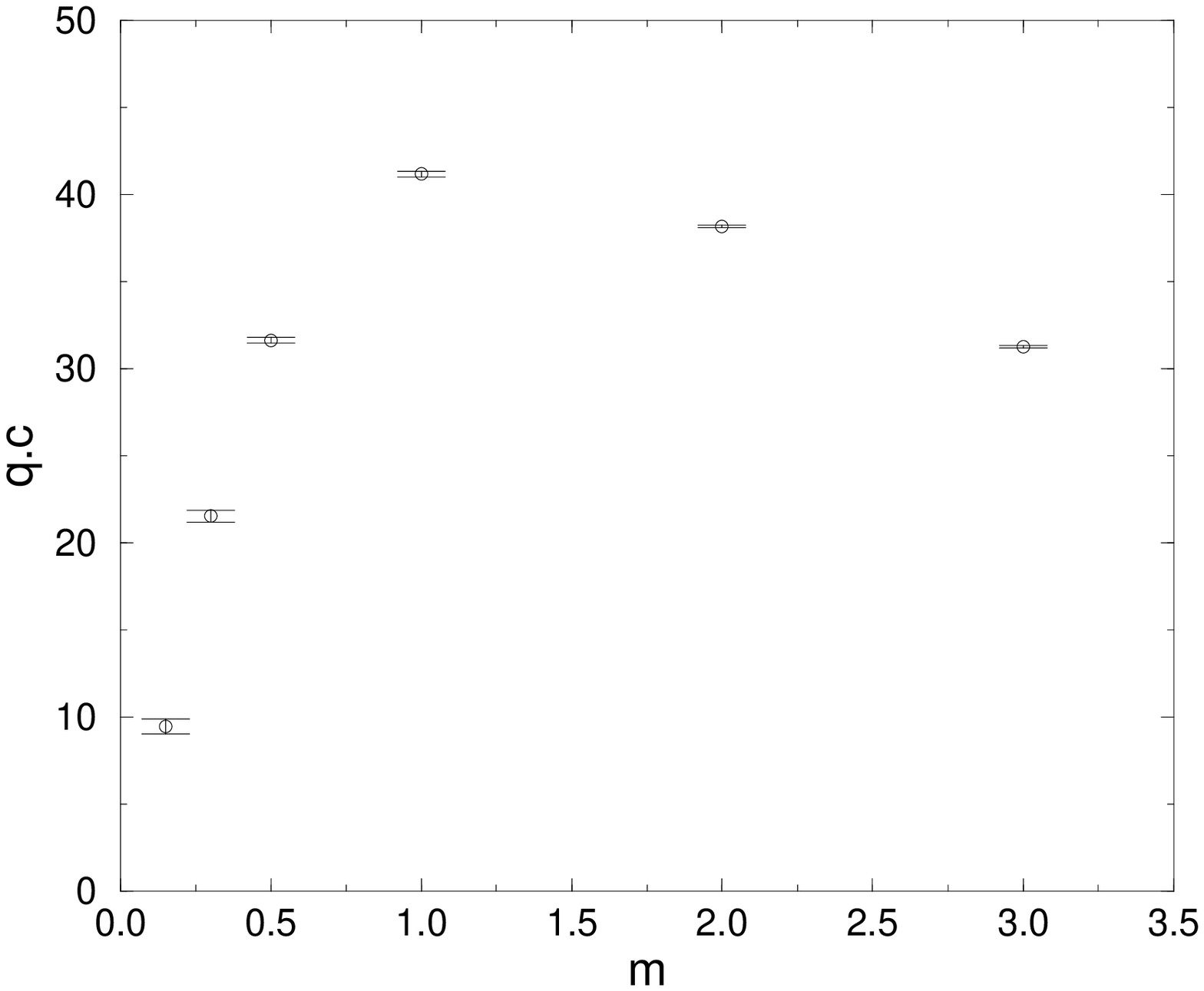}
\end{center}
\caption{$N_{f}=2$, Fixed $N_{T},\ V=1$. The quark condensate \ssi\ as
a function of the quark mass $m$.}
\label{fig:qc_rx2}
\end{figure}
\begin{figure}[tb]
\begin{center}
\leavevmode
\epsfxsize=80mm
\epsfbox{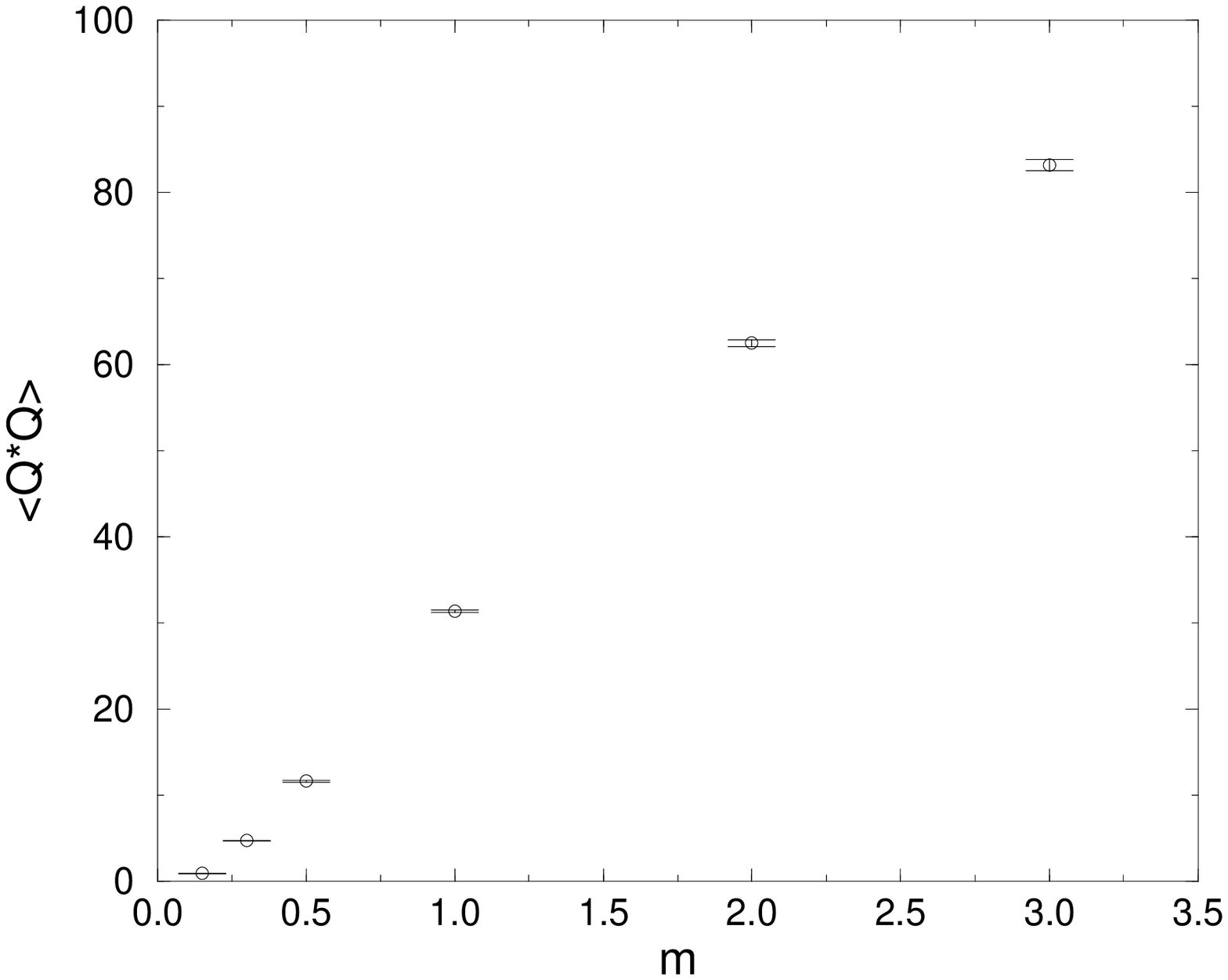}
\end{center}
\caption{$N_{f}=2$, Fixed $N_{T},\ V=1$.  The second moment of the
winding number distribution $\langle Q^{2}\rangle$ as a function of
the quark mass $m$.}
\label{fig:qq_rx2}
\end{figure}

If we integrate the spectra for the various masses then we get
figure~\ref{fig:qc_rx2}. This is a wonderful figure for us, it is
precisely what we had hoped to see. We see at the larger quark masses
the sure sign of chiral symmetry breakdown; as we move from $m=3.0$ to
$m=1.0$, the quark condensate increases, and, a linear extrapolation
leads to a finite non-zero condensate. As we reduce the mass further
however, we get a totally different behaviour, the quark condensate
collapses towards zero, behaviour consistent with the trivial
restoration of chiral symmetry due to finite volume effects. The
corresponding plot of $\langle Q^{2}(m)\rangle$ given in
figure~\ref{fig:qq_rx2} is also pleasing, we see approximately linear
behaviour for the larger quark masses, as we would expect for chiral
symmetry breakdown, and, approximately quadratic behaviour for the
smaller quark masses as before (see~\ref{fig:qq_rm2}).

We even find the behaviour of the $\eta^{'}$ mass to be as before, it
is massive in the chiral limit (see table~\ref{tab:partmass}). We
cannot extract masses so easily for the $\sigma$ as this relies on
correlations of the number of objects in slices through our spacetime,
and, we have now fixed the total number of objects in every
configuration.

\section{Discussion}
We have managed to extract a large amount of physics from our
model. Important results include the fact that we expect a spectral
density which is divergent at finite quark mass even in full QCD. The
parameters of the divergence are mass dependent however, and, we
obtain a finite quark condensate, as we must. We find that chiral
symmetry is broken for $N_{f}=1$ but restored for $N_{f}=2$. We have
strong evidence however, that the restoration is due to finite volume
effects, and, that chiral symmetry is broken within the framework of
our model, even for the $N_{f}=2$ case. Our model further agrees with
predictions of the form for the winding number distribution as a
function of quark mass, both for when chiral symmetry is broken, and,
when it is restored. This is highly non-trivial and most
unexpected. The presence of the fermion determinant generates
non-trivial correlations between objects. Interpreting these
correlations in terms of particles, we obtain masses for the
$\eta^{'}$ and the $\sigma$ which remain massive in the chiral limit.

\begin{table}[tbh]
\begin{center}
\begin{tabular}{|c|c|c|l|r|l|c|}
\hline
Set & $\overline{\lambda}_{NZ}$ & $N_{f}$ & $m$ & $\langle Q\rangle$ & $\langle
Q^{2}\rangle$ & $\langle N_{I}\rangle$\\
\hline
A & 2.0 & 1 & 0.5 & 0.152 & 25.65$\pm$0.45 & 35.53\\
B & 2.0 & 1 & 0.3 & -0.078 & 14.28$\pm$0.16 & 34.91\\
C & 2.0 & 1 & 0.15 & -0.209 & 8.13$\pm$0.04 & 35.75\\
D & 2.0 & 2 & 0.5 & -0.014 & 5.04$\pm$0.04 & 21.35\\
E & 2.0 & 2 & 0.3 & -0.059 & 2.43$\pm$0.03 & 23.17\\
F & 2.0 & 2 & 0.15 & -0.016 & 0.72$\pm$0.01 & 30.85\\
G & - & 1 & 3.0 & -0.403 & 98.615$\pm$0.78 & 62.80\\
H & - & 1 & 2.0 & 0.061 & 80.097$\pm$0.92 & 63.03\\
I & - & 1 & 1.0 & -0.111 & 60.383$\pm$0.49 & 62.95\\
J & - & 1 & 0.5 & -0.046 & 34.22$\pm$0.26 & 62.98\\
K & - & 1 & 0.3 & -0.015 & 22.00$\pm$0.19 & 62.99\\
L & - & 1 & 0.15 & 0.027 & 11.66$\pm$0.07 & 63.01\\
M & - & 2 & 3.0 & 0.139 & 83.20$\pm$0.64 & 63.07\\
N & - & 2 & 2.0 & -0.193 & 62.49$\pm$0.40 & 62.90\\
O & - & 2 & 1.0 & -0.147 & 31.37$\pm$0.13 & 62.93\\
P & - & 2 & 0.5 & -0.033 & 11.61$\pm$0.10 & 62.98\\
Q & - & 2 & 0.3 & -0.089 & 4.71$\pm$0.05 & 62.96\\
R & - & 2 & 0.15 & -0.043 & 0.89$\pm$0.02 & 62.98\\
\hline
\end{tabular}
\end{center}
\caption{Some information about the synthetic ensembles analysed in
this chapter. A ``-'' indicates that the parameter is inapplicable, in
this case, the sets are with fixed $N_{T}$ and hence, do not require a
replacement eigenvalue $\overline{\lambda}_{NZ}$.}
\label{tab:unq}
\end{table}

\begin{table}[tbh]
\begin{center}
\begin{tabular}{|c|c|c|c|}
\hline
Set & b & d & $\chi^{2}/N_{DF}$\\
\hline
A & 2.904$\pm$0.158 & 0.668$\pm$0.008 & 1.09\\
B & 1.956$\pm$0.181 & 0.699$\pm$0.017 & 0.94\\
C & 0.491$\pm$0.058 & 0.861$\pm$0.022 & 2.41\\
D & 0.228$\pm$0.031 & 0.882$\pm$0.021 & 1.15\\
E & 0.043$\pm$0.009 & 0.979$\pm$0.032 & 1.44\\
F & - & - & -\\
G & 30.917$\pm$3.324 & 0.300$\pm$0.018 & 0.99\\
H & 28.766$\pm$2.058 & 0.307$\pm$0.013 & 1.35\\
I & 36.517$\pm$4.636 & 0.252$\pm$0.019 & 1.41\\
J & 24.048$\pm$2.936 & 0.288$\pm$0.019 & 2.05\\
K & 17.599$\pm$2.632 & 0.309$\pm$0.023 & 3.27\\
L & 4.395$\pm$0.586 & 0.478$\pm$0.024 & 4.20\\
M & 30.410$\pm$3.919 & 0.296$\pm$0.020 & 1.17\\
N & 30.990$\pm$2.520 & 0.285$\pm$0.014 & 1.57\\
O & 17.665$\pm$1.714 & 0.343$\pm$0.017 & 1.41\\
P & 4.404$\pm$0.267 & 0.519$\pm$0.013 & 1.99\\
Q & 0.580$\pm$0.051 & 0.788$\pm$0.019 & 1.55\\
R & 0.006$\pm$0.002 & 1.307$\pm$0.043 & 2.30\\
\hline
\end{tabular}
\end{center}
\caption{The spectral densities for the ensembles given in
table~\ref{tab:unq}. A ``-'' indicates that a satisfactory fit was not
possible.}
\label{tab:sp}
\end{table}

\begin{table}[tbh]
\begin{center}
\begin{tabular}{|c|c|c|c|c|}
\hline
Set & $\eta^{'}$ & $\chi_{\eta^{'}}^{2}/N_{DF}$ & $\sigma$ &
$\chi_{\sigma}^{2}/N_{DF}$\\
\hline
A & 11.180$\pm$0.466 & 10.6 & 11.589$\pm$2.612 & 0.566\\
B & 11.880$\pm$0.600 & 14.61 & 11.083$\pm$3.149 & 0.229\\
C & 12.271$\pm$0.548 & 15.00 & 9.426$\pm$2.673 & 1.44\\
D & 18.849$\pm$0.775 & 5.94 & 14.691$\pm$3.589 & 0.640\\
E & 18.904$\pm$0.892 & 8.01 & 19.011$\pm$3.291 & 0.708\\
F & 19.265$\pm$0.873 & 8.75 & 29.955$\pm$3.592 & 5.899\\
G & 9.503$\pm$2.328 & 0.55 & - & -\\
H & 6.618$\pm$0.908 & 2.10 & - & -\\
I & 11.147$\pm$0.490 & 4.59 & - & -\\
J & 11.313$\pm$0.589 & 12.40 & - & -\\
K & 10.969$\pm$0.278 & 14.51 & - & -\\
L & 11.620$\pm$0.290 & 16.44 & - & -\\
M & 11.895$\pm$1.275 & 1.524 & - & -\\
N & 12.642$\pm$0.951 & 3.405 & - & -\\
O & 13.780$\pm$0.454 & 7.900 & - & -\\
P & 15.190$\pm$0.465 & 11.17 & - & -\\
Q & 15.884$\pm$0.518 & 8.54 & - & -\\
R & 17.359$\pm$0.665 & 11.40 & - & -\\
\hline
\end{tabular}
\end{center}
\caption{Particle masses derived from ensembles given in
table~\ref{tab:unq}. $\chi_{\eta^{'}}^{2}/N_{DF}$ refers to the
chi-square of the exponential fit to the eta-correlation function,
$\chi_{\sigma}^{2}/N_{DF}$ is the corresponding fit to the
sigma-correlation function. A ``-'' indicates that a mass could not be
extracted (see text).}
\label{tab:partmass}
\end{table}

%% file: conc.tex
\chapter{Conclusions}
\label{ch:conc}

We established a framework which allowed us to construct a
representation of the Dirac operator for a given configuration of
instantons. The representation operated on the subspace of the Hilbert
space spanned by the zero modes from the individual objects. We also
approximated the Dirac operator by a simpler structure which we felt
kept the bare essentials of the underlying theory. This has been a
recurring idea throughout this work, to strip out as many details as
possible whilst maintaining the symmetries and certain other
properties of the underlying theory (the results in certain limits
etc.).

It is currently not possible to carry out the simulations which could
verify parts of this work. Some of the difficulties are conceptual
(for instance, how can we disentangle instanton effects from other
effects, such as the confinement mechanism, when we deal with the full
field theory), others are practical (limitations of algorithms and
computer power). However, we tested our model for ``qualitative
universality'' and found this to hold: we get similar results
regardless of the details of the type of wavefunction we used and the
exact ansatz for the presence of the Dirac operator. A direct
calculation on a lattice would be preferable, but our more limited
checks at least satisfy the ``necessary'' requirement for making
predictions for quenched and full QCD.

Whilst the approximations required to proceed with this work have
effectively precluded quantitative predictions, we have been able to
discern a number of qualitative features which are of interest. We
have found strong evidence for chiral symmetry breaking. It seems that
generic instanton configurations (which have no other dynamical
effects at all - each configuration is random) are enough to break
chiral symmetry. We have further seen that a divergence in the
spectral density is almost ubiquitous for instanton gases. This leads
to the prediction of a divergence in the chiral condensate in quenched
QCD. We do not see such an effect from direct calculations of the
Dirac spectra on lattice gauge configurations, presumably because the
spectrum near $\lambda \rightarrow 0$ is distorted by lattice
artefacts. We have found that the divergence can be parameterized for
small $\lambda$ by a power law $\overline{\nu}(\lambda) = a +
b\lambda^{-d}$. The power of the divergence $d$ is inversely related
to the packing fraction of the gas, in particular the divergence is
negligible for dense gases (though admittedly our methodology is
doubtful in this limit). It is therefore possible, even in quenched
QCD, for a dense gas of objects to break chiral symmetry and have a
finite condensate. A detailed analysis found evidence that the
divergence was not as a result of isolated barely overlapping dipoles
of instanton and anti-instantons, but was in fact the result of
complicated many body effects.

We found evidence of a ``separation of scales'' between large and
small objects co-existing in a gas. A simple argument indicated that
it was possible for the small objects in a gas to drive a divergent
peak, even if they were overlapping with larger objects. This provides
a mechanism whereby some of the successful phenomenological models,
which require relatively low densities of objects (such as the
instanton liquid model described in~\cite{Shuryak-RMP}) can be
reconciled with lattice calculations indicating a far higher packing
fraction~\cite{Smith}. We also found evidence for the screening of
topological charge in the $SU(3)$ gauge theory, and, for scaling of
spectra (if we altered the number of cools and $\beta$, the lattice
coupling, in an appropriate fashion). An important ``negative result''
of our work is the conclusion that the Dirac spectrum for small
$\lambda$ is strongly dependent upon the number of cooling sweeps
undertaken to move from a ``hot'' lattice configuration to a
collection of overlapping instantons. This leads to real ambiguities
in the interpretation of the lattice data, how long should we cool
for, should we extrapolate back to zero cools, etc.~? Some of these
questions have been addressed in an interesting paper
recently~\cite{Ringwald} but undoubtedly more work needs to be done in
this area. This result is in marked contrast to the na\"{\i}ve hope
that the spectral density, like for instance the topological
susceptibility, would be relatively invariant to the local cooling
process.

We have been able to generate ensembles of objects which incorporate
both a gauge and a fermion weighting. These ensembles are our
equivalent of full QCD instanton ensembles. We have found evidence
that chiral symmetry is broken, and as before, we have a power law
divergence for small $\lambda$. This would have been disastrous for
our model (as we know QCD is a proper field theory and is certainly
not pathological) if it wasn't for the result that the divergence is
quark mass dependent. In fact the spectra change in such a way that we
obtain chiral symmetry breaking with a perfectly acceptable quark
condensate.  We believe this to be novel result which certainly merits
further investigation.  We found these results for both $N_{f}=1$ and
$N_{f}=2$ (though we found strong evidence of finite size effects
entering into our calculation for $N_{f}=2$ for our lowest quark
masses). We found the correct behaviour for the topological
susceptibility for both the broken phase (linear with respect to the
quark mass), and, the symmetric phase (where restoration was due to
finite size effects). Remarkably we have also been able to extract
masses corresponding to the $\eta^{'}$ and the $\sigma$. We have found
both of these to be massive in the chiral limit.

%% file: ovlap.tex
\chapter{Calculation of overlap integrals.}
\label{app:ovlap}

\section{Hard Sphere}
The wavefunctions are given by:

\begin{equation}
\begin{array}{lcll}
\langle x|\psi_{i}\rangle & = & 1 & \quad |x - x_{i}| \leq r_{i}\\
& = & 0 & \quad {\rm otherwise ,}
\end{array}
\end{equation}
where $i = 1,2$ labels the two objects in question. We wish to compute
the integral:

\begin{equation}
\langle\psi_{1}|\psi_{2}\rangle = \int_{\mathbb M}\
d^{4}x\langle\psi_{1}|x\rangle\langle\psi_{2}|x\rangle ,
\end{equation}
where ${\mathbb M}$ is either ${\mathbb T}^{4}$ or ${\mathbb
R}^{4}$. So we wish to calculate the volume of intersection of the two
spheres. For simplicity, let us first consider the problem for
${\mathbb R}^{4}$. In this case we work out the separation $s$ of the
objects in the usual Euclidean fashion. The two trivial cases are
where $s > r_{1} + r_{2}$, in which case the intersection is zero, or
where $s \leq |r_{1} - r_{2}|$, in which case one sphere is inside the
other and the intersection is simply the volume of the smaller
$(\pi^{2}/2)\min(r_{1},r_{2})^{4}$.
\begin{figure}[tb]
\begin{center}
\leavevmode
\epsfxsize=80mm
\epsfbox{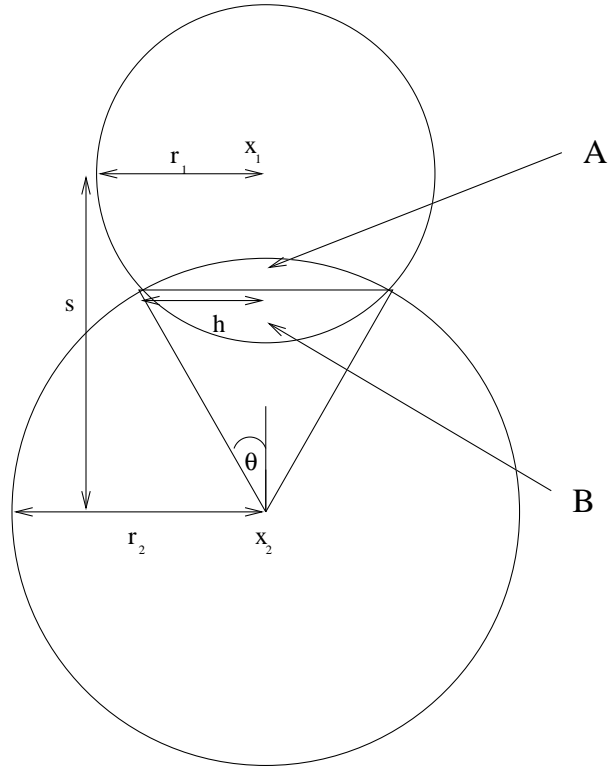}
\end{center}
\caption{Intersection of two spheres in ${\mathbb R}^{4}$.}
\label{sph_int}
\end{figure}
The non-trivial case $|r_{1}-r_{2}| < s \leq r_{1}+r_{2}$ is depicted
in figure~\ref{sph_int} and its evaluation is a simple excercise in
geometry. It consists of adding up the volumes $A$ and $B$. We will
calculate the volume of $A$, as the calculation of $B$ is similar. The
angle $\theta$ as shown in figure~\ref{sph_int} is given by:

\begin{equation}
\cos\theta = \frac{s^{2} + r_{2}^{2} - r_{1}^{2}}{2r_{2}s}\ .
\end{equation}
We first calculate the volume of the cone which originates at $x_{2}$:

\begin{equation}
\int_{0}^{r_{2}}dr\int_{0}^{2\pi}d\phi\int_{0}^{\theta}d\theta_{2}\int_{0}^{\pi}d\theta_{1}
r^{3}\sin^{2}\theta_{2}\sin\theta_{1} = \frac{r_{2}^{4}\pi}{2}\left(\theta
- \frac{sin(2\theta)}{2}\right)
\label{coneseg}
\end{equation}
The volume $A$ is given by~\ref{coneseg} take away the volume of the
right-angled cone which originates at $x_{2}$. The volume of the
right-angled cone is simply
$(\pi/3)r_{2}^{4}\cos\theta\sin^{3}\theta$. Therefore:

\begin{eqnarray}
Vol.(A) & = & V(\theta)\nonumber\\
& = & \frac{r_{2}^{4}\pi}{2}\left(\theta -
\sin2\theta\left(\frac{1}{2} + \frac{1}{3}\sin^{2}\theta\right)\right)
\end{eqnarray}
The volume of our intersection is given by $V(\theta) + V(\omega)$,
where $\omega$ is defined analogously to $\theta$.

The overlap when the space is ${\mathbb T}^{4}$ is only slightly more
complicated. In this work all objects obey the condition $r_{i} < l/2\
\forall i$, where $l$ is the length of the side of the periodic
box. This simplifies matters considerably (and also make sense from a
physics viewpoint as otherwise we would be almost certain to be
suffering from finite size effects). A little thought will confirm
that the following algorithm works for the case of the periodic box
(for simplicity consider the unit periodic box - the generalisation is
obvious).

Consider the overlap of two objects, A and B in a periodic box. Do
nothing with object A. Consider the set of 81 objects $B_{S},\
S=1,\ldots,81$ given by translating object B to all unit boxes
adjacent to the original i.e. $x_{B} \rightarrow x_{B} + v$ where the
components of the shift vector $v^{i} \in \{-1,0,1\}$. The overlap
integral is given by

\begin{equation}
\langle\psi_{A}|\psi_{B}\rangle = \sum_{S=1}^{81}Vol.(A \cap B_{S}) ,
\end{equation}
where the volume of intersection is calculated in ${\mathbb R}^{4}$ as
above.

\section{Gaussian}
We wish to construct a smooth function which is periodic with period
the unit box (again for simplicity). We do so by first defining:
\begin{eqnarray}
G(x;x_{j}^{\pm},\sigma_{j}^{\pm},l) & = &
\frac{1}{\sqrt{2\pi}\sigma_{j}^{\pm}}\exp\left(\frac{-(x-x_{j}^{\pm}-l)^{2}}{2\sigma_{j}^{\pm2}}\right).
\end{eqnarray}
where $x_{j}^{\pm}$ lies in the unit four box with $l \in {\mathbb
Z}^{4}$. The Gaussian zero mode wavefunction is given by:

\begin{eqnarray}
\langle x|\psi_{j}^{\pm}\rangle & = &
N\sum_{l \in {\mathbb Z}^{4}}G(x;x_{j}^{\pm},\sigma_{j}^{\pm},l),
\end{eqnarray}
where $N$ is a suitable normalization constant. If we now wish to
integrate this function on ${\mathbb T}^{4}$ then all we do is
integrate over the unit box. It is relatively simple to perform this
integral, the results being:

\begin{equation}
\int_{{\mathbb T}^{4}} d^{4}x
G(x;x_{1},\sigma_{1})G(x;x_{2},\sigma_{2}) =
\frac{1}{2\pi^{2}(\sigma_{1}^{2}+\sigma_{2}^{2})}\sum_{m_{\mu} \in
{\mathbb Z}}
\exp\left(\frac{(x_{1}-x_{2}+m)^{2}}{2(\sigma_{1}^{2}+\sigma_{2}^{2})}\right)
\end{equation}
This is an infinite series which does not have an obvious analytical
solution (though it certainly converges). In practice we truncate the
series after some number of terms, we have found for $\sigma \leq 0.2$
that $m_{\mu} \in \{-1,0,1\}$ leads to negligible errors.

\section{Classical zero mode}
This wavefunction is defined by:

\begin{equation}
\langle x|\psi_{i}\rangle = \frac{\sqrt{2}}{\pi}\frac{\rho_{i}}{(\rho_{i}^{2}
+ (x-x_{i})^{2})^{3/2}}
\end{equation}
We wish to evaluate the inner product of two such functions over
${\mathbb R}^{4}$. This can be done by the standard method of Feynman
parameters to yield the following integral solution:

\begin{equation}
\langle\psi_{i}\psi_{j}\rangle =
\frac{8}{\pi}\frac{\rho_{i}}{\rho_{j}}\int_{0}^{1}dq\frac{\sqrt{q(1-q)}}{1
+ q(\frac{s^{2}}{\rho_{j}^{2}} + \frac{\rho_{i}^{2}}{\rho_{j}^{2}}) -
q^{2}\frac{s^{2}}{\rho_{j}^{2}}} .
\end{equation}
The lack of a closed form expression for this overlaps makes the
calculation far slower than when we use the hard sphere (or even the
Gaussian) wavefunction. One can indeed evaluate this integral in terms
of the hypergeometric function of two variables, but the quickest way
of evaluating this function is to calculate the above integral~!

%% file: jackk.tex
\chapter{Error estimation \& Best Fits}
\label{app:jk}

\section{Error estimation}
All errors quoted in this work have been calculated using
``jack-knife'' methodology. This is similar to conventional methods
except more robust when data is limited (as for instance in the case
of configurations of instantons derived from lattice data). The idea
is as follows:

\begin{itemize}
\item{Split up the data into $N$ sets $d_{1},\ldots,d_{N}$, each of
which containts $(N-1)/N$ of all the data.}
\item{Calculate $N$ estimates of the desired quantity
$\Phi_{1},\ldots,\Phi_{N}$, where $\Phi_{i}$ uses the data from
$d_{i}$. So for instance, consider the case of $N=10$ and the example
of calculating the spectral density at a point $\lambda =
\lambda_{0}$. In this case we would get $10$ estimates of the spectral
density, each of which would use nine-tenths of the available
data. The most important thing to note about these estimates is that
they are not independent.}
\item{An unbiased estimate of the mean is given by the usual formula:
\begin{equation}
\hat{\mu} = \frac{1}{N}\sum_{i=1}^{N}\Phi_{i}
\end{equation}}
\item{An unbiased estimate of the variance is again given by the usual
formula:
\begin{equation}
\hat{s}^{2} = \frac{1}{N-1}\sum_{i=1}^{N}(\Phi_{i}-\hat{\mu})^{2}
\end{equation}}
\item{The distribution the mean is different however, the variance of
the distribution is not given by $\hat{s}^{2}/N$ but by
$N\hat{s}^{2}$. So the error estimate is given as:
\begin{equation}
\hat{\mu} \pm N\hat{s}^{2} .
\end{equation}}
\end{itemize}

The advantage of this method is really in its robustness. If data is
limited, for example, then the usual method of splitting data into $N$
sets each of which is independent and contains $1/N$ of the total data
become unstable. (Consider the lattice data with 50 configurations in
total: what sort of spectrum could we obtain is we used only 5
configurations as opposed to 45~?)

\section{Best Fits}
We have a set of numerical data $\{x_{j},E(x_{j}),\sigma(x_{j})\}$
where $x_{j}$ denotes the points, $E(x_{j})$ denotes the values
obtained and $\sigma(x_{j})$, the errors around those values. Consider
a fit function $f(x;\alpha_{i})$ defined by a set of $K$ parameters
$\{\alpha_{i},\ i=1,\ldots,K\}$. The ``best fit'' is given by finding
the parameters $\{\alpha_{i}^{B}\}$ which minimize the $\chi^{2}$ of
the fit:

\begin{equation}
\chi^{2}(\alpha_{1},\ldots,\alpha_{K}) = \sum_{j=1}^{N}
\frac{(f(x_{j};\alpha_{i}) - E(x_{j}))^{2}}{\sigma_{j}^{2}}
\end{equation}
There is no general procedure for finding these best parameters. When
we are dealing with fit functions which are non-linear (in the
parameters), such as the power law fit function~\ref{power-law}, then
the problem is not simple at all. Inevitably one uses a ``canned
package'', the one used in this thesis is the standard
Levenberg-Marquardt method. These do not find the absolute minima of
the $\chi^{2}$ function, they do find {\em a} minima, at least most of
the time.

In practice, the quality of the data for the ``synthetic''
configurations is so high that the fitting program is robust. The
situation is not so good for the lattice data. One way to test how
good the fit is, is if one mutiplies all the $\sigma_{j}$ by a
constant factor $\sigma_{j} \rightarrow c\sigma_{j}$. If one carries
out the fit again, then the resultant parameters should be identical,
only the $\chi^{2}$ should be altered, $\chi^{2} \rightarrow
\chi^{2}/c^{2}$. This is always so for the synthetic data (to within
hundreths of a percent), but not for the lattice data: the algorithm
has wandered to a different minima in parameter space. These
difficulties are to some extent inevitable if one is forced to use
non-linear fits with limited data. We rely on comparisons with the
synthetic data to give us confidence about our results for the lattice
data.